\documentclass[pra,superscriptaddress,showpacs,papersize=a4paper,twocolumn,floatfix]{revtex4-1}

\usepackage[pdftex]{graphicx}
\usepackage{bm}
\usepackage{etoolbox}
\usepackage[usenames,dvipsnames]{color}
\usepackage{mathtools}
\usepackage{amsmath}
\usepackage{amsfonts}
\usepackage{amssymb}

\DeclarePairedDelimiter\ket{\lvert}{\rangle}
\DeclarePairedDelimiterX\braket[2]{\langle}{\rangle}{#1 \delimsize\vert #2}

\newcommand{\bg}{ \begin{gather} }
\newcommand{\eg}{\end{gather}}
\newcommand{\be}{ \begin{equation} }
\newcommand{\ee}{\end{equation}}
\newcommand{\bea}{ \begin{eqnarray} }
\newcommand{\eea}{\end{eqnarray}}

\def\Tr{\mathop{\rm Tr}}

\renewcommand{\Im}{\mathop{\rm Im}}

\AtEndEnvironment{thebibliography}{}
\begin{document}

\title{Eigenstate correlations around many-body localization transition}

\author{K.\,S.~Tikhonov}
\affiliation{Skolkovo Institute of Science and Technology, Moscow, 121205, Russia}
\affiliation{L.\,D.~Landau Institute for Theoretical Physics RAS, 119334 Moscow, Russia}

\author{A.\,D.~Mirlin}
\affiliation{Institute for Quantum Materials and Technologies, Karlsruhe Institute of Technology, 76021 Karlsruhe, Germany}
\affiliation{Institute for Condensed Matter Theory, Karlsruhe Institute of Technology, 76128 Karlsruhe, Germany}
\affiliation{L.\,D.~Landau Institute for Theoretical Physics RAS, 119334 Moscow, Russia}
\affiliation{Petersburg Nuclear Physics Institute,188300 St.\,Petersburg, Russia.}

\begin{abstract}
We explore correlations of eigenstates around the many-body localization (MBL) transition in their dependence on the energy difference (frequency) $\omega$ and disorder $W$. 
In addition to the genuine many-body problem, XXZ spin chain in random field, we consider localization on random regular graphs (RRG) that serves as a toy model of the MBL transition. Both models show a very similar behavior. On the localized side of the transition, the eigenstate correlation function $\beta(\omega)$ shows a power-law enhancement of correlations with lowering $\omega$; the corresponding exponent depends on $W$. The correlation between adjacent-in-energy eigenstates exhibits a maximum at the transition point $W_c$, visualizing the drift of $W_c$ with increasing system size towards its thermodynamic-limit value. The correlation function $\beta(\omega)$ is related, via Fourier transformation, to the Hilbert-space return probability. We discuss measurement of such (and related) eigenstate correlation functions on state-of-the-art quantum computers and simulators. 
\end{abstract}
\maketitle


\section{Introduction}
\label{s1}

The many-body localization (MBL) in disordered interacting systems \cite{gornyi2005interacting,basko2006metal}  is one of active directions of the modern condensed-matter physics research (see recent reviews \cite{abanin2019colloquium,gopalakrishnan2020dynamics}). The MBL is of fundamental importance as it can break ergodicity and suppress low-temperature transport in a great variety of complex systems. The prediction of the MBL transition \cite{gornyi2005interacting,basko2006metal} has been corroborated by numerous subsequent analytical and computational studies, see, in particular, Refs.~\onlinecite{oganesyan07,monthus10,kjall14,gopalakrishnan14,luitz15,nandkishore15,karrasch15,imbrie16,imbrie16a,gornyi-adp16,Abanin2017a,Doggen2018a,Thiery2017a, Alet2018a,Dumitrescu2019a, Goremykina2019a}. While the MBL can be destabilized in the thermodynamic limit (system size $L \to \infty$, with other parameters fixed) due to long-range interactions \cite{burin06,Demler14,Burin15,Gutman16,tikhonov18,safavi-naini19,Gopalakrishnan2019a}, spatial dimensionality $d > 1$ \cite{Gopalakrishnan2019a,doggen2020slow}, or continuum character of the model \cite{nandkishore14continuum,gornyi17continuum},  the MBL transition is well defined also in these cases but with critical disorder $W_c(L)$ depending on the system size.  
On the experimental side, the evidence of the MBL transition was reported and the associated physics was studied in a variety of structures.
These include systems of cold atoms and ions in optical traps\cite{schreiber2015observation,choi2016exploring,kondov2015disorder,smith2016many,bordia15,lueschen16,Zhang17,Bordia2017a,Rubio-Abadal2019a}, of spin defects in a solid state\cite{Choi17a,Choi16,choi2017observation,Ho17,choi2017observation}, and of superconducting qubits\cite{roushan17,Chiaro2019a}, as well as InO films~\cite{ovadyahu1,ovadyahu2,ovadia2015evidence}.

The MBL can be viewed as an extension of the Anderson localization \cite{anderson58} from single-particle to many-body setting. Correspondingly, the MBL transitions are counterparts of Anderson localization transitions between localized and delocalized phases \cite{evers08} of a quantum particle subjected to a random potential in $d>2$ dimensions (or $d=2$ for some symmetry classes). A hallmark of Anderson transitions is the multifractality of eigenstates \cite{evers08} that implies strong fluctuations of eigenfunction amplitudes at criticality and around the transition point, with a non-trivial power-law scaling of the corresponding moments (inverse participation ratios).  Furthermore, the multifractality implicates a complex pattern of enhancement of correlations between eigenstate amplitudes, both in the coordinate and the energy spaces \cite{chalker90scaling,mirlin00,cuevas07multifrac,evers08}. 
To understand the physics of the MBL, it is of central importance to explore eigenstate correlations at and around  the MBL transitions. This is the main goal of the present work. More specifically, we focus on correlations between eigenstates in the Hilbert space as a function of energy separation and disorder---the problem that can be posed very generally, for any spatial structure of the system. A related but different question was addressed in Ref.~\cite{serbyn2017thouless} which considered matrix elements of local operators.

To explore the eigenstates correlations, we use two models. First, we consider the Anderson model on random regular graphs (RRG), which has emerged as a toy model of MBL.
Second, we study a genuine many-body problem, the XXZ spin chain in a random field, which has become a paradigmatic model for the MBL transition. 

The RRG are finite-size graphs that have locally tree-like structure with fixed coordination number but do not have boundary (i.e., have large-scale loops). The structure of these graphs mimics that of Hamiltonians of interacting systems in the many-body Hilbert space. The idea that single-particle models on a tree (Bethe lattice) can be useful for the analysis of many-body problems was put forward in Ref.~\onlinecite{altshuler1997quasiparticle} in the context of a quasiparticle decay in a quantum dot. Later work has demonstrated that one can think about tree-like graphs more generally as approximately modelling the  Hilbert-space structure of a finite many-body system and that the appropriate graphs are then not Bethe lattices but rather RRG. This analogy between many-body Hilbert space and the RRG lattice has led to a recent surge of interest in properties of delocalized phase of the disordered hopping problem on RRG \cite{biroli2012difference,de2014anderson,tikhonov2016anderson,garcia-mata17,metz2017level,Biroli2017,kravtsov2018non,biroli2018,tikhonov19statistics,tikhonov19critical,PhysRevResearch.2.012020}. The RRG model oversimplifies the many-body problem by discarding matrix-element correlations in Hilbert space states resulting from the fact that the number of independent parameters in the Hamiltonian is much smaller than the number of non-zero matrix elements. 
One important consequence is that, in the localized phase, the inverse participation ratios of eigenstates in the RRG model are of order unity \cite{tikhonov19statistics}, while in the MBL models they exhibit a multifractal scaling with respect to the Hilbert-space volume  \cite{luitz15,gornyi-adp16,tikhonov18,mace19multifractal}. 
Despite this, there are remarkable analogies between the localization transitions in the RRG and true MBL models. In particular, in both models (i) the critical point has a localized character, (ii) there are strong finite-size effects with a drift of the apparent transition point towards stronger disorder, (iii) the ``correlation volume'' of the Hilbert space grows exponentially when the transition is approached, (iv) the delocalized phase is ergodic. For the RRG model these results have been analytically proven \cite{tikhonov19statistics} (see also Refs.~\onlinecite{mirlin1991universality,fyodorov1991localization,fyodorov1992novel} where a related sparse-random-matrix model was studied) and numerically verified \cite{tikhonov2016anderson,garcia-mata17,metz17,tikhonov19statistics,biroli2018}. For the MBL models, analytical arguments are of less rigorous character but still lead to analogous conclusions, in consistency with numerical simulations (see, in particular, the MBL papers cited above and references therein). The connection between 
the Anderson model on RRG and the MBL transition is especially close in models with long-range interaction (decaying as a power-law of distance), see Ref.~\onlinecite{tikhonov18}. 

In view of a close similarity between the RRG and MBL problems, and since the RRG model is much more amenable to the analytical treatment, it is advantageous to perform a numerical study in parallel for both models whenever the MBL observable can also be defined in the RRG problem. This is exactly the case for the eigenstates correlations studied in the present work. Let us illustrate one of advantages of using RRG as a benchmark model. For the RRG model, we know exactly the position of the thermodynamic-limit transition point $W_c$ (as well as values of critical exponents and various other observables) \cite{tikhonov19critical}. This allows us to determine the magnitude of finite-size effects in exact-diagonalization computation, which turn out to be rather strong. As a result, one gets a lower bound for finite-size effects in the MBL problem (which can only be stronger due to additional, rare-event fluctuations related to a smaller number of independent parameter in MBL as comparison to RRG). 

As an additional motivation for this work, it is worth pointing out that  correlations between many-body eigenstates can be measured in quantum computers and simulators. In particular, the eigenstate correlation function $\beta(\omega)$ studied in this paper (see Sec. \ref{s2} for precise definition) is related via the Fourier transformation to the probability $p(t)$ of return to an initial many-body state after time $t$. Such probabilities can be measured in state-of-the-art engineered many-body systems. Recent examples of experimentally implemented systems on which related measurements were performed include one-dimensional (1D) arrays of  53 trapped ions \cite{zhang2017observation} and 51 atoms \cite{bernien2017probing} as well as 1D and two-dimensional (2D) arrays of superconducting qubits  (with up to 21 qubits) \cite{Chiaro2019a}. Furthermore, very recently a quantum processor with 53 superconducting qubits was used to demonstrate the quantum supremacy \cite{google_quantum_supremacy}. The key observable in this demonstration is the fidelity defined as a correlation function of two many-body wave functions (corresponding to the idealized and perturbed Hamiltonians, respectively). While it is somewhat different from the quantity we study in the present work---correlation function of two eigenfunctions of the same Hamiltonian with different energies, Ref.~\cite{google_quantum_supremacy} makes evident the importance of Hilbert-space correlations between many-body wave functions for quantum-information physics and quantum technologies. 
 
The structure of the article is as follows. Section \ref{s2} deals with the eigenstate correlations across the localization transition in the RRG model.
In Section \ref{s3} an analogous study is carried out for the XXZ spin chain.
  Section~\ref{s4} contains a summary of our findings as well as a discussion of their implications and of prospective research directions.

\section{Random regular graphs}
\label{s2}

We study a model of non-interacting spinless quantum particles hopping over a random regular graph (RRG) with connectivity $p = m+1$ (number of sites adjacent to any given site)  in a potential disorder,
\begin{equation}
\label{H}
\mathcal{H}=\sum_{\left<i, j\right>}\left(c_i^\dagger c_j + c_j^\dagger c_i\right)+\sum_i \epsilon_i c_i^\dagger c_i \,.
\end{equation}
Here the index $i =1, \ldots, N$ labels sites of the graph and the sum in the first term is over the pairs of nearest-neighbor sites of the RRG. The energies $\epsilon_i$ are independent random variables sampled from a uniform distribution on $[-W/2,W/2]$. In the definition of all correlation functions introduced below the averaging $\langle \ldots \rangle$ goes over the random structure of the underlying graph and over the random potential $\epsilon_i$. 

An important statistical characteristic of a disordered system is the correlations of different (but relatively close in energy) eigenstates with a given energy separation $\omega$. Formally, we define the corresponding correlation function as follows:
\bea
\label{sigmadef}
&& \beta \left(\omega \right) = \Delta
^{2}R^{-1}\left( \omega \right) \nonumber\\  && \times\left\langle \sum_{k\neq l}\left\vert \psi
_{k}\left( j\right) \psi _{l}\left( j\right) \right\vert ^{2}\delta
\left( E- \frac{\omega}{2}-E_{k}\right) \delta \left( E+\frac{\omega}{2} -E_{l}\right)
\right\rangle. \nonumber \\
\eea
Here $\psi_k$ are eigenstates and $E_{k}$ the corresponding energy levels,  $E$ is the energy at which the statistics is studied, $\Delta=1/\nu(E)N$ is the mean level spacing, $\nu(E)=N^{-1}\left<\Tr\delta(E-\hat H)\right>$ is the density of states, and $R(\omega)$ the level correlation function
\be
\label{R-omega}
R(\omega)=\frac{1}{\nu^2}\left<\nu(E-\omega/2)\nu(E+\omega/2)\right>.
\ee
The argument $j$ in Eq.~(\ref{sigmadef}) is the lattice site; since all sites are equivalent, the r.h.s. does not actually depend on $j$ upon ensemble averaging. In the numerical computations below, we average also over $j$.  In the sequel, it will be convenient to present results for  $\beta(\omega)$ multiplied by $N^2$. For two completely uncorrelated wave functions one has $N^2\beta(\omega)=1$. 

The correlation function $\beta(\omega)$ in the delocalized phase and at the critical point on RRG has been studied in Ref. \cite{tikhonov19statistics} with the following results:
\be
\label{betascres}
N^2\beta(\omega)\sim
\left\{
\begin{array}{cc} \displaystyle
 N_\xi, & \quad \omega<\omega_{\xi},\\[0.4cm]
 \displaystyle 
\frac{1}{\omega \ln^{3/2}1/\omega}, & \quad \omega>\omega_{\xi}.
\end{array}
\right.
\ee
In this equation, $N_\xi$ (which depends on $W$) stands for the correlation volume and $\omega_\xi\sim 1/N_\xi$ for the associated level spacing. The correlation volume $N_\xi$ exhibits on RRG the following critical behavior when the disorder $W$ approaches from the delocalized side the critical point $W_c$ \cite{tikhonov19critical}:
\be
\ln N_\xi \sim (W_c-W)^{-1/2}.
\label{Nxi}
\ee
More specifically, for the ``minimal'', $p=3$, RRG model and in the center of the band ($E=0$), the critical disorder is $W_c=18.17$ and the scaling of $\ln N_\xi$ reads (with a subleading term included) \cite{tikhonov19critical}
\be
\label{Nxicrit}
1/\ln N_\xi=c_1(W_c-W)^{1/2}+c_2(W_c-W)^{3/2},
\ee
where $c_1=0.0313$ and $c_2= 0.00369$. Equation (\ref{Nxicrit}) is valid with a good accuracy in the range $12 < W < W_c$. 

Let us briefly comment on the physical significance of Eq.~(\ref{betascres}). The first line of this equation describes eigenstates correlations in the ``metallic'' regime. The factor $N_\xi$ in this formula implies that the correlations get enhanced when the system approaches the transition point. This is related to strong spatial fluctuations (multifractality) of eigenstates near criticality. Independence of this formula of $\omega$ demonstrates that eigenstates separated by a sufficiently small energy $\omega < \omega_\xi$ exhibit essentially the same ``multifractal pattern'' (despite its randomness). The second line of Eq.~(\ref{betascres}) describes critical correlations, which is why it does not depend on $N_\xi$. The distance to the critical point enters only via the range of validity, $\omega > \omega_\xi$. Exactly at critical point, $W=W_c$, we have $\omega_\xi=0$, so that the second line of Eq.~(\ref{betascres})  holds in the whole range of frequencies (limited only by the level spacing $\Delta \sim 1/N$). 

The goal of this section is to extend the study of the RRG correlation function $\beta(\omega)$ to the localized phase and thus 
to obtain a full description of eigenstates correlations around the localization transition on RRG. 
We begin by presenting qualitative arguments concerning expected behavior of $\beta(\omega)$ on the localized side of the transition. First, in the limit of very strong disorder, $W\to \infty$, individual eigenstates are essentially localized on different sites and do not overlap. We thus expect $N^2 \beta(\omega) \to 0$ in the limit $W\to \infty$. 
Combining this with Eq.~(\ref{betascres}), we conclude that, for a small $\omega$,  the correlation function $N^2 \beta(\omega)$ should be a non-monotonic function of $W$ that shows a maximum in the vicinity of the critical point $W=W_c$.

Second, for a given $W > W_c$,  two localized states will be typically located in remote regions of the system and overlap very weakly in view of the exponential decay of the localized wave functions. However, there is a certain probability that two such states turn out to be in resonance, which then strongly enhances the overlap.  The probability of a resonance is enhanced for small energy separation $\omega$, so that  $N^2 \beta(\omega)$ is expected to decay with $\omega$ in the localized phase. For the single-particle problem in $d$ dimensions, this decay was studied in Ref.~\cite{cuevas07multifrac}, with the result
\be
\label{finite-d-loc}
N^2 \beta(\omega) \sim \xi^{d-d_2} \ln^{d-1} (\delta_\xi / \omega)\,, \qquad \omega < \delta_\xi \,,
\ee
where $\xi$ is the localization length, $\delta_\xi \sim \xi^{-d}$ the level spacing in the localization volume, and $d_2$ the multifractal exponent. It was pointed out in Ref.~\cite{cuevas07multifrac} that the logarithmic enhancement of correlations with lowering $\omega$ in Ref.~(\ref{finite-d-loc}) is closely related to the Mott's behavior law for the ac conductivity.

What kind of behavior of $N^2 \beta(\omega)$ can one expect on this basis in the localized phase on RRG?  The RRG model can be in a certain sense viewed as a $d\to\infty$ limit of the $d$-dimensional Anderson model; this limit is, however, highly singular \cite{tikhonov19statistics}.  Equation (\ref{finite-d-loc}) suggests that the enhancement of correlations for small $\omega$ on RRG should be faster than a power of $\ln \omega$. It is even more difficult to guess what the dependence on disorder [encoded in the localization length $\xi$ and the corresponding spacing $\delta_\xi$ in Eq.(\ref{finite-d-loc})]  transforms into when the RRG model is considered. As we show below, the eigenstate correlation function $N^2 \beta(\omega)$ has a power-law dependence on $\omega$ in the localized phase of the RRG model, with an exponent that is a function of disorder. We will also see that such a behavior holds also for a genuine MBL problem.

\begin{figure}[tbp]
\includegraphics[width=0.5\textwidth]{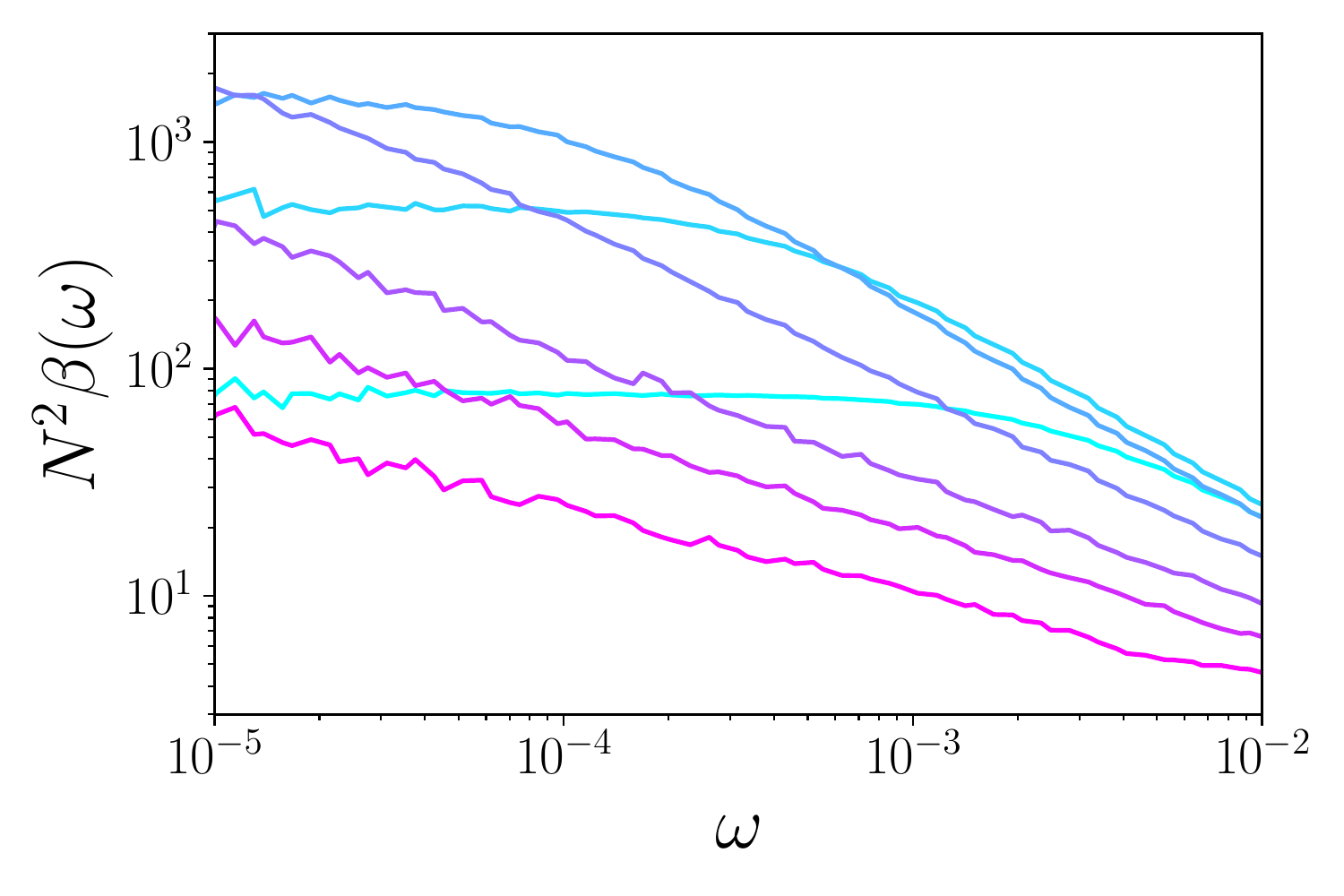}
\caption{Eigenstate correlation function $\beta(\omega)$ for RRG with disorder strengths $W= 10, 12, 14, 18, 24, 30, 42$ (from cyan to magenta). The first three values are on the delocalized side of the transition, $W=18$ is essentially the critical point, and the three largest values are on the localized side.  The system size is $N=32768$.}
\label{beta_rrg_full}
\end{figure}

We have computed $\beta(\omega)$ numerically by exact diagonalization of the RRG model with the connectivity $p=3$, focussing on the vicinity of the band center, $E=0$.  We consider system sizes $N$ in the range from $2^{12}$ to $2^{16}$. For each realization of disorder, we average over $N/32$ states near the band center. 
In addition, we average over disorder realizations; their number ranges from $50000$ for smaller systems to $50$ for the largest systems. 

\begin{figure*}[tbp]
\minipage{0.5\textwidth}\includegraphics[width=\textwidth]{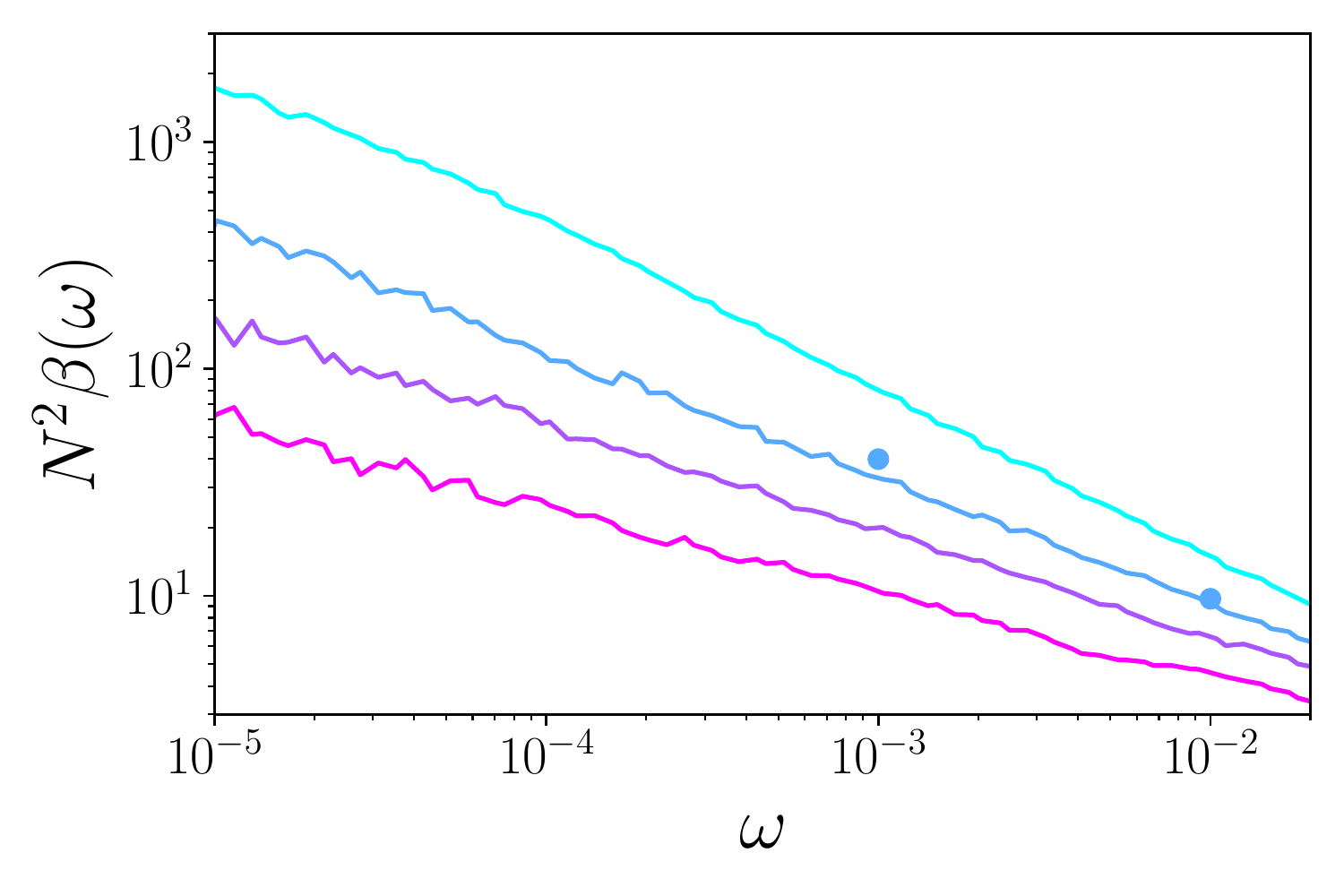}\endminipage
\minipage{0.5\textwidth}\includegraphics[width=\textwidth]{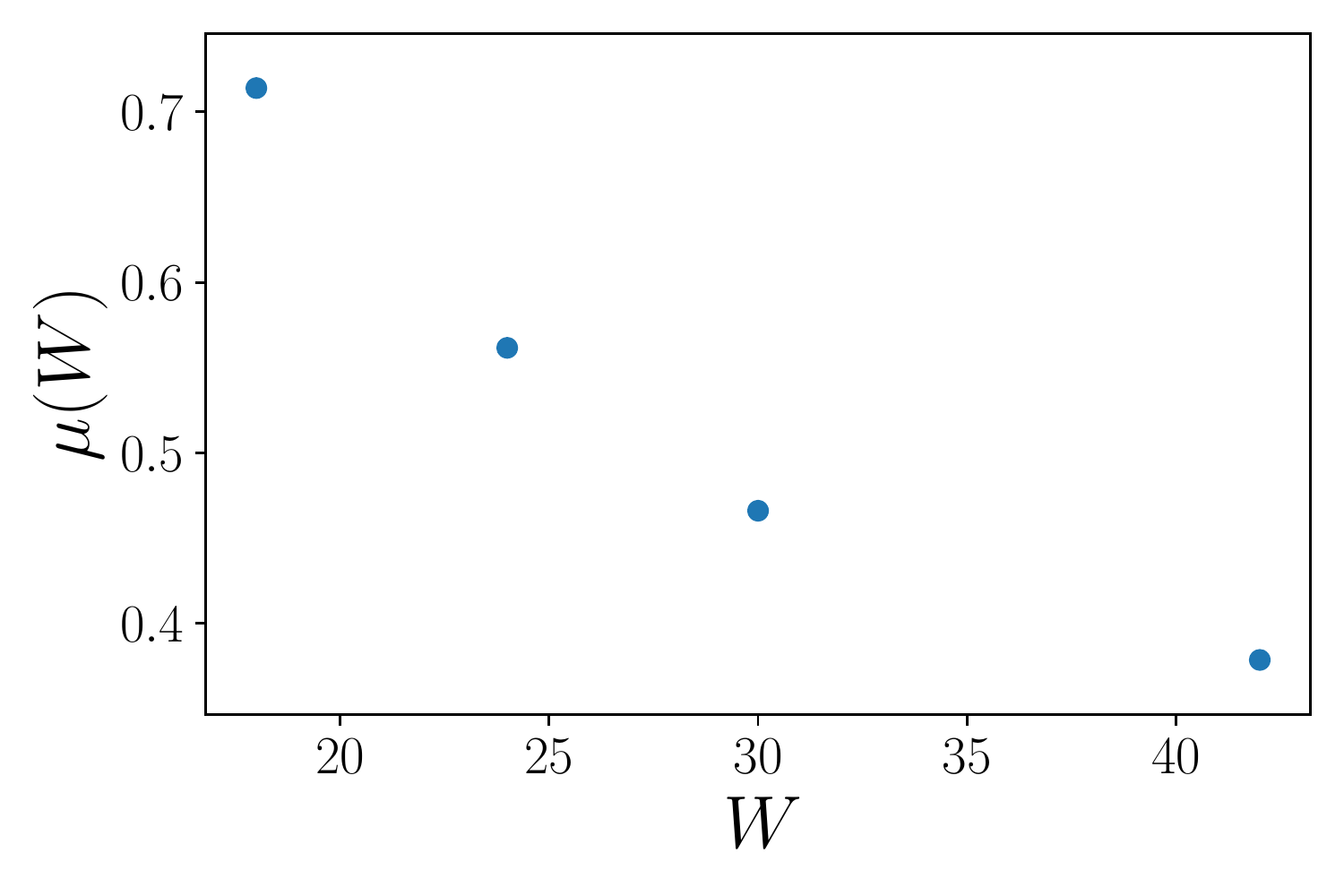}\endminipage
\caption{Eigenstate correlations in the localized phase on RRG. The system size is $N=32768$. {\it Left:} $\beta(\omega)$ for disorders $W = 18$ (essentially the critical point), 24, 30, and 42 (from cyan to magenta). The straight lines on the log-log scale imply a power-law dependence of $\beta(\omega)$ on $\omega$ in the localized phase, with a disorder-dependent exponent $\mu(W)$. Dots show results from the population dynamics for $W=24$ (see Supplementary Material \cite{SM} for details); they have the same (blue) color as the corresponding line of exact-diagonalization data. {\it Right:} Exponent $\mu(W)$ characterizing the frequency scaling of $\beta(\omega)$, see Eq. (\ref{betafit}).}
\label{beta_rrg}
\end{figure*}

In Fig.~\ref{beta_rrg_full} we show the results of exact-diagonalization study for $N=32768$ and several disorder values ranging from the delocalized phase ($W=10,12,14$)  through the critical point ($W=18$) to the localized phase $W=24, 30, 42$.  (Note that the difference between $W=18$ and the exact critical value $W_c=18.17$ is immaterial for system sizes amenable for exact diagonalization.) In the delocalized phase, $W=10$, 12, and 14, the behavior (\ref{betascres}) is clearly observed: the power-law ($1/\omega$) behavior at high frequencies and a saturation at lower frequencies. The saturation frequency $\omega_\xi$ becomes smaller when the disorder increases, i.e., the system approaches the critical point. At criticality, $W=18$, the power-law (approximately $1/\omega$) behavior is indeed observed in the whole range of frequencies. Remarkably, the power-law behavior of $\beta(\omega)$  survives in the localized phase $W=24$, 30, and 42, where it is characterized by a disorder--dependent exponent $\mu(W)$.
The numerical results thus unambiguously suggest the power-law scaling of eigenstate correlations in the localized phase:
\be
\label{betafit}
N^2\beta(\omega) \sim \omega^{-\mu(W)}.
\ee
At criticality, $\mu(W=W_c)=1$, while in the localized phase, $W>W_c$, the exponent  $\mu(W)$ is less than unity and gradually decreases towards zero as $W$ grows. 
In Fig. \ref{beta_rrg}, we highlight the correlation function $\beta(\omega)$ in the localized phase. The right panel shows the numerically determined exponent $\mu(W)$ obtained from the data presented in the left panel.

It is worth emphasizing that the exponent $\mu(W)$ is, strictly speaking, defined in the limit of large system size $N$ and low frequency $\omega$. For finite $N$ there are corrections to an apparent position of the critical point (that are discussed in more detail below), which also influence the numerically determined values of  $\mu(W)$.  Also, there might be in principle a subleading (e.g. logarithmic) factor in $\omega$-dependence in Eq.~(\ref{betafit}).  We know that such a logarithmic factor does exist at criticality, see second line of Eq.~(\ref{betascres}). Its emergence at criticality has a clear physical reason, as it provides convergence of the time-dependent correction to return probability, see Ref.~\cite{tikhonov19statistics}. On the localized side, such a convergence is already guaranteed by the power-law dependence (\ref{betafit}) with $\mu(W) < 1$, so that we do not have any arguments in favor of such a factor. Also, the corresponding lines in the left panel of Fig.~\ref{beta_rrg} are rather straight (up to fluctuations), without any clear indication of such a factor. When fitting the numerical data to extract the exponent $\mu$, we thus assume a pure power-law dependence (\ref{betafit}), without any subleading prefactors. 

We turn now to the analytical approach to eigenstate correlations. In Ref.~\cite{tikhonov19statistics}, we have shown how various observables characterizing the RRG model can be expressed in terms of a solution of a saddle-point equation for the effective action of the problem. This equation is equivalent to a self-consistency equation for the distribution of local Green functions on an infinite Bethe lattice (BL). This is a non-linear integral equation, and a full analysis of its solution is by no means an easy task. In the localized phase, the solution is singular in the low-frequency limit. Leading contribution to the correlation function that is needed for our purposes determines properties of individual eigenstates (e.g., the participation ratio). To obtain the correlation function of different eigenstates, one needs a subleading term (see Ref.~\cite{tikhonov19statistics}), which makes the analysis much more difficult. We first discuss the numerical solution of the self-consistency equation; below we discuss an analytical solution in the limit of strong disorder $W$.
To solve numerically this equation, one can use the population-dynamics approach, see also Refs.~\cite{biroli2010anderson,metz17}. 
In particular, we calculated in this way in Ref.~\cite{tikhonov19statistics} the correlation function $\beta(\omega)$ as well its Fourier transform, the return probability $p(t)$, in the delocalized phase and demonstrated perfect agreeement with exact-diagonalization results. Here we demonstrate that the population-dynamics approach to the solution of the self-consistency equation can also be used for computing $\beta(\omega)$ on the localized side, even though it turns out to require much more efforts. 

Determining $\beta(\omega)$ in this way amounts to evaluation of the correlation function of local densities of states $\rho(\epsilon; j)$ on an infinite Bethe lattice 
\be
\label{Komega}
K(\omega)=\frac{\left<\rho(E+\omega/2; j)\rho(E-\omega/2; j)\right>_{\rm BL}}{\left<\nu(E)\right>^2} \,,
\ee 
where 
\be
\label{rhoepsilon}
\rho(\epsilon; j) = - \frac{1}{\pi} {\rm Im}\: G_R(j,j,\epsilon) \,,
\ee
and $G_R(j,j,\epsilon)$ is the retarded Green function at energy $\epsilon$ with equal spatial arguments $j$. Calculation of $\beta(\omega)$
 by population-dynamics approach (see Supplementary Material \cite{SM} for more details) requires introducing a finite imaginary part of frequency $\omega \to \omega+i\eta$, evaluation of $K(\omega)$ for a complex frequency and then considering the limit of $\eta\to+0$,
 \be
 \label{betawithK}
 N^2\beta(\omega) = \lim_{\eta\to +0 }K(\omega).
 \ee
 This is rather non-trivial in the localized phase, as taking the limit requires considering very small values of $\eta$. In Fig. \ref{rrg_pool_omegadep} of the Supplementary Material \cite{SM}, we show the  $\eta$-dependence of $K(\omega)$ for $W=24$ and two values of $\omega$ as obtained by population dynamics. It is seen that $K(\omega)$ does have a finite limit at $\eta \to 0$ but the saturation of $\eta$ dependence takes place at very small $\eta$. The resulting values of  $N^2\beta(\omega)$ are shown by dots in the left panel of the Fig. \ref{beta_rrg} and are in good agreement with results of the exact diagonalization. 
 
In order to shed light on the physical origin of the power-law scaling of the correlation function $\beta(\omega)$ in the localized phase on RRG, we perform its analysis by assuming a limit of strong disorder $W$. In this limit, almost every single-particle state is localized within a small localization length $\zeta$ around a certain lattice site (localization center).  Typically, two localized states are located a distance of order of system size $L = \ln N / \ln m$ apart and have an exponentially small overlap $\propto e^{-L / \zeta}$. However, there are rare resonant events: two states located far apart may form a resonant pair and strongly hybridise, so that the resulting states will have amplitudes of order unity at both localization centers. Such a resonant pair will thus realize the strongest possible overlap and therefore give a maximal possible contribution to the correlation function $\beta(\omega)$. Even though such resonance events are rare, they determine the average value $\beta(\omega)$ in the case of $d$-dimensional system with $d>1$, see Ref.~\cite{cuevas07multifrac}.
Specifically, the factor $\ln^{d-1} (\delta_\xi / \omega)$ in Eq.~(\ref{finite-d-loc}) represents resonant enhancement of eigenstate correlations. Clearly, its role increases with increasing $d$. More generally, it is known that the role of resonances is particularly important in the RRG and MBL models, in view of the effectively  infinite-dimensional character of the corresponding Hilbert space.  It is thus natural to expect that the power-law scaling of $\beta(\omega)$ on RRG can be understood in the framework of the resonance mechanism. This turns out to be indeed the case. 

To shed light on the origin of the power-law scaling, it is instructive to make the following simplistic estimate. It is known that, upon averaging, localized eigenstates on a tree-like graph decay in the following way \cite{zirnbauer1986localization,mirlin1991localization,tikhonov19statistics,PhysRevResearch.2.012020}:
\be
\langle |\psi^2(r)| \rangle \sim m^{-r} \exp\{-r/\zeta(W) \} \,,
 \label{eigenstate-exp-decay}
 \ee
where $r$ is the distance from the ``localization center'' of the state and $\zeta(W)$ is the localization length that diverges at the transition point as $\zeta(W) \sim (W-W_c)^{-1}$ and diminishes slowly at strong disorder, $\zeta(W) \sim 1/ \ln (W / W_c)$. While Eq.~(\ref{eigenstate-exp-decay}) yields the average, let us assume  that all eigenstates decay in this way; this will be sufficient to understand the $W$-dependent power-law in eigenstate correlations. 
Consider two such eigenstates separated by a distance $R$.  The corresponding overlap matrix element is then $M \sim m^{-R} \exp\{-R/\zeta(W)\}$. The optimal condition of the Mott-like resonance for two eigenstates with the energy difference $\omega$ is $M \sim \omega$. Under this condition, two considered eigenstates (let's call them $\psi_k$ and $\psi_l$) get strongly hybridized, so that 
\be 
\sum_j |\psi_k(j) \psi_l(j)|^2 \sim 1.
\label{resonance-overlap}
\ee  
Expressing the distance between the eigenstates centers through the frequency, we find
\be
R(\omega)  \simeq \frac{\ln (1/\omega)}{\ln m + \zeta^{-1}(W)} \,.
\label{R-omega}
\ee
The total number of states whose centers are separated by distance $R$ from that of the state $\psi_k$ is
\be
N_{R(\omega)} \sim m^{R(\omega)} \sim  \omega^{-\mu(W)} \,,
\ee
where
\be
\mu(W) = \frac{\zeta(W) \ln m}{\zeta(W) \ln m+1} \,.
\label{mu-omega}
\ee
The formation of the resonance requires that one of these states is separated by an energy difference $\sim \omega$ from the state $\psi_k$. Thus, the probability $p_\omega$ of the resonance in the frequency interval $[\omega, 2 \omega]$ involving the given state $\psi_k$  is equal to a ratio of the frequency $\omega$ to the level spacing $\sim N^{-1}_{R(\omega)}$, 
\be
p_\omega \sim \omega N_{R(\omega)} \sim \omega^{1- \mu(W)} \,.
\label{p-omega}
\ee
Using the definition (\ref{sigmadef}), we get
\bea
N^2 \omega \beta(\omega) &\sim& N^2 \int_\omega^{2\omega} d\omega' \beta(\omega') \nonumber \\
&= &  \sum_{l \,:\:   \omega < | E_k - E_l | < 2\omega}
\left\langle \sum_j |\psi_k(j) \psi_l(j)|^2 \right\rangle  \nonumber \\[0.3cm]
& \sim &  \omega^{1- \mu(W)} \,.
\label{beta-omega-res}
\eea
In the second line of  Eq.~(\ref{beta-omega-res}) the state $k$ is fixed; the summation goes over states $l$ with the energy difference in the $[\omega, 2 \omega]$ interval. 
In the last line, we used Eq.~(\ref{p-omega}) for the probability of a resonance in this interval and Eq.~(\ref{resonance-overlap}) for the resonant overlap.  Comparing the starting and the final expressions in Eq.~(\ref{beta-omega-res}), we finally come to the result,  Eq.~(\ref{betafit}), for the scaling of $N^2 \beta(\omega)$ with frequency $\omega$, where the exponent $\mu(\omega)$ is given by Eq.~(\ref{mu-omega}). 

Inspecting Eq.~(\ref{mu-omega}) for the exponent $\mu(W)$, we find the following asymptotic behavior. When the disorder $W$ approaches the critical point (from the localized side), 
Eq.~(\ref{mu-omega}) yields 
\be 
\mu(W) \to 1 \,,  \qquad W \to W_c +0 \,.
\label{mu-critical}
\ee
This matches the critical scaling $\beta(\omega) \propto 1/\omega$ (up to a logarithmic correction), see second line of Eq.~\eqref{betascres}.
In the opposite limit of large $W$,  we get, by using the asymptotic behavior of the localization length,
\be
\mu(W) \sim \frac{1}{ \ln(W/W_c)}\,, \qquad W\gg W_c \,.
\label{mu-strong-disorder}
\ee
Thus, $\mu(W)$ decays to zero at $W \to \infty$ but this decay is logarithmically slow.  These results are in good agreement with the numerical observations presented above.

As we have already mentioned, Eq. (\ref{eigenstate-exp-decay}) describes the average decay of a wavefunction. At the same time, wavefunctions fluctuate strongly; in particular, decay of the typical wavefunction amplitude is described by a different localization length \cite{PhysRevResearch.2.012020}. In the Supplementary Material \cite{SM} we present a more accurate version of the above resonance-counting analysis, which takes into account strong fluctuations of eigenstates  around the average (\ref{eigenstate-exp-decay}). It confirms the power-law scaling (\ref{betafit}) and yields qualitatively the same results for the behavior of the exponent $\mu(W)$.

\begin{figure*}[tbp]
\minipage{0.5\textwidth}\includegraphics[width=\textwidth]{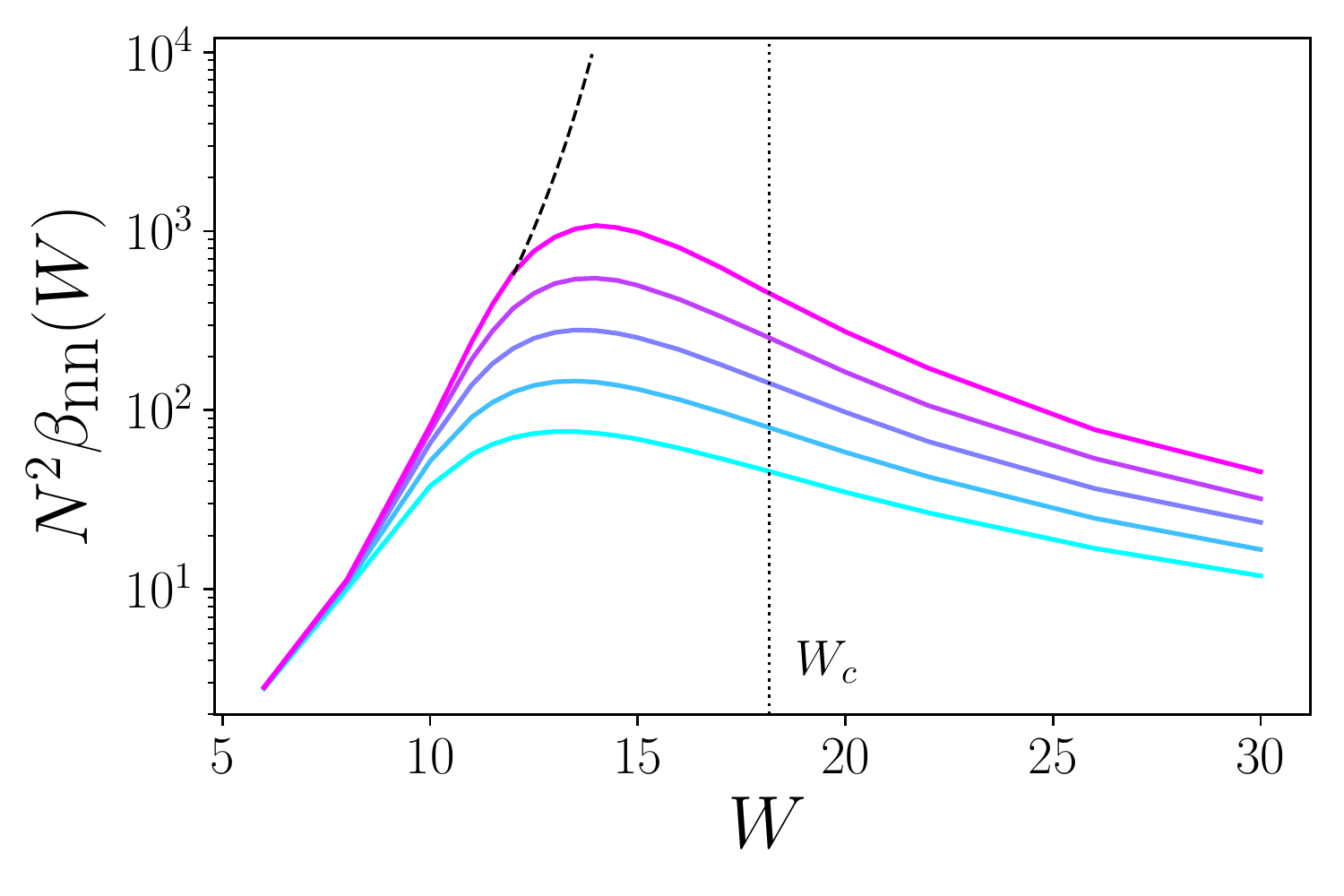}\endminipage
\minipage{0.5\textwidth}\includegraphics[width=\textwidth]{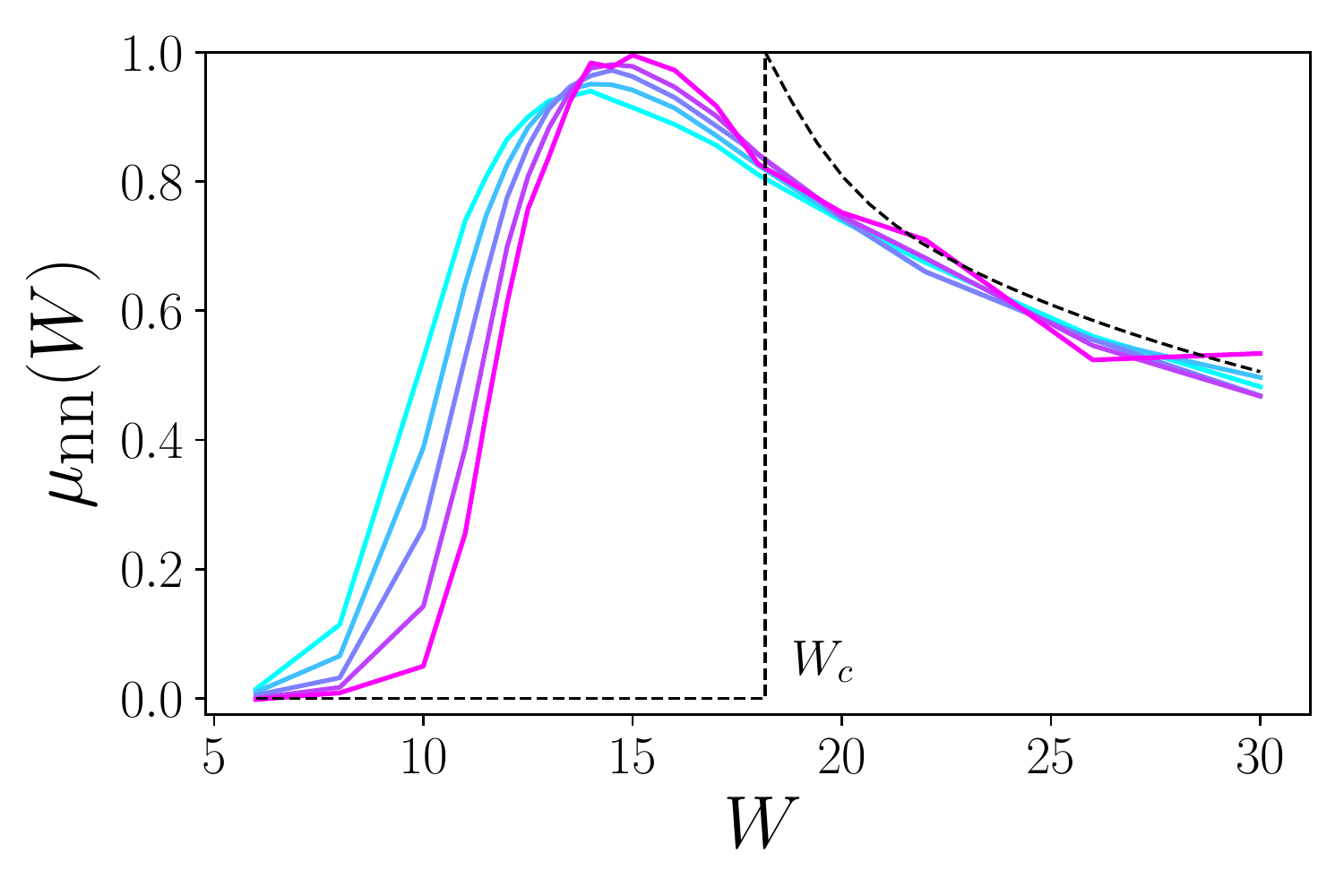}\endminipage
\caption{Correlation of adjacent wavefunctions on RRG. System sizes are $N=4096, 8192, 16384, 32768, 65536$ (from cyan to magenta).  {\it Left:} Correlation function $\beta_{\textrm{nn}}(W)$. Dashed line is the expected asymptotic behaviour of $N^2\beta_{\textrm{nn}}$ on the delocalized side, see Eqs. (\ref{betann}) and (\ref{Nxicrit}). Vertical dotted line marks the critical point of the localization transition,  $W_c=18.17$.  {\it Right:} Exponent $\mu_{\textrm{nn}}$ characterizing the $N$ scaling of adjacent-state correlations, see Eq. (\ref{alphafit}). Dashed line shows theoretically expected $N\to\infty$ behaviour, see Eq.~(\ref{betann-del}) for the delocalized phase and Eq.~(\ref{betann}) for the localized phase (this part of the dashed line is schematic.) }
\label{alpha_rrg}
\end{figure*}

It is useful to introduce a correlator that is closely related to $\beta(\omega)$---a correlation function of adjacent-in-energy eigenstates:
\be
\beta_{\textrm{nn}} = \Delta\left<\sum_k\delta(E_k-E)\left\vert \psi
_{k}\left( j\right) \psi _{k+1}\left( j\right) \right\vert ^{2}\right>.
\label{beta-nn}
\ee
Here the subscript ``nn'' stands for ``nearest neighbor'' (in energy space).  Clearly, $\beta_{\textrm{nn}} \simeq \beta(\omega \sim \Delta)$, where $\Delta$ is the level spacing.
Thus, in the delocalized phase we and in the large-$N$  limit (the condition is $N\gg N_\xi$) we have
\be
\label{betann-del}
N^2\beta_{\textrm{nn}} = N_\xi/3
\ee
The numerical coefficient in this formula depends, of course, on precise definition of the correlation volume $N_\xi$. The value $1/3$ in Eq.~(\ref{betann-del}) holds if this normalization is fixed by the condition that the average inverse participation ratio of an eigenstate,
\be
P_2 = \left\langle \sum_j |\psi_k(j)|^4 \right\rangle \,,
\label{ipr}
\ee
 is $P_2 \simeq N_\xi / N$ at $N \gg N_\xi$ \cite{tikhonov19statistics}. 
In the localized phase, we have, according to Eq.~(\ref{betafit}),
\be
\label{betann}
N^2\beta_{\textrm{nn}} \sim N^{\mu(W)}.
\ee

In the left panel of Fig. \ref{alpha_rrg}, we show results of numerical simulations for the correlator $N^2\beta_{\textrm{nn}}$ for several system sizes. 
For $W<W_c$, the lines clearly approach a limiting ($N\to \infty$) curve, in agreement with Eq. (\ref{betann-del}). According to Eq. (\ref{betann-del}), this limiting curve is determined by the disorder dependence of the correlation volume, $N_\xi(W)$. Indeed, we observe a perfect agreement with the asymptotic behavior of $N_\xi(W)$ given by Eq. (\ref{Nxicrit}) (shown by dashed line). For any given $N$, the curve $N^2\beta_{\textrm{nn}}(W)$ deviates from the limiting curve upon increasing $W$, since the condition $N\gg N_\xi$ ceases to be satisfied, shows a maxmum at certan size-dependent disorder $W_\mathrm{peak}(N)$, and then decays. The non-monotonic behavior of $N^2\beta_{\textrm{nn}}(W)$ is a general feature of a system that undergoes a localization transition, see qualitative discussion above. The position of the maximum $W_\mathrm{peak}(N)$
can be viewed as a size-dependent apparent critical point which drifts to larger $W$ with growing system size. In the limit of $N\to\infty$ the drift stops at the limiting value $W_\mathrm{peak}(N\to\infty)=W_c \simeq 18.17$. 

In Table \ref{table-W-peak}, we list the numerically obtained values of $W_\mathrm{peak}(N)$ for several system sizes $N$. The drift towards $W_c$ is evident but it is rather slow. 
Looking at these slowly drifting values, one could naively think, that they are close to the actual value of $W_c$. This is not true, however: these values are still rather far from the true critical point. (One indication of this is absense of a clear trend to saturation.)
For the RRG model, we have a luxury of knowing  the true critical point with a high precision, $W_c=18.17$, which is obtained by a very different approach---investigation of stability of the solution of the saddle-point equation corresponding to the localized phase \cite{tikhonov19critical}. Therefore, the RRG model, as a toy model of MBL, is very useful for benchmarking exact-diagonalization studies of MBL problems. We see from Table  \ref{table-W-peak} that if only exact-diagonalization data would be available, it would be very hard to determine the position of the $N\to \infty$ critical point with a reasonable accuracy. An even more difficult task for the exact-diagonalization numerics is to find the true (asymptotic) value of the critical exponent $\nu$ of the correlation length $\xi = \ln N_\xi / \ln m$. 
Asymptotically, the drift of the peak can be characterized by the critical exponent via the scaling relation
\be
\ln\ln N = -\nu\ln\left[W_c-W_\mathrm{peak}(N)\right].
\ee
While this equation is valid in the limit of $N\to\infty$, it is convenient to introduce an apparent finite-size exponent via 
\be
\frac{1}{\nu(N)}=-\frac{\partial \ln\left[W_c-W_\mathrm{peak}(N)\right]}{\partial\ln\ln N}.
\label{nu-N} 
\ee
The last line of the Table \ref{table-W-peak} presents values of $\nu(N)$ obtained by numerical differentiation according to Eq.~(\ref{nu-N}). We see a strong variation of $\nu(N)$ towards the true asymptotic value $\nu=1/2$  \cite{tikhonov19critical}, see Eq.~(\ref{Nxi}). The fact that the ``flowing exponent'' $\nu(N)$ approaches its asymptotic value 1/2 from above, and that values of $N$ much larger than those amenable to exact diagonalization are needed to obtain numerically 1/2 with a good accuracy, was demonstrated in detail in Ref.~ \cite{tikhonov19critical}. Our findings are in full agreement with these previous results. It should be stressed that when calculating $\nu(N)$ in Table \ref{table-W-peak}, we used the high-precision value of the critical disorder, $W_c=18.17$. For the MBL problems, $W_c$ is found from numerical simulations with a much lower precision (see the discussion above), which further increases uncertainty of numerical determination of the critical exponent $\nu$.

\begin{table}
\begin{center}
 \begin{tabular}{|c|c c c c c|c|} 
 \hline
$\log_2 N$ & $12$ & $13$ & $14$ & $15$ & $16$ & $\infty$\\
\hline
$W_\mathrm{peak}(N)$ &$13.70$ & $13.78$ & $13.89$ & $14.06$ & $14.28$ & $18.17$\\
$\nu(N)$ &$4.31$ & $3.52$ & $2.22$ & $1.42$ & $0.96$ & $1/2$\\
 \hline
\end{tabular}
\end{center} 
\caption{Position $W_\mathrm{peak}(N)$ of the maximum of $N^2\beta_{\textrm{nn}}(W)$ curves that can be viewed as an $N$-dependent apparent critical point.  Upon increasing $N$, it shows a slow drift towards the limiting ($N\to \infty$) value $W_c=18.17$.  The lower line of the table shows the ``flowing ($N$-dependent) critical exponent" extracted from $W_\mathrm{peak}(N)$ according to Eq.~(\ref{nu-N}). It evolves to the asymptotic ($N\to \infty$) value $\nu=1/2$.}
\label{table-W-peak}
\end{table}

To characterize the evolution of $\beta_\textrm{nn}$ with the system size, we define  a disorder-  and size-dependent exponent:
\be
\label{alphafit}
\mu_{\textrm{nn}}(W, N) = \frac{\partial\ln\left(N^2\beta_{\textrm{nn}}\right)}{\partial\ln N}.
\ee
On the delocalized side, $ N^2\beta_{\textrm{nn}}$ is independent on $N$ at large $N$, which implies that $\mu_{\textrm{nn}}(W<W_c,N) \to 0$  at $N\gg N_\xi(W)$. At the critical point, $W=W_c$, we have $\mu_{\textrm{nn}}(W_c) \to 1$  at $N\to \infty$. On the localized side,   Eq.~(\ref{betann}) yields $\mu_{\textrm{nn}}(W>W_c, N) \to \mu(W)$ in the large-$N$ limit. 
In the right panel of Fig. \ref{alpha_rrg}, we show numerical results for $\mu_{\textrm{nn}}(W, N)$. As expected, for $W<W_c$  the $\mu_{\textrm{nn}}(N)$ curves gradually drift downwards, towards zero, with increasin $N$. Closer to $W_c$, this drift is in fact non-monotonic (first upward, then downward), see Ref.~\cite{tikhonov2016anderson} for a discussion of the physical origin of such behavior on RRG. For $15 \lesssim W < W_c$ we observe only upward drift; one needs much larger $N$ to see that it will be eventually superseded by a downward drift with the ultimate large-$N$ limit $\mu_{\textrm{nn}}(W) \to 0$. On the localized side, $W > W_c$, we find a nearly $N$-indepedent $\mu_{\textrm{nn}}(W, N)$, in consistency with the expected limiting behavior  $\mu_{\textrm{nn}}(W>W_c, N) \to \mu(W)$ and in agreement with numerical data for $\mu(W)$ (which are, of course, also subjected to finite-size corrections)  in the right panel of Fig. \ref{beta_rrg}.

\begin{figure}[tbp]
\includegraphics[width=0.5\textwidth]{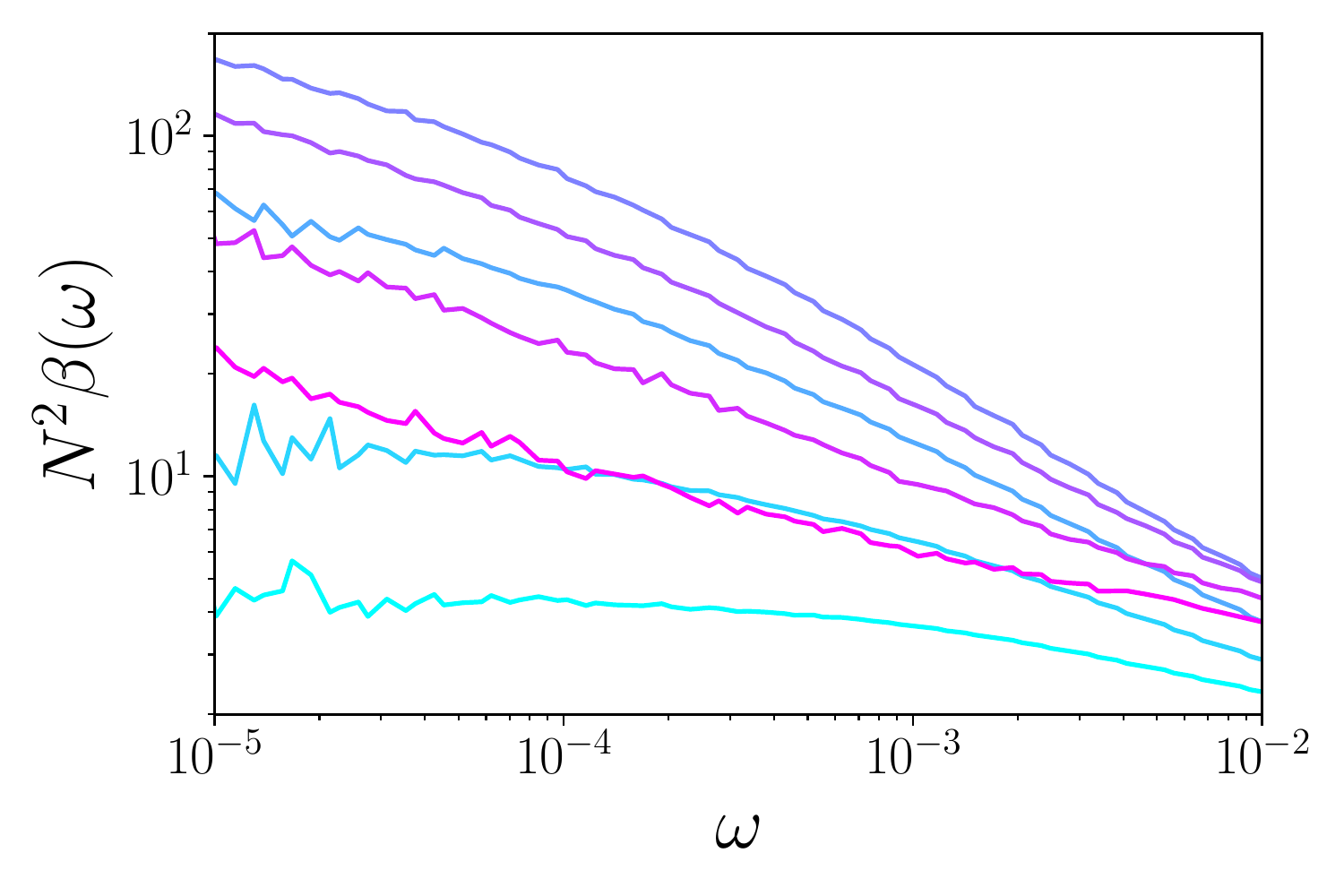}
\caption{Eigenstate correlation function $\beta(\omega)$ for spin chain of length $L=16$. The disorder strengths are  $W = 1.5, 1.7, 2, 3, 4, 6, 10$ (from cyan to magenta).
The first three values are deeply in the ergodic phase. The next two values are also on the delocalized side of the transition but correspond to the critical regime for relatively small systems available for exact diagonalization. The largest two values are on the localized side of the transition. This figure is a spin-chain counterpart of Fig.~\ref{beta_rrg} for the RRG model.}
\label{beta_spins_full}
\end{figure}

\section{Spin chain}
\label{s3}
In this section, we apply a similar methodology to study the model of the $S=\frac{1}{2}$ Heisenberg chain in a random magnetic field, 
governed by the Hamiltonian
\be 
H=\sum_{i\in [1,L]} {\bf S}_i \cdot {\bf  S}_{i+1} -h_iS_i^z, 
\label{eq:H}
\ee
with $h_i$ drawn from a uniform distribution $[-W,W]$  and with periodic boundary conditions, ${\bf  S}_{L+1} \equiv {\bf  S}_{1}$. 
(Note that the total magnetization $S^z = \sum_i S^z_i$ is conserved.) This model has become one of paradigmatic models for the investigation  of  the  MBL physics \cite{Oganesyan2007,PhysRevB.82.174411}. Numerically, systems of sizes up to $L=  22$ \cite{luitz15} and $L=24$ \cite{mace19multifractal} were  investigated  via  exact  diagonalization, which yielded estimates $W_c=3.7$\:--\:$3.8$ for the critical disorder in the middle of the many--body spectrum. However, it has been understood that an accurate determination of the critical point from exact diagonalization data is complicated due to a rather slow convergence towards thermodynamic limit, see Ref.~\cite{abanin2019distinguishing} for a recent discussion. The experience with RRG teaches us that the actual (thermodynamic-limit) critical disorder is considerably larger, due to finite-size effects, than the value suggested by the exact diagonalization. Indeed, for RRG the exact diagonalization would suggest $W_c \approx 15$, while the actual value is $W_c = 18.17$, i.e., about 20\% higher. In genuine interacting MBL models (like the spin-chain model considered in this section), the finite-size effects are expected to be still stronger due to effects of rare spatial regions.  This has been supported by an analysis based on the time-dependent variational principle with  matrix product states which was used to study the dynamics (relaxation of spin imbalance) in Ref.~\cite{Doggen2018a} in much larger systems, up to $L=100$. This study has demonstrated a strong drift of apparent (size-dependent) $W_c$ with system size $L$,  suggesting the critical value $W_c\approx 5$\:--\:$5.5$ for the thermodynamic-limit transition between the ergodic and MBL phases.

\begin{figure*}[tbp]
\minipage{0.5\textwidth}\includegraphics[width=\textwidth]{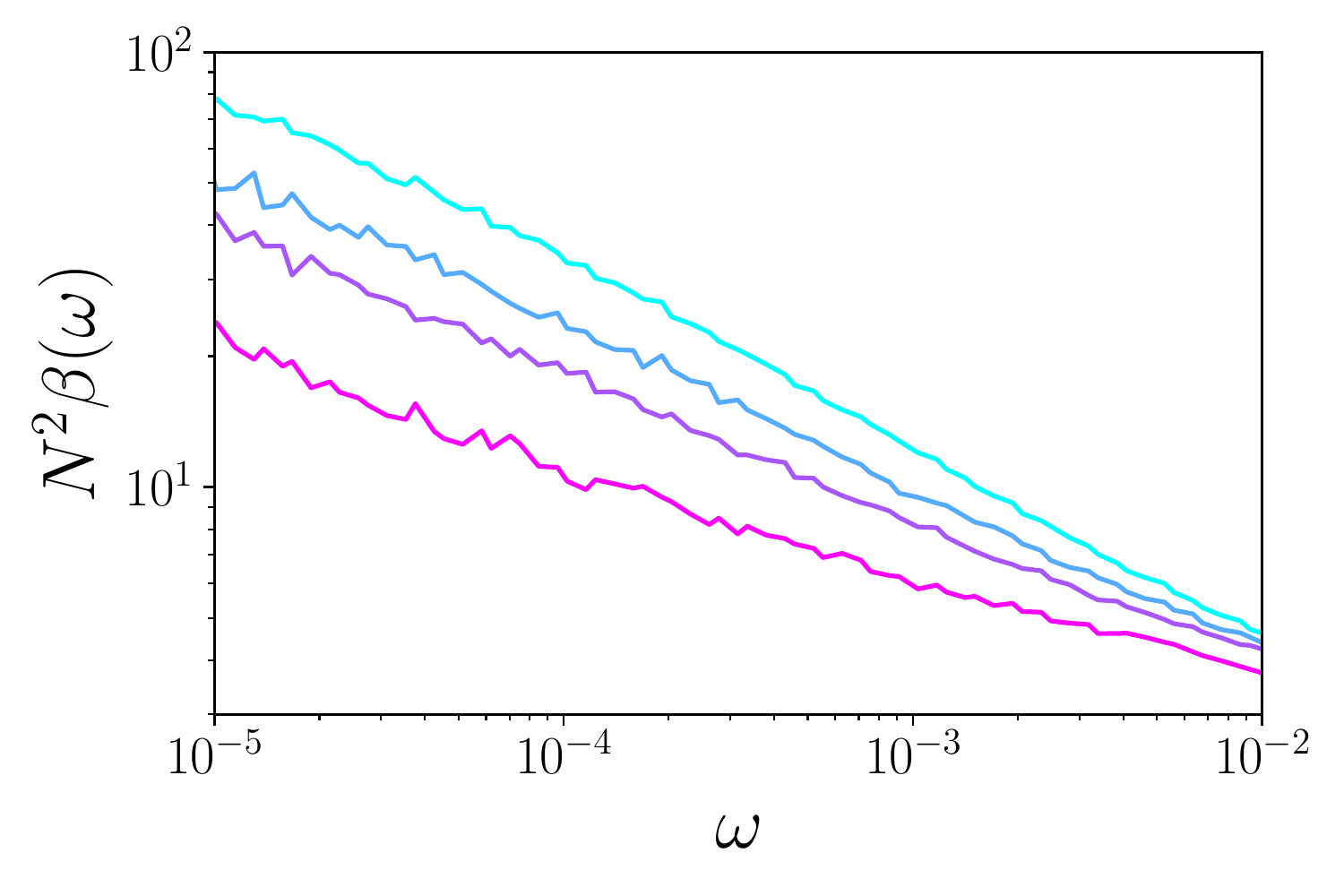}\endminipage
\minipage{0.5\textwidth}\includegraphics[width=\textwidth]{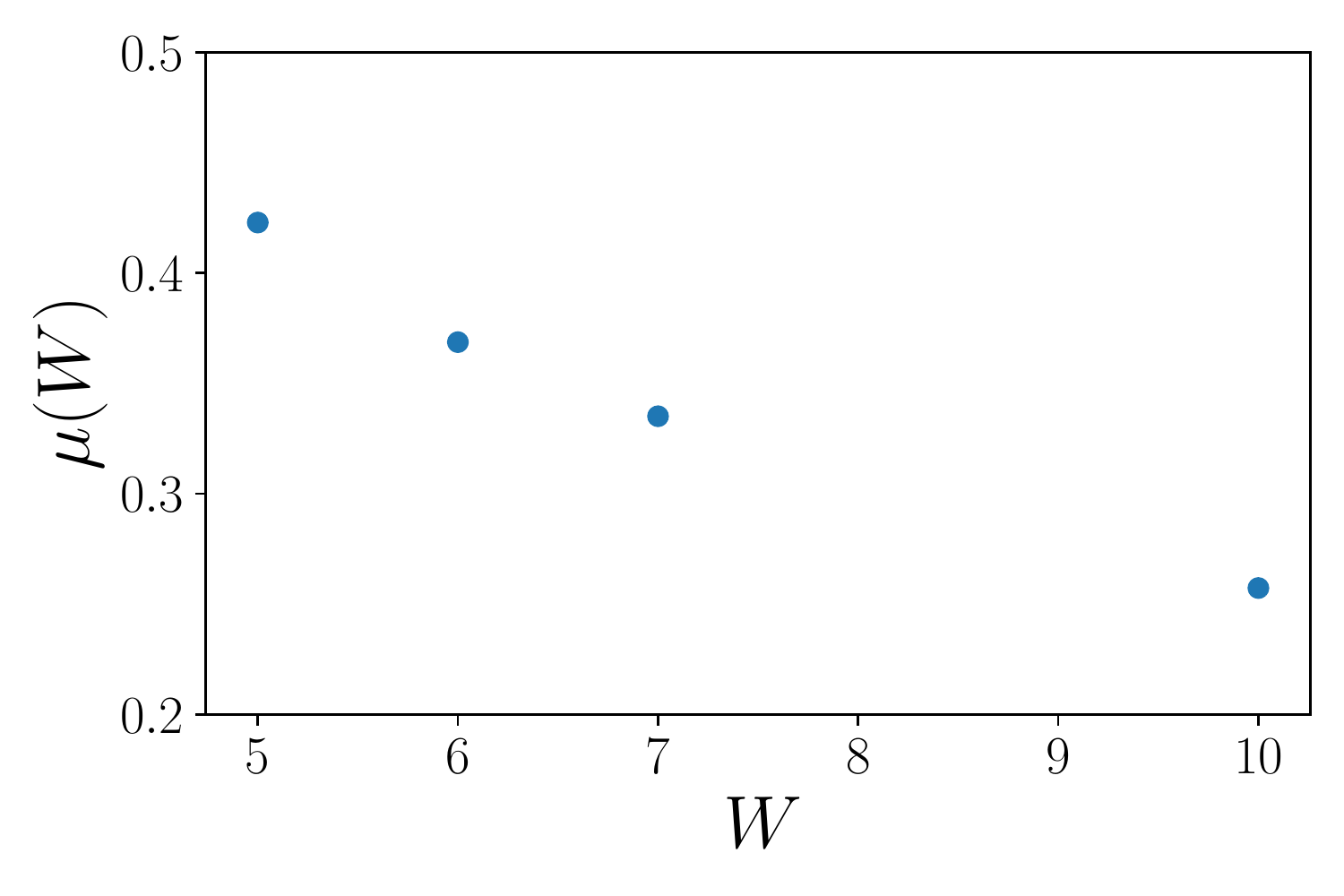}\endminipage
\caption{Dynamical eigenstate correlations for the spin chain of length $L=16$. {\it Left:}  Correlation function $\beta(\omega)$ for disorders $W = 5, 6, 7, 10$ (from cyan to magenta). {\it Right:}  Exponent $\mu(W)$ characterizing the power-law frequency scaling $\beta(\omega) \propto \omega^{-\mu(W)}$.
This figure is a spin-chain counterpart of Fig.~\ref{beta_rrg} for the RRG model.
}
\label{beta_spins}
\end{figure*}

For the spin-chain model (\ref{eq:H}), we study below the same correlation functions as for the RRG model: the finite--frequency correlation function $\beta(\omega)$,
Eq.~(\ref{sigmadef}), and the correlation function of closest-in-energy wave functions, $\beta_{\textrm{nn}}$, Eq.~(\ref{beta-nn}).  Let us emphasize that these quantities now characterize exact many-bony eigenstates $\psi_k(j)$. Here the index $k$ labels eigenstates and the argument $j$ runs over basis states of the Hilbert space which are eigenstates of $S_i^z$ for all $i$ (and thus are eigenstates of the Hamiltonian in the extreme-localization limit $W = \infty$). 

Before presenting our results for the frequency-dependent wave function correlations, we briefly recall the existing knowledge about the average inverse participation ratio which characterizes statistical properties of an individual many-body eigenstate, Eq.~(\ref{ipr}). In the delocalized phase, it was found \cite{mace19multifractal} that eigenfunctions in the model of Eq. (\ref{eq:H}) are ergodic in the sense that
\be
\label{mbl:erg}
P_2 \propto 1/N.
\ee 
Here $N$ is the volume of the many-body space, i.e. the dimensionality of the subspace of the full Hilbert space that is allowed by conservation laws. In the model of Eq. (\ref{eq:H}) and in the zero magnetization sector $S_z=0$, which we consider below (we limit ourselves to even $L$ only), one has 
\be 
N=\frac{L!}{ [(L/2)!]^2} \simeq 2^L\sqrt\frac{2}{\pi L}.
\ee 
This ergodic behavior of the inverse participation ratio in the spin-chain model is fully analogous to that in the delocalized phase  of the RRG model  (see Sec.~\ref{s1}).  At the same time, on the localized side of the transition, there is some difference in the scaling of the inverse participation ratio in the RRG model and in the genuine many-body problem (like a spin chain). While $P_2 \sim 1$ in the localized phase on RRG, one finds a fractal scaling
\be
P_2 \sim N^{-\tau(W)}
\label{mbl:loc_frac}
\ee
in the MBL phase of spin-chain models  \cite{Luitz2016a,gornyi-adp16,tikhonov18,mace19multifractal}, with a disorder-dependent exponent $\tau(W)$. It was shown
\cite{tikhonov18} that in the strong disorder regime (large $W$), the exponent $\tau(W)$ scales as 
\be 
\tau(W) \propto 1/W
\label{tau-W}
\ee
 with disorder. Further, at criticality one also finds the fractal scaling (\ref{mbl:loc_frac}), with the exponent $\tau(W_c)$ that is equal to the limiting value of $\tau(W)$ at $W \to W_c +0$. (This is another manifestation of the fact that the critical point in the MBL problem has properties of the localized phase.)   The non-trivial scaling in Eq. (\ref{mbl:loc_frac}) originates from a finite density (in real space) of local resonances that are not able to establish a global delocalization  but lead  to an exponential increase of the support of the many-body wavefunction in the spin configuration basis of the Hamiltonian Eq. (\ref{eq:H}). A detailed analysis of the model  (\ref{eq:H}) revealed \cite{mace19multifractal} that the scaling (\ref{tau-W}) is valid with a good accuracy up to the critical point and yielded $\tau(W_c) \approx 0.2$.

Let us now estimate the contribution ot Mott--like resonances to the correlation function $\beta(\omega)$ in the MBL phase of the spin chain. Consider a given basis state $\ket{j}$, i.e., an eigenstate of all $S_i^z$. For strong disorder (deeply in the MBL phase), $\ket{j}$ is close to an exact eigenstate $\psi_k$.
More precisely, there exists a small density $\sim 1/W$ of lowest-order resonant processes (flips of  pairs $s_{r},s_{r+1}$ of adjacent spins) which ``dress'' the state $\ket{j}$, leading to fractality of the inverse participation ratio discussed above. To estimate the number of higher-order resonant processes, we consider $n$--th order of the perturbation theory in interaction. It is important that involved spins should form a connected cluster (of maximal length $2n$) in order to guarantee that this tentative
resonant process does not decouple into independent pieces. This is clear already in case of $n=2$: consider a process involving
spin flips in two remote pairs $s_{1},s_{2}$ and $s_{r},s_{r+1}$ such that $r>3$. This process is not a resonant one, even if
the energies of initial and final states are arbitrarily close. In the perturbation theory, 
inability of such a process to create a resonance happens due to cancellation between two amplitudes, corresponding to flipping the disconnected pairs $1,2$ and $r,r+1$ in two distinct orders\,
see Ref. \cite{gornyi-adp16} and references therein. As a result, the number of processes that can actually lead to resonance in the $n$--th order of perturbation theory scales as $N_{n;L}\sim L\rho(n)$ with $\rho(n)$ independent on the system length $L$.  In other words, $\rho(n)$ is the spatial density of $n$-th order processes which may potentially lead to resonances.
Crucially, $\rho(n)\sim m^n$ grows exponentially with $n$, in analogy with the RRG problem \cite{gornyi-adp16}, and $m$ is independent on $n$.
 In a conventional spin chain, we thus have the branching factor $m=O(1)$. (One can have a parametrically large $m$ in a chain of coupled ``spin quantum dots'' with large number of spins per dot \cite{gornyi-adp16}.) 

The density $\rho(n)$ was considered in the context of ac conductivity in Ref.~\cite{gopalakrishnan2015low}, where it was denoted $e^{s(\gamma)n}$ and $s(\gamma)$ was termed ``configuration entropy per flipped spin of the possibly resonant clusters''. The argument $\gamma$ was introduced to emphasize that $s$ is actually a fluctuating quantity.  Our effective branching number $m$ thus corresponds to $e^s$ of Ref. \cite{gopalakrishnan2015low}; the fluctuations of $m$ are discarded in our simplified argument.

The number of ``potentially resonant'' processes for a given initial state scales therefore as 
\be
N_{n;L}\sim L m^n \,,
\ee
with $m$ of order unity, which is the same behavior as on RRG, up to an overall factor $L$. This behavior is responsible for the MBL transition in a spin chain taking place at a disorder $W_c$ of order unity (i.e., independent on $L$.  We can now repeat, with minor modifications, the simplified analysis performed for RRG in Sec.\ref{s2}, which yields [cf. Eq.~\eqref{beta-omega-res}]
\bea
N^2\omega\beta(\omega)
& \sim &  \sum_{l \,:\:   \omega < | E_k - E_l | < 2\omega}
\left\langle \sum_j |\psi_k(j) \psi_l(j)|^2 \right\rangle  \nonumber \\[0.3cm]
& \sim & (N\Delta) \: p_\omega P_{\textrm{res}},
\label{mbl_beta_derivation}
\eea
where $\Delta$ is the many-body level spacing, $p_\omega$ is the number of resonances (for a given state $k$) in the band $\left[\omega,2\omega\right]$, and $P_{\textrm{res}}$ is the resonant overlap,
$$
P_{\textrm{res}}=\sum_j |\psi_k(j)|^2 |\psi_l(j)|^2 \,.
$$ 
 Let us estimate three factors $N\Delta$, $p_\omega$, and $P_{\textrm{res}}$  in Eq. (\ref{mbl_beta_derivation}): 
 \begin{enumerate}
 \item[(i)] The typical energy of an eigenstate of all $S_i^z$ of $L$ spins is $\sim\sqrt{L}$, hence $N\Delta \sim \sqrt{L}$. 
  \item[(ii)] 
 The average number of processes that represent potential resonances in the band $[\omega,2\omega]$ can be estimated,
 in analogy with Eq.~\eqref{p-omega},
  as 
  \be
  p_\omega\sim \omega N_{R(\omega);L} \sim \omega L m^{R(\omega)}=L \omega^{1-\mu(W)} \,
  \ee
   with $R(\omega)$ as in Eq.~\eqref{R-omega} and $\mu(W)$ as in Eq. (\ref{mu-omega}). 
  \item[(iii)] 
  The resonant overlap scales in the same way as the inverse participation ratio \eqref{mbl:loc_frac}, i.e., 
  \be
  P_{\textrm{res}}\sim N^{-\tau(W)}.
  \label{P-res}
  \ee
  In this respect, the spin-chain problem differs from the RRG model, for which $P_{\textrm{res}} \sim 1$, see Eq.\eqref{resonance-overlap}.
 \end{enumerate}

Combining the above estimates and using $L\approx \log_2 N$, we finally get
\be
N^2\beta(\omega)\sim \omega^{-\mu(W)}\frac{(\log_2 N)^{3/2}}{N^{\tau(W)}}.
\label{mbl-final-scaling}
\ee
This derivation should be viewed as substantially oversimplified since fluctuations were not fully taken account. Nevertheless, this treatment is sufficient to understand the emergence of power--law dynamical scaling (revealed by numerical results below) with continuously varying exponent in the MBL phase. 

The result \eqref{mbl-final-scaling} is largely the same as Eq.~(\ref{betafit}) for RRG; the most important factor is the power-law frequency dependence $\omega^{-\mu(W)}$. The only difference is in the additional $N$-dependent factor. We note, however, that the exponent $\tau(W)$ is parametrically small ($\sim 1/W$) in the MBL phase, remaining numerically quite small at the critical point. Also, for realistic $N$, the factor $N^{\tau(W)}$ in the denominator is essentially compensated by the logarithmic factor in the numerator. So, in practice, the difference between the results for RRG and for the spin chain is relatively minor.

We now present the numerical results for the dynamical eigenstate correlations in the spin-chain model (\ref{eq:H}).  To evaluate the correlation function $\beta(\omega)$ we computed, via exact diagonalization, eigenstates of Eq. (\ref{eq:H}) in the $S_i^z$ basis. We studied systems of sizes $L$ in the range $12 \: \textup{--}\: 18$ and averaged over $5\cdot 10^5$ (for smallest systems) -- $5\cdot 10^2$ (for largest systems) realizations of disorder. For each disorder realization, we determined the middle of the band $E$ by the condition  $(E-E_\textrm{min})/(E_\textrm{max}-E_\textrm{min}) = 0.5$, where $E_\textrm{min}$ and $E_\textrm{max}$ are the lowest and the largest eigenstate energy, and considered $1/32$ fraction of all states around the middle of the band. The correlation function $\beta(\omega)$ for various strengths of disorder, from delocalized to the MBL phase, is  shown  in Fig. \ref{beta_spins_full}, which is a direct spin-chain counterpart of Fig.~\ref{beta_rrg} for the RRG model. The observed behavior is fully analogous to that found in Fig.~\ref{beta_rrg}. For sufficiently weak disorder, $W=1.5$, 1.7, 2, we see a power-law critical behavior at higher frequencies with a saturation at low frequencies. As we show below, the saturation value confirms the ergodicity of the delocalized phase. For the intermediate disorder values, $W=3$ and 4, the tendency towards saturation is also achieved but we are still far from reaching the full saturation.  This is an indication of the fact that these two values are also on the delocalized side of the MBL transition (in the thermodynamic limit) but the system sizes are too small to observe ergodicity. In other words, these values correspond to the critical regime for system sizes that can be studied via exact diagonalization. 
For strong disorder, $W=6$ and 10, the data exhibit a power-law behavior in the full range of frequencies, which is a hallmark of the MBL phase. 

In Fig.~ \ref{beta_spins} we show the results for  strong disorder, from  $W=5$ (approximately the critical point) till $W=10$ (deeply in the MBL phase), cf. the analogous figure \ref{beta_rrg} for the RRG model. In the left panel, we see once more that $\beta(\omega)$ in the MBL phase shows a power-law frequency scaling, $\beta(\omega) \propto \omega^{-\mu(W)}$, with the disorder-dependent exponent
(slope on the log-log scale). In the right panel, the corresponding exponent $\mu(W)$ is plotted as a function of disorder. 
This figure is again similar to the right panel of Fig.~\ref{beta_rrg} although numerical values of the exponent $\mu(W)$  are somewhat smaller than in the RRG model. 

\begin{figure*}[tbp]
\minipage{0.5\textwidth}\includegraphics[width=\textwidth]{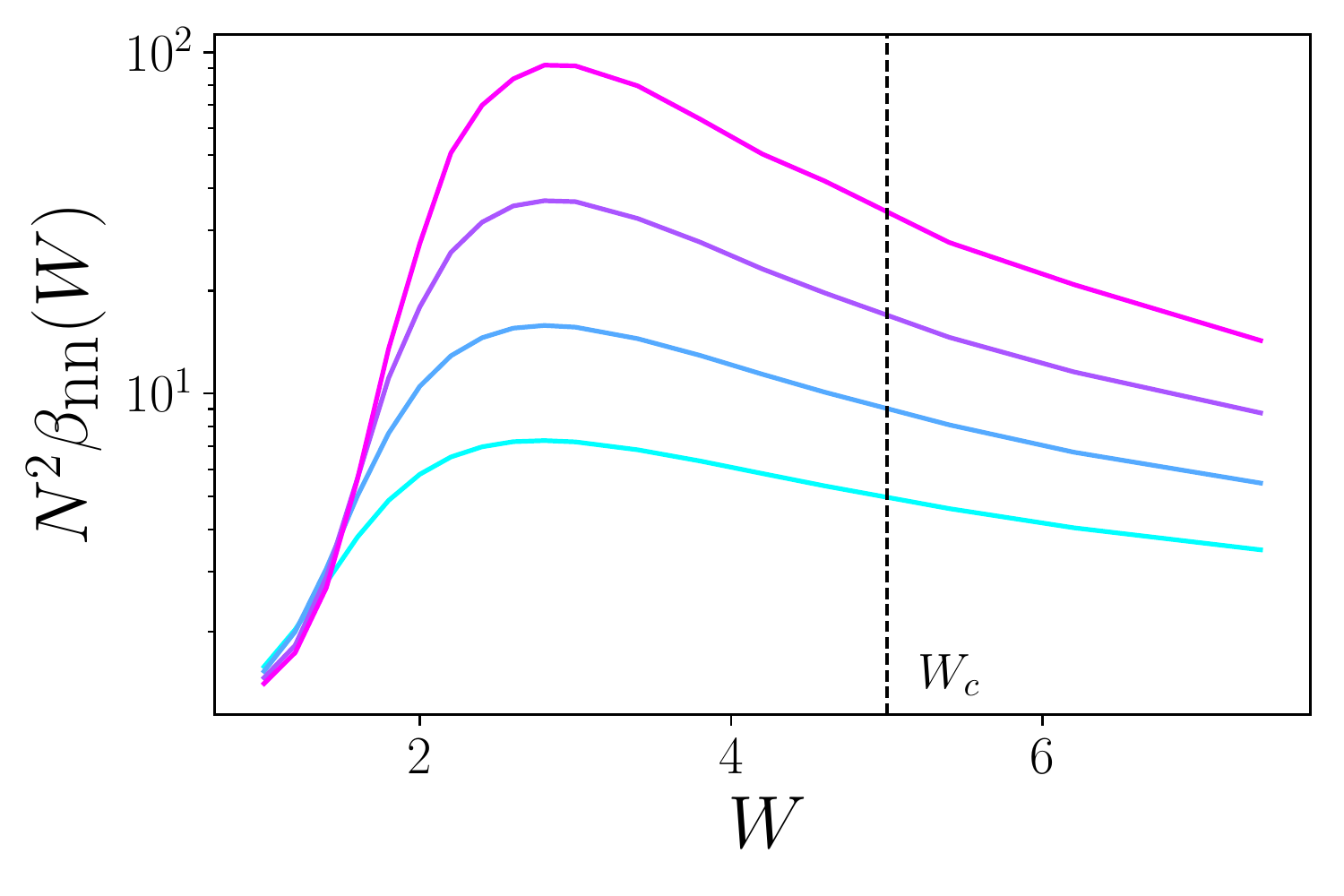}\endminipage
\minipage{0.5\textwidth}\includegraphics[width=\textwidth]{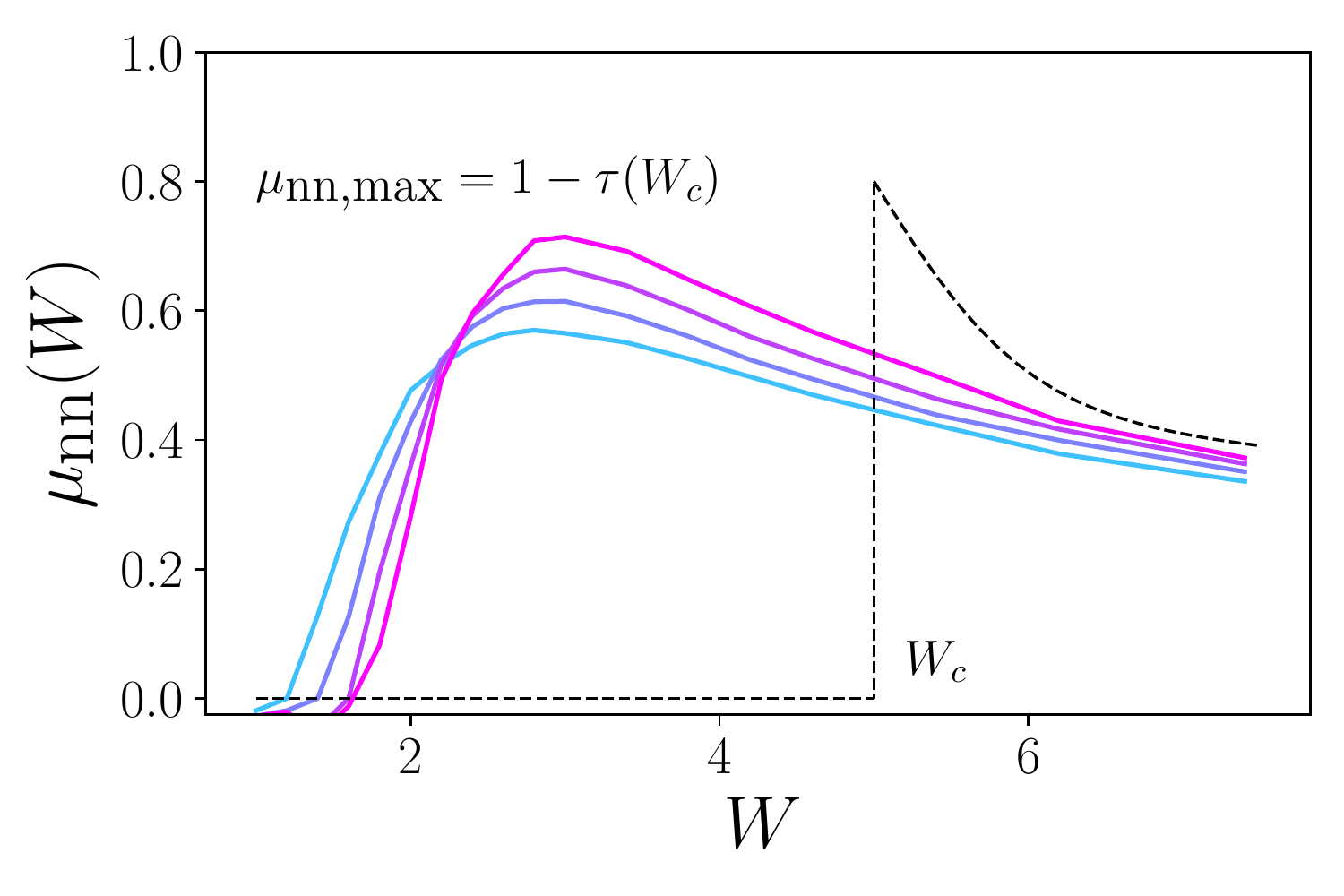}\endminipage
\caption{Correlation of adjacent wavefunctions for spin chain. 
System sizes are $n=12, 14, 16, 18$ (from cyan to magenta).
{\it Left:}  Correlation function $\beta_{\textrm{nn}}(W)$. 
Vertical dashed line marks an approximate value of the critical point of the MBL transition,  $W_c \simeq 5$, as obtained from the quantum dynamics in large chains \cite{Doggen2018a}.
{\it Right:}  Exponent $\mu_{\textrm{nn}}$ characterizing the $N$ scaling of adjacent-state correlations, see Eq. (\ref{alphafit}).  Dashed line shows theoretically expected $N\to\infty$ behaviour, see Eq.~(\ref{betann-del}) for the delocalized phase and Eq.~(\ref{betann}) for the localized phase (this part of the dashed line is schematic.) 
This figure is a spin-chain counterpart of Fig.~\ref{alpha_rrg} for the RRG model.
}
\label{alpha_spins}
\end{figure*}

A related quantity---the adjacent-state correlation function $\beta_{\textrm{nn}}$ defined by Eq.~(\ref{beta-nn})---is shown in  Fig.~\ref{alpha_spins}. The results are very similar to their RRG analog, Fig.~\ref{alpha_rrg}.  The left panel of Fig.~\ref{alpha_spins} displays the correlation function $\beta_{\textrm{nn}}(W)$ in a broad range of disorder strengths for several values of the system size $L$. Like in the case of RRG, $W < W_c$, the curves gradually approach, with increasing $L$, a limiting curve, thus demonstrating ergodicity of the delocalized phase. For system sizes available for exact diagonalization the ergodic (large-$L$) behavior is reached for $W \lesssim 2$.  In full analogy with the RRG model, the $\beta_{\textrm{nn}}(W)$ curves exhibit a maximum near $W \approx 3$ that serves as a finite-size estimate for the critical point and slowly drifts towards the actual ($L \to \infty$) value of $W_c$.

The right panel  shows the  flowing exponent $\mu_{\textrm{nn}}$ defined by Eq. (\ref{alphafit}).  In general, the curves are rather similar to those for RRG in the right panel of Fig.~\ref{alpha_rrg}. However, it is worth noticing a difference in the maximum value of $\mu_{\textrm{nn}}$ for the largest system size. While for the RRG model this maximum value is equal to unity with a good accuracy, in the spin-chain case the maximum value is approximately 0.75. One reason for this is finite-size effects which are considerably stronger for the spin chain than for RRG.  In fact, there is also a deeper reason for this difference, which should remain also in the limit $N \to \infty$. Indeed, let us consider a system at criticality ($W= W_c$) in the large-$N$ limit. Recall that the inverse participation ratio $P_2$ at the critical point shows ``fractal'' scaling (\ref{mbl:loc_frac}). The overlap of two adjacent states at criticality is expected to follow the same power-law scaling, $N \beta_{\rm nn} \sim N^{-\tau(W_c)}$   [cf. Eq.~\eqref{P-res}], which implies (at $N\to \infty$)
\be
\mu_{\textrm{nn}}(W_c)=1-\tau(W_c) \,,
\ee
yielding $\mu_{\textrm{nn}}(W_c) \approx 0.8$. The same result is obtained from Eq.~\eqref{mbl-final-scaling}  if one extends it from the MBL phase to the critical point and sets $\mu(W_c)= 1$ (as on RRG).  More generally, Eq.~\eqref{mbl-final-scaling} suggests a relation between the exponents in the MBL phase,
\be
\label{exponent-relation}
\mu_{\textrm{nn}}(W)= \mu(W) - \tau(W) \,.
\ee
At large $W$ the exponent $\tau(W)$ is small ($\sim 1/W$), so that $\tau(W) \ll \mu(W)$ and thus
\be
\label{exponent-relation-approx}
\mu_{\textrm{nn}}(W) \approx \mu(W) \,.
\ee
This remains valid with reasonable accuracy up to the critical point since $\tau(W_c)$ is quite small.  It is also worth noting that logarithmic corrections to scaling, like the logarithmic factor in Eq.~\eqref{mbl-final-scaling}, and further strong finite-size effects substantially influence numerical values of exponents characterizing the MBL phase as obtained by means of exact diagonalization.

The expected extrapolation of $\mu_{\textrm{nn}}(W)$ to the thermodynamic limit $L\to\infty$ is shown by a dashed line in the right panel of the Fig. \ref{alpha_spins}. 
Similarly to the $\beta_{\textrm{nn}}$ peak, the position of the peak in $\mu_{\textrm{nn}}$ provides a finite-size estimate for the position of the transition and drifts, with increasing $L$, towards $W_c$, see Table \ref{table-W-peak-mbl}. The drift is approximately linear with number of spins $L$;  these system sizes are clearly too small to allow for a reliable estimate of the thermodynamic-limit critical disorder $W_c$ As in the RRG model, a considerable part of the delocalized phase gives rise to a  broad critical regime, $2.5 \lesssim W \lesssim 5$,  for system sizes available for exact diagonalization.

\begin{table}
\begin{center}
 \begin{tabular}{|c|c c c c|} 
 \hline
$L$ & $12$ & $14$ & $16$ & $18$\\
\hline
$W_\mathrm{peak}(L)$ &$2.66$ & $2.84$ & $3.01$ & $3.11$\\
 \hline
\end{tabular}
\end{center} 
\caption{Position $W_\mathrm{peak}(L)$ of the maximum of $\mu_{\textrm{nn}}(W)$ curves that can be viewed as an $L$-dependent apparent critical point.  Upon increasing $L$, it shows a slow drift towards the limiting ($L\to \infty$) value.}
\label{table-W-peak-mbl}
\end{table}

\section{Summary}
\label{s4}

In this paper, we have studied dynamical eigenstate correlations across the MBL transition. This was done for two models: (i) the RRG model that serves as  a toy-model of the MBL transition and (ii) a spin chain representing a genuine many-body problem. The results for both models were found to be very similar. The main observables that we have considered are the frequency-dependent eigenstate correlation function $\beta(\omega)$ and the adjacent-state correlation function $\beta_{\textrm{nn}}$. For both of them, we explored dependences on disorder $W$ and on the system size. We have introduced the exponent $\mu(W)$ controlling the scaling of $\beta(\omega)$ with frequency $\omega$ and the running exponent
$\mu_{\textrm{nn}}(W, N)$ characterizing the scaling of $N^2 \beta_{\textrm{nn}}$ with $N$. Our key findings are briefly summarized below.

\begin{itemize}

\item[(i)]  For $W < W_c$ our results confirm the ergodicity of the delocalized phase.  In particular, the correlation function $N \beta_{\textrm{nn}}$ shows at large $N$ the ergodic $1/N$ scaling, in analogy with the inverse participation ratio $P_2$. Equivalently, the exponent $\mu_{\textrm{nn}}(W, N)$ tends to zero at $N \to \infty$.  

\item[(ii)] Dynamical eigenstate correlations in the localized phase $W > W_c$ are characterized,  in the large-$N$ limit, by fractal scaling, $N^2\beta(\omega) \sim \omega^{-\mu(W)}$ and
$N^2\beta_{\textrm{nn}} \sim N^{\mu_{\rm nn}(W)}$, with disorder-dependent exponents $\mu(W)$ and $\mu_{\rm nn}(W)$.  
The source of the power-law enhancement of correlations with lowering $\omega$ is Mott-like resonances between distant localized states. 
For finite $N$ (as in the exact-diagonalization numerics), the exponents are subjected to finite-size corrections.  Our analytical arguments (for $N \to \infty$) suggest that for RRG model $\mu(W) = \mu_{\rm nn}(W)$, while for the spin-chain problem there is a small difference between these exponents due to fractal scaling of the inverse participation ratio in the MBL phase. 
Since the critical point has a localized character, the value $\mu(W_c)$ is equal to the limit of $\mu(W)$ at $ W \to W_c +0$, and similarly for $\mu_{\rm nn}(W_c)$. On RRG we find (again on the basis of analytical arguments that assume the large-$N$ limit) $\mu_{\rm nn}(W_c)=\mu(W_c)=1$, while for the spin chain  $\mu(W_c)=1$ and  $\mu_{\rm nn}(W_c) = 1- \tau(W_c) \approx 0.8$, where $\tau(W)$ is the exponent characterizing the fractality of the inverse participation ratio in the localized phase. With increasing disorder, the exponents $\mu(W)$ and $\mu_{\rm nn}(W)$ decay rather slowly, $\mu(W) \approx \mu_{\rm nn}(W) \sim 1/ [\ln (W/W_c)]$.

\item[(iii)]  The correlation function $\beta_{\textrm{nn}}$ and the corresponding exponent $\mu_{\textrm{nn}}$ exhibit, as functions of disorder $W$, a maximum that serves as an indication of the MBL transition. With increasing $N$, the positions of this maxima drift towards $W_c$. This drifts is, however, rather slow, so that the position of the maximum remains quite far from the actual $W_c$ for all system sizes amenable to exact diagonalization. This is a manifestation of strong finite-size effects in the MBL problems \cite{bera2017density,Doggen2018a,abanin2019distinguishing,sierant2020polynomially}, which make extremely difficult a reliable determination of the critical point of the MBL transition and of the associated critical behavior on the basis of exact diagonalization.  A closely related observation is a rather broad critical regime on the delocalized side of $W_c$, where the system sizes that can be treated by exact diagonalization are too small in order to reach (even approximately) the ergodic behavior. 

\end{itemize}

Let us now discuss a possible way to experimentally measure the frequency-dependent eigenstate correlations. In this connection, consider the return probability $p(t)$ to a many-body state $\psi(0)$:
\be
p(t) = | \langle \psi(t) | \psi(0) \rangle |^2 \,,
\label{pt}
\ee
where $\psi(t) = e^{-iHt} \psi(0)$ follows the dynamics determined by the Hamiltonian $H$. Let us choose one of the basis states as the initial state, $|\psi(0) \rangle = | j \rangle$. Expanding $\psi(0)$ and $\psi(t)$ in terms of the eigenstates $\psi_k$, we get
\be
p(t) = \sum_{kl} e^{-i(E_k-E_l)t} |\psi_k(j)|^2  |\psi_l(j)|^2 \,.
\label{pt-expanded}
\ee
In the limit $t \to \infty$, the return probability is determined by the diagonal ($k=l$) terms in Eq.~(\ref{pt-expanded}), which yields
\be
p(t \to \infty) \equiv p_\infty = \sum_k |\psi_k(j)|^4 = P_2^{(j)} \,.
\label{pt-limiting}
\ee
Here $P_2^{(j)}$ is the inverse participation ratio (\ref{ipr}), with a slight difference that the summation goes over $k$ rather than over $j$ (i.e.  it characterizes the expansion of a basis state over exact eigenstates). This difference is not essential (and disappears completely upon averaging). Focussing on the MBL phase and the critical point, we have thus
\be
p_\infty \sim N^{-\tau(W)} \,.
\ee

The dynamical part of the return probability $p(t)$ is given by non-diagonal terms in Eq.~(\ref{pt-expanded}):
\bea
p(t) - p_\infty &=&  \sum_{k \ne l} e^{-i(E_k-E_l)t} |\psi_k(j)|^2  |\psi_l(j)|^2 
\nonumber \\
&=& \int dE \, \nu^2(E) \int d\omega \, e^{-i \omega t} N^2 \beta(\omega) \,.
\label{pt-dynamical}
\eea
where $\beta(\omega)$ is the eigenstate correlation function (\ref{sigmadef}). (Note that $\beta(\omega)$ implicitly depends on $E$.) The many-body density of states $\nu(E)$ is  sharply peaked near the middle of the band, so that the integral in Eq.~(\ref{pt-dynamical}) is governed by the vicinity of the corresponding value of $E$.  Using Eq.~(\ref{betafit}) or Eq.(\ref{mbl-final-scaling}),  we get a power-law temporal decay of the many-body return probability in the localized phase and at the critical point:
\be
p(t) - p_\infty \sim t^{-1+ \mu(W)} \,.
\label{pt-power-law}
\ee
Here we have discarded the $N$-dependent factor in Eq.~(\ref{mbl-final-scaling}) that is not that important in practice in view of smallness of the exponent $\tau(W)$ and of large compensation between the logarithmic and power-law factor for realistic $N$, see comment after Eq.~(\ref{mbl-final-scaling}). 

As has been already pointed out in the Introduction, Sec.~\ref{s1}, the return probability $p(t)$ can be efficiently measured in engineered many-body systems (quantum simulators or quantum processors), such as arrays of trapped ions, atoms, and supeconducting qubits \cite{zhang2017observation,bernien2017probing,Chiaro2019a,google_quantum_supremacy}. The state-of-the-art devices contain $\approx 50$ elements [``qubits'' analogous to spins in the Hamiltonian (\ref{eq:H})]; it is expected that this number will grow up to $\approx 100$ in near future. Clearly, the full quantum-state tomography is impossible in such devices, in view of the huge size of the many-body Hilbert space, $N \sim 2^{50}$\:--\:$2^{100} \approx 10^{15}$\:--\:$10^{30}$.  At the same time, the measurement of the many-body return probability $p(t)$ [i.e., of the Fourier transform of the dynamical eigenstate correlation function $\beta(\omega)$] is absolutely feasible. This is done by measuring the evolved state $\psi(t)$ in the non-interacting basis $j$ (i.e. measuring all $S_i^z$). If the measurement is performed, e.g.,  $ \sim 10^4$ times (as in Refs.~\cite{zhang2017observation,bernien2017probing}), one can determine $p(t)$ as long as it is $\gtrsim 10^{-4}$. For a rather slow, power-law decay (\ref{pt-power-law}), this allows one to determine $p(t)$ up to very long times $t > 10^4$ (in microscopic units set by characteristic magnitude of parameters in the Hamiltonian). In a related experiment on quantum processors \cite{google_quantum_supremacy}, the number of measurements was $\sim 10^6$, implying a possibility to proceed up to $p(t)$ as small as $\sim 10^{-6}$. Experimental investigation of the eigenstate correlations across the MBL transition is thus a very promising avenue for future research.

Finally, it is worth mentioning connections between our results for the eigenstate correlation function $\beta(\omega)$ and the behavior of other dynamical observables. In particular, Ref.~\cite{serbyn2017thouless} studied frequency dependence of matrix elements of local (in real space) operator $S_z^i$, where $i$ is a given site of the lattice (see also a similar study in a recent Ref.~\cite{sels2020dynamical}). In our notations, this means the following correlation function
\be
\langle | \left(S_z^i \right)_{kl} |^2 \rangle    \equiv 
\langle | \langle \psi_k | S_z^i |\psi_l \rangle |^2 \rangle \,,
\label{local-spin-corr}
\ee
considered as a function of the frequency $\omega = E_k - E_l$.   Here $\psi_k$ and $\psi_l$ are exact many-body eigenstates, and $E_k$ and $E_l$ the corresponding energies, as in Eq.~(\ref{sigmadef}).  The $\omega \to t$ Fourier transform of Eq.~(\ref{local-spin-corr}) can be viewed as a return probability in real space, which is in general very different from the return probability in the many-body space $p(t)$ given by the Fourier transform of the correlation function $\beta(\omega)$ studied in the present work. At the same time, there is a remarkable similarity in the behavior of both correlation functions in the localized phase (and at criticality): they both show a power-law dependence on frequency, with a continuously changing exponent. A related power-law scaling of the ac conductivity, $\sigma(\omega) \sim \omega^\alpha$, with $1 < \alpha < 2$ in the MBL phase,  was found in Ref.~\cite{gopalakrishnan2015low}. A better understanding of connections between the exponents governing the scaling of the observables characterizing dynamics in the real space [like the correlation functions of the type  (\ref{local-spin-corr}) or the conductivity] and in the many-body Hilbert space [the correlation function $\beta(\omega)$ studied in this work and its Fourier transform $p(t)$] remains an interesting goal for future research.

\section{Acknowledgments}

We thank G. Lemari\'e and M. Serbyn for useful discussions. This research was financially supported by the DFG-RFBR Grant [No. MI 658/12-1  (DFG) and No. 20-52-12034 (RFBR)].

\bibliography{rrg}

\begin{thebibliography}{87}%
\makeatletter
\providecommand \@ifxundefined [1]{%
 \@ifx{#1\undefined}
}%
\providecommand \@ifnum [1]{%
 \ifnum #1\expandafter \@firstoftwo
 \else \expandafter \@secondoftwo
 \fi
}%
\providecommand \@ifx [1]{%
 \ifx #1\expandafter \@firstoftwo
 \else \expandafter \@secondoftwo
 \fi
}%
\providecommand \natexlab [1]{#1}%
\providecommand \enquote  [1]{``#1''}%
\providecommand \bibnamefont  [1]{#1}%
\providecommand \bibfnamefont [1]{#1}%
\providecommand \citenamefont [1]{#1}%
\providecommand \href@noop [0]{\@secondoftwo}%
\providecommand \href [0]{\begingroup \@sanitize@url \@href}%
\providecommand \@href[1]{\@@startlink{#1}\@@href}%
\providecommand \@@href[1]{\endgroup#1\@@endlink}%
\providecommand \@sanitize@url [0]{\catcode `\\12\catcode `\$12\catcode
  `\&12\catcode `\#12\catcode `\^12\catcode `\_12\catcode `\%12\relax}%
\providecommand \@@startlink[1]{}%
\providecommand \@@endlink[0]{}%
\providecommand \url  [0]{\begingroup\@sanitize@url \@url }%
\providecommand \@url [1]{\endgroup\@href {#1}{\urlprefix }}%
\providecommand \urlprefix  [0]{URL }%
\providecommand \Eprint [0]{\href }%
\providecommand \doibase [0]{http://dx.doi.org/}%
\providecommand \selectlanguage [0]{\@gobble}%
\providecommand \bibinfo  [0]{\@secondoftwo}%
\providecommand \bibfield  [0]{\@secondoftwo}%
\providecommand \translation [1]{[#1]}%
\providecommand \BibitemOpen [0]{}%
\providecommand \bibitemStop [0]{}%
\providecommand \bibitemNoStop [0]{.\EOS\space}%
\providecommand \EOS [0]{\spacefactor3000\relax}%
\providecommand \BibitemShut  [1]{\csname bibitem#1\endcsname}%
\let\auto@bib@innerbib\@empty
\bibitem [{\citenamefont {Gornyi}\ \emph {et~al.}(2005)\citenamefont {Gornyi},
  \citenamefont {Mirlin},\ and\ \citenamefont
  {Polyakov}}]{gornyi2005interacting}%
  \BibitemOpen
  \bibfield  {author} {\bibinfo {author} {\bibfnamefont {I.}~\bibnamefont
  {Gornyi}}, \bibinfo {author} {\bibfnamefont {A.}~\bibnamefont {Mirlin}}, \
  and\ \bibinfo {author} {\bibfnamefont {D.}~\bibnamefont {Polyakov}},\
  }\href@noop {} {\bibfield  {journal} {\bibinfo  {journal} {Physical Review
  Letters}\ }\textbf {\bibinfo {volume} {95}},\ \bibinfo {pages} {206603}
  (\bibinfo {year} {2005})}\BibitemShut {NoStop}%
\bibitem [{\citenamefont {Basko}\ \emph {et~al.}(2006)\citenamefont {Basko},
  \citenamefont {Aleiner},\ and\ \citenamefont {Altshuler}}]{basko2006metal}%
  \BibitemOpen
  \bibfield  {author} {\bibinfo {author} {\bibfnamefont {D.}~\bibnamefont
  {Basko}}, \bibinfo {author} {\bibfnamefont {I.}~\bibnamefont {Aleiner}}, \
  and\ \bibinfo {author} {\bibfnamefont {B.}~\bibnamefont {Altshuler}},\
  }\href@noop {} {\bibfield  {journal} {\bibinfo  {journal} {Annals Of
  Physics}\ }\textbf {\bibinfo {volume} {321}},\ \bibinfo {pages} {1126}
  (\bibinfo {year} {2006})}\BibitemShut {NoStop}%
\bibitem [{\citenamefont {Abanin}\ \emph
  {et~al.}(2019{\natexlab{a}})\citenamefont {Abanin}, \citenamefont {Altman},
  \citenamefont {Bloch},\ and\ \citenamefont {Serbyn}}]{abanin2019colloquium}%
  \BibitemOpen
  \bibfield  {author} {\bibinfo {author} {\bibfnamefont {D.~A.}\ \bibnamefont
  {Abanin}}, \bibinfo {author} {\bibfnamefont {E.}~\bibnamefont {Altman}},
  \bibinfo {author} {\bibfnamefont {I.}~\bibnamefont {Bloch}}, \ and\ \bibinfo
  {author} {\bibfnamefont {M.}~\bibnamefont {Serbyn}},\ }\href@noop {}
  {\bibfield  {journal} {\bibinfo  {journal} {Reviews of Modern Physics}\
  }\textbf {\bibinfo {volume} {91}},\ \bibinfo {pages} {021001} (\bibinfo
  {year} {2019}{\natexlab{a}})}\BibitemShut {NoStop}%
\bibitem [{\citenamefont {Gopalakrishnan}\ and\ \citenamefont
  {Parameswaran}(2020)}]{gopalakrishnan2020dynamics}%
  \BibitemOpen
  \bibfield  {author} {\bibinfo {author} {\bibfnamefont {S.}~\bibnamefont
  {Gopalakrishnan}}\ and\ \bibinfo {author} {\bibfnamefont {S.}~\bibnamefont
  {Parameswaran}},\ }\href@noop {} {\bibfield  {journal} {\bibinfo  {journal}
  {Physics Reports}\ } (\bibinfo {year} {2020})}\BibitemShut {NoStop}%
\bibitem [{\citenamefont {Oganesyan}\ and\ \citenamefont
  {Huse}(2007{\natexlab{a}})}]{oganesyan07}%
  \BibitemOpen
  \bibfield  {author} {\bibinfo {author} {\bibfnamefont {V.}~\bibnamefont
  {Oganesyan}}\ and\ \bibinfo {author} {\bibfnamefont {D.~A.}\ \bibnamefont
  {Huse}},\ }\href@noop {} {\bibfield  {journal} {\bibinfo  {journal} {Phys.
  Rev. B}\ }\textbf {\bibinfo {volume} {75}},\ \bibinfo {pages} {155111}
  (\bibinfo {year} {2007}{\natexlab{a}})}\BibitemShut {NoStop}%
\bibitem [{\citenamefont {Monthus}\ and\ \citenamefont
  {Garel}(2010)}]{monthus10}%
  \BibitemOpen
  \bibfield  {author} {\bibinfo {author} {\bibfnamefont {C.}~\bibnamefont
  {Monthus}}\ and\ \bibinfo {author} {\bibfnamefont {T.}~\bibnamefont
  {Garel}},\ }\href@noop {} {\bibfield  {journal} {\bibinfo  {journal} {Phys.
  Rev. B}\ }\textbf {\bibinfo {volume} {81}},\ \bibinfo {pages} {134202}
  (\bibinfo {year} {2010})}\BibitemShut {NoStop}%
\bibitem [{\citenamefont {Kj{\"a}ll}\ \emph {et~al.}(2014)\citenamefont
  {Kj{\"a}ll}, \citenamefont {Bardarson},\ and\ \citenamefont
  {Pollmann}}]{kjall14}%
  \BibitemOpen
  \bibfield  {author} {\bibinfo {author} {\bibfnamefont {J.~A.}\ \bibnamefont
  {Kj{\"a}ll}}, \bibinfo {author} {\bibfnamefont {J.~H.}\ \bibnamefont
  {Bardarson}}, \ and\ \bibinfo {author} {\bibfnamefont {F.}~\bibnamefont
  {Pollmann}},\ }\href@noop {} {\bibfield  {journal} {\bibinfo  {journal}
  {Physical Review Letters}\ }\textbf {\bibinfo {volume} {113}},\ \bibinfo
  {pages} {107204} (\bibinfo {year} {2014})}\BibitemShut {NoStop}%
\bibitem [{\citenamefont {Gopalakrishnan}\ and\ \citenamefont
  {Nandkishore}(2014)}]{gopalakrishnan14}%
  \BibitemOpen
  \bibfield  {author} {\bibinfo {author} {\bibfnamefont {S.}~\bibnamefont
  {Gopalakrishnan}}\ and\ \bibinfo {author} {\bibfnamefont {R.}~\bibnamefont
  {Nandkishore}},\ }\href@noop {} {\bibfield  {journal} {\bibinfo  {journal}
  {Phys. Rev. B}\ }\textbf {\bibinfo {volume} {90}},\ \bibinfo {pages} {224203}
  (\bibinfo {year} {2014})}\BibitemShut {NoStop}%
\bibitem [{\citenamefont {Luitz}\ \emph {et~al.}(2015)\citenamefont {Luitz},
  \citenamefont {Laflorencie},\ and\ \citenamefont {Alet}}]{luitz15}%
  \BibitemOpen
  \bibfield  {author} {\bibinfo {author} {\bibfnamefont {D.~J.}\ \bibnamefont
  {Luitz}}, \bibinfo {author} {\bibfnamefont {N.}~\bibnamefont {Laflorencie}},
  \ and\ \bibinfo {author} {\bibfnamefont {F.}~\bibnamefont {Alet}},\
  }\href@noop {} {\bibfield  {journal} {\bibinfo  {journal} {Phys. Rev. B}\
  }\textbf {\bibinfo {volume} {91}},\ \bibinfo {pages} {081103} (\bibinfo
  {year} {2015})}\BibitemShut {NoStop}%
\bibitem [{\citenamefont {Nandkishore}\ and\ \citenamefont
  {Huse}(2015)}]{nandkishore15}%
  \BibitemOpen
  \bibfield  {author} {\bibinfo {author} {\bibfnamefont {R.}~\bibnamefont
  {Nandkishore}}\ and\ \bibinfo {author} {\bibfnamefont {D.~A.}\ \bibnamefont
  {Huse}},\ }\href@noop {} {\bibfield  {journal} {\bibinfo  {journal} {Annu.
  Rev. Condens. Matter Phys.}\ }\textbf {\bibinfo {volume} {6}},\ \bibinfo
  {pages} {15} (\bibinfo {year} {2015})}\BibitemShut {NoStop}%
\bibitem [{\citenamefont {Karrasch}\ and\ \citenamefont
  {Moore}(2015)}]{karrasch15}%
  \BibitemOpen
  \bibfield  {author} {\bibinfo {author} {\bibfnamefont {C.}~\bibnamefont
  {Karrasch}}\ and\ \bibinfo {author} {\bibfnamefont {J.~E.}\ \bibnamefont
  {Moore}},\ }\href@noop {} {\bibfield  {journal} {\bibinfo  {journal} {Phys.
  Rev. B}\ }\textbf {\bibinfo {volume} {92}},\ \bibinfo {pages} {115108}
  (\bibinfo {year} {2015})}\BibitemShut {NoStop}%
\bibitem [{\citenamefont {Imbrie}\ \emph {et~al.}()\citenamefont {Imbrie},
  \citenamefont {Ros},\ and\ \citenamefont {Scardicchio}}]{imbrie16}%
  \BibitemOpen
  \bibfield  {author} {\bibinfo {author} {\bibfnamefont {J.~Z.}\ \bibnamefont
  {Imbrie}}, \bibinfo {author} {\bibfnamefont {V.}~\bibnamefont {Ros}}, \ and\
  \bibinfo {author} {\bibfnamefont {A.}~\bibnamefont {Scardicchio}},\
  }\href@noop {} {\bibinfo  {journal} {Annalen der Physik}\ }\BibitemShut
  {NoStop}%
\bibitem [{\citenamefont {Imbrie}(2016)}]{imbrie16a}%
  \BibitemOpen
\bibfield  {journal} {  }\bibfield  {author} {\bibinfo {author} {\bibfnamefont
  {J.~Z.}\ \bibnamefont {Imbrie}},\ }\href@noop {} {\bibfield  {journal}
  {\bibinfo  {journal} {Physical Review Letters}\ }\textbf {\bibinfo {volume}
  {117}},\ \bibinfo {pages} {027201} (\bibinfo {year} {2016})}\BibitemShut
  {NoStop}%
\bibitem [{\citenamefont {Gornyi}\ \emph
  {et~al.}(2017{\natexlab{a}})\citenamefont {Gornyi}, \citenamefont {Mirlin},
  \citenamefont {Polyakov},\ and\ \citenamefont {Burin}}]{gornyi-adp16}%
  \BibitemOpen
  \bibfield  {author} {\bibinfo {author} {\bibfnamefont {I.~V.}\ \bibnamefont
  {Gornyi}}, \bibinfo {author} {\bibfnamefont {A.~D.}\ \bibnamefont {Mirlin}},
  \bibinfo {author} {\bibfnamefont {D.~G.}\ \bibnamefont {Polyakov}}, \ and\
  \bibinfo {author} {\bibfnamefont {A.~L.}\ \bibnamefont {Burin}},\ }\href
  {\doibase 10.1002/andp.201600360} {\bibfield  {journal} {\bibinfo  {journal}
  {Annalen der Physik}\ }\textbf {\bibinfo {volume} {529}},\ \bibinfo {pages}
  {1600360} (\bibinfo {year} {2017}{\natexlab{a}})}\BibitemShut {NoStop}%
\bibitem [{\citenamefont {Abanin}\ and\ \citenamefont
  {Papi\'c}(2017)}]{Abanin2017a}%
  \BibitemOpen
  \bibfield  {author} {\bibinfo {author} {\bibfnamefont {D.~A.}\ \bibnamefont
  {Abanin}}\ and\ \bibinfo {author} {\bibfnamefont {Z.}~\bibnamefont
  {Papi\'c}},\ }\href {\doibase 10.1002/andp.201700169} {\bibfield  {journal}
  {\bibinfo  {journal} {Ann. Phys. (Berl.)}\ }\textbf {\bibinfo {volume}
  {529}},\ \bibinfo {pages} {1700169} (\bibinfo {year} {2017})}\BibitemShut
  {NoStop}%
\bibitem [{\citenamefont {Doggen}\ \emph {et~al.}(2018)\citenamefont {Doggen},
  \citenamefont {Schindler}, \citenamefont {Tikhonov}, \citenamefont {Mirlin},
  \citenamefont {Neupert}, \citenamefont {Polyakov},\ and\ \citenamefont
  {Gornyi}}]{Doggen2018a}%
  \BibitemOpen
  \bibfield  {author} {\bibinfo {author} {\bibfnamefont {E.~V.~H.}\
  \bibnamefont {Doggen}}, \bibinfo {author} {\bibfnamefont {F.}~\bibnamefont
  {Schindler}}, \bibinfo {author} {\bibfnamefont {K.~S.}\ \bibnamefont
  {Tikhonov}}, \bibinfo {author} {\bibfnamefont {A.~D.}\ \bibnamefont
  {Mirlin}}, \bibinfo {author} {\bibfnamefont {T.}~\bibnamefont {Neupert}},
  \bibinfo {author} {\bibfnamefont {D.~G.}\ \bibnamefont {Polyakov}}, \ and\
  \bibinfo {author} {\bibfnamefont {I.~V.}\ \bibnamefont {Gornyi}},\ }\href
  {\doibase 10.1103/PhysRevB.98.174202} {\bibfield  {journal} {\bibinfo
  {journal} {Phys. Rev. B}\ }\textbf {\bibinfo {volume} {98}},\ \bibinfo
  {pages} {174202} (\bibinfo {year} {2018})}\BibitemShut {NoStop}%
\bibitem [{\citenamefont {Thiery}\ \emph {et~al.}(2018)\citenamefont {Thiery},
  \citenamefont {Huveneers}, \citenamefont {M\"uller},\ and\ \citenamefont
  {De~Roeck}}]{Thiery2017a}%
  \BibitemOpen
  \bibfield  {author} {\bibinfo {author} {\bibfnamefont {T.}~\bibnamefont
  {Thiery}}, \bibinfo {author} {\bibfnamefont {F.}~\bibnamefont {Huveneers}},
  \bibinfo {author} {\bibfnamefont {M.}~\bibnamefont {M\"uller}}, \ and\
  \bibinfo {author} {\bibfnamefont {W.}~\bibnamefont {De~Roeck}},\ }\href
  {\doibase 10.1103/PhysRevLett.121.140601} {\bibfield  {journal} {\bibinfo
  {journal} {Phys. Rev. Lett.}\ }\textbf {\bibinfo {volume} {121}},\ \bibinfo
  {pages} {140601} (\bibinfo {year} {2018})}\BibitemShut {NoStop}%
\bibitem [{\citenamefont {Alet}\ and\ \citenamefont
  {Laflorencie}(2018)}]{Alet2018a}%
  \BibitemOpen
  \bibfield  {author} {\bibinfo {author} {\bibfnamefont {F.}~\bibnamefont
  {Alet}}\ and\ \bibinfo {author} {\bibfnamefont {N.}~\bibnamefont
  {Laflorencie}},\ }\href {\doibase 10.1016/j.crhy.2018.03.003} {\bibfield
  {journal} {\bibinfo  {journal} {C. R. Phys.}\ }\textbf {\bibinfo {volume}
  {19}},\ \bibinfo {pages} {498} (\bibinfo {year} {2018})}\BibitemShut
  {NoStop}%
\bibitem [{\citenamefont {Dumitrescu}\ \emph {et~al.}(2019)\citenamefont
  {Dumitrescu}, \citenamefont {Goremykina}, \citenamefont {Parameswaran},
  \citenamefont {Serbyn},\ and\ \citenamefont {Vasseur}}]{Dumitrescu2019a}%
  \BibitemOpen
  \bibfield  {author} {\bibinfo {author} {\bibfnamefont {P.~T.}\ \bibnamefont
  {Dumitrescu}}, \bibinfo {author} {\bibfnamefont {A.}~\bibnamefont
  {Goremykina}}, \bibinfo {author} {\bibfnamefont {S.~A.}\ \bibnamefont
  {Parameswaran}}, \bibinfo {author} {\bibfnamefont {M.}~\bibnamefont
  {Serbyn}}, \ and\ \bibinfo {author} {\bibfnamefont {R.}~\bibnamefont
  {Vasseur}},\ }\href {\doibase 10.1103/PhysRevB.99.094205} {\bibfield
  {journal} {\bibinfo  {journal} {Phys. Rev. B}\ }\textbf {\bibinfo {volume}
  {99}},\ \bibinfo {pages} {094205} (\bibinfo {year} {2019})}\BibitemShut
  {NoStop}%
\bibitem [{\citenamefont {Goremykina}\ \emph {et~al.}(2019)\citenamefont
  {Goremykina}, \citenamefont {Vasseur},\ and\ \citenamefont
  {Serbyn}}]{Goremykina2019a}%
  \BibitemOpen
  \bibfield  {author} {\bibinfo {author} {\bibfnamefont {A.}~\bibnamefont
  {Goremykina}}, \bibinfo {author} {\bibfnamefont {R.}~\bibnamefont {Vasseur}},
  \ and\ \bibinfo {author} {\bibfnamefont {M.}~\bibnamefont {Serbyn}},\ }\href
  {\doibase 10.1103/PhysRevLett.122.040601} {\bibfield  {journal} {\bibinfo
  {journal} {Phys. Rev. Lett.}\ }\textbf {\bibinfo {volume} {122}},\ \bibinfo
  {pages} {040601} (\bibinfo {year} {2019})}\BibitemShut {NoStop}%
\bibitem [{\citenamefont {Burin}(2006)}]{burin06}%
  \BibitemOpen
  \bibfield  {author} {\bibinfo {author} {\bibfnamefont {A.}~\bibnamefont
  {Burin}},\ }\href@noop {} {\bibfield  {journal} {\bibinfo  {journal} {arXiv:
  cond-mat/0611387}\ } (\bibinfo {year} {2006})}\BibitemShut {NoStop}%
\bibitem [{\citenamefont {Yao}\ \emph {et~al.}(2014)\citenamefont {Yao},
  \citenamefont {Laumann}, \citenamefont {Gopalakrishnan}, \citenamefont
  {Knap}, \citenamefont {Mueller}, \citenamefont {Demler},\ and\ \citenamefont
  {Lukin}}]{Demler14}%
  \BibitemOpen
  \bibfield  {author} {\bibinfo {author} {\bibfnamefont {N.~Y.}\ \bibnamefont
  {Yao}}, \bibinfo {author} {\bibfnamefont {C.~R.}\ \bibnamefont {Laumann}},
  \bibinfo {author} {\bibfnamefont {S.}~\bibnamefont {Gopalakrishnan}},
  \bibinfo {author} {\bibfnamefont {M.}~\bibnamefont {Knap}}, \bibinfo {author}
  {\bibfnamefont {M.}~\bibnamefont {Mueller}}, \bibinfo {author} {\bibfnamefont
  {E.~A.}\ \bibnamefont {Demler}}, \ and\ \bibinfo {author} {\bibfnamefont
  {M.~D.}\ \bibnamefont {Lukin}},\ }\href@noop {} {\bibfield  {journal}
  {\bibinfo  {journal} {Physical Review Letters}\ }\textbf {\bibinfo {volume}
  {113}},\ \bibinfo {pages} {243002} (\bibinfo {year} {2014})}\BibitemShut
  {NoStop}%
\bibitem [{\citenamefont {Burin}(2015)}]{Burin15}%
  \BibitemOpen
  \bibfield  {author} {\bibinfo {author} {\bibfnamefont {A.~L.}\ \bibnamefont
  {Burin}},\ }\href@noop {} {\bibfield  {journal} {\bibinfo  {journal} {Phys.
  Rev. B}\ }\textbf {\bibinfo {volume} {91}},\ \bibinfo {pages} {094202}
  (\bibinfo {year} {2015})}\BibitemShut {NoStop}%
\bibitem [{\citenamefont {Gutman}\ \emph {et~al.}(2016)\citenamefont {Gutman},
  \citenamefont {Protopopov}, \citenamefont {Burin}, \citenamefont {Gornyi},
  \citenamefont {Santos},\ and\ \citenamefont {Mirlin}}]{Gutman16}%
  \BibitemOpen
  \bibfield  {author} {\bibinfo {author} {\bibfnamefont {D.}~\bibnamefont
  {Gutman}}, \bibinfo {author} {\bibfnamefont {I.}~\bibnamefont {Protopopov}},
  \bibinfo {author} {\bibfnamefont {A.}~\bibnamefont {Burin}}, \bibinfo
  {author} {\bibfnamefont {I.}~\bibnamefont {Gornyi}}, \bibinfo {author}
  {\bibfnamefont {R.}~\bibnamefont {Santos}}, \ and\ \bibinfo {author}
  {\bibfnamefont {A.}~\bibnamefont {Mirlin}},\ }\href@noop {} {\bibfield
  {journal} {\bibinfo  {journal} {Phys. Rev. B}\ }\textbf {\bibinfo {volume}
  {93}},\ \bibinfo {pages} {245427} (\bibinfo {year} {2016})}\BibitemShut
  {NoStop}%
\bibitem [{\citenamefont {Tikhonov}\ and\ \citenamefont
  {Mirlin}(2018)}]{tikhonov18}%
  \BibitemOpen
  \bibfield  {author} {\bibinfo {author} {\bibfnamefont {K.~S.}\ \bibnamefont
  {Tikhonov}}\ and\ \bibinfo {author} {\bibfnamefont {A.~D.}\ \bibnamefont
  {Mirlin}},\ }\href {\doibase 10.1103/PhysRevB.97.214205} {\bibfield
  {journal} {\bibinfo  {journal} {Phys. Rev. B}\ }\textbf {\bibinfo {volume}
  {97}},\ \bibinfo {pages} {214205} (\bibinfo {year} {2018})}\BibitemShut
  {NoStop}%
\bibitem [{\citenamefont {Safavi-Naini}\ \emph {et~al.}(2019)\citenamefont
  {Safavi-Naini}, \citenamefont {Wall}, \citenamefont {Acevedo}, \citenamefont
  {Rey},\ and\ \citenamefont {Nandkishore}}]{safavi-naini19}%
  \BibitemOpen
  \bibfield  {author} {\bibinfo {author} {\bibfnamefont {A.}~\bibnamefont
  {Safavi-Naini}}, \bibinfo {author} {\bibfnamefont {M.~L.}\ \bibnamefont
  {Wall}}, \bibinfo {author} {\bibfnamefont {O.~L.}\ \bibnamefont {Acevedo}},
  \bibinfo {author} {\bibfnamefont {A.~M.}\ \bibnamefont {Rey}}, \ and\
  \bibinfo {author} {\bibfnamefont {R.~M.}\ \bibnamefont {Nandkishore}},\
  }\href {\doibase 10.1103/PhysRevA.99.033610} {\bibfield  {journal} {\bibinfo
  {journal} {Phys. Rev. A}\ }\textbf {\bibinfo {volume} {99}},\ \bibinfo
  {pages} {033610} (\bibinfo {year} {2019})}\BibitemShut {NoStop}%
\bibitem [{\citenamefont {Gopalakrishnan}\ and\ \citenamefont
  {Huse}(2019)}]{Gopalakrishnan2019a}%
  \BibitemOpen
  \bibfield  {author} {\bibinfo {author} {\bibfnamefont {S.}~\bibnamefont
  {Gopalakrishnan}}\ and\ \bibinfo {author} {\bibfnamefont {D.~A.}\
  \bibnamefont {Huse}},\ }\href {\doibase 10.1103/PhysRevB.99.134305}
  {\bibfield  {journal} {\bibinfo  {journal} {Phys. Rev. B}\ }\textbf {\bibinfo
  {volume} {99}},\ \bibinfo {pages} {134305} (\bibinfo {year}
  {2019})}\BibitemShut {NoStop}%
\bibitem [{\citenamefont {Doggen}\ \emph {et~al.}(2020)\citenamefont {Doggen},
  \citenamefont {Gornyi}, \citenamefont {Mirlin},\ and\ \citenamefont
  {Polyakov}}]{doggen2020slow}%
  \BibitemOpen
  \bibfield  {author} {\bibinfo {author} {\bibfnamefont {E.~V.~H.}\
  \bibnamefont {Doggen}}, \bibinfo {author} {\bibfnamefont {I.~V.}\
  \bibnamefont {Gornyi}}, \bibinfo {author} {\bibfnamefont {A.~D.}\
  \bibnamefont {Mirlin}}, \ and\ \bibinfo {author} {\bibfnamefont {D.~G.}\
  \bibnamefont {Polyakov}},\ }\href@noop {} {\enquote {\bibinfo {title} {Slow
  many-body delocalization beyond one dimension},}\ } (\bibinfo {year}
  {2020}),\ \Eprint {http://arxiv.org/abs/2002.07635} {arXiv:2002.07635
  [cond-mat.dis-nn]} \BibitemShut {NoStop}%
\bibitem [{\citenamefont {Nandkishore}(2014)}]{nandkishore14continuum}%
  \BibitemOpen
  \bibfield  {author} {\bibinfo {author} {\bibfnamefont {R.}~\bibnamefont
  {Nandkishore}},\ }\href {\doibase 10.1103/PhysRevB.90.184204} {\bibfield
  {journal} {\bibinfo  {journal} {Phys. Rev. B}\ }\textbf {\bibinfo {volume}
  {90}},\ \bibinfo {pages} {184204} (\bibinfo {year} {2014})}\BibitemShut
  {NoStop}%
\bibitem [{\citenamefont {Gornyi}\ \emph
  {et~al.}(2017{\natexlab{b}})\citenamefont {Gornyi}, \citenamefont {Mirlin},
  \citenamefont {Müller},\ and\ \citenamefont {Polyakov}}]{gornyi17continuum}%
  \BibitemOpen
  \bibfield  {author} {\bibinfo {author} {\bibfnamefont {I.~V.}\ \bibnamefont
  {Gornyi}}, \bibinfo {author} {\bibfnamefont {A.~D.}\ \bibnamefont {Mirlin}},
  \bibinfo {author} {\bibfnamefont {M.}~\bibnamefont {Müller}}, \ and\
  \bibinfo {author} {\bibfnamefont {D.~G.}\ \bibnamefont {Polyakov}},\ }\href
  {\doibase 10.1002/andp.201600365} {\bibfield  {journal} {\bibinfo  {journal}
  {Annalen der Physik}\ }\textbf {\bibinfo {volume} {529}},\ \bibinfo {pages}
  {1600365} (\bibinfo {year} {2017}{\natexlab{b}})}\BibitemShut {NoStop}%
\bibitem [{\citenamefont {Schreiber}\ \emph {et~al.}(2015)\citenamefont
  {Schreiber}, \citenamefont {Hodgman}, \citenamefont {Bordia}, \citenamefont
  {L{\"u}schen}, \citenamefont {Fischer}, \citenamefont {Vosk}, \citenamefont
  {Altman}, \citenamefont {Schneider},\ and\ \citenamefont
  {Bloch}}]{schreiber2015observation}%
  \BibitemOpen
  \bibfield  {author} {\bibinfo {author} {\bibfnamefont {M.}~\bibnamefont
  {Schreiber}}, \bibinfo {author} {\bibfnamefont {S.~S.}\ \bibnamefont
  {Hodgman}}, \bibinfo {author} {\bibfnamefont {P.}~\bibnamefont {Bordia}},
  \bibinfo {author} {\bibfnamefont {H.~P.}\ \bibnamefont {L{\"u}schen}},
  \bibinfo {author} {\bibfnamefont {M.~H.}\ \bibnamefont {Fischer}}, \bibinfo
  {author} {\bibfnamefont {R.}~\bibnamefont {Vosk}}, \bibinfo {author}
  {\bibfnamefont {E.}~\bibnamefont {Altman}}, \bibinfo {author} {\bibfnamefont
  {U.}~\bibnamefont {Schneider}}, \ and\ \bibinfo {author} {\bibfnamefont
  {I.}~\bibnamefont {Bloch}},\ }\href@noop {} {\bibfield  {journal} {\bibinfo
  {journal} {Science}\ }\textbf {\bibinfo {volume} {349}},\ \bibinfo {pages}
  {842} (\bibinfo {year} {2015})}\BibitemShut {NoStop}%
\bibitem [{\citenamefont {Choi}\ \emph {et~al.}(2016)\citenamefont {Choi},
  \citenamefont {Hild}, \citenamefont {Zeiher}, \citenamefont {Schau{\ss}},
  \citenamefont {Rubio-Abadal}, \citenamefont {Yefsah}, \citenamefont
  {Khemani}, \citenamefont {Huse}, \citenamefont {Bloch},\ and\ \citenamefont
  {Gross}}]{choi2016exploring}%
  \BibitemOpen
  \bibfield  {author} {\bibinfo {author} {\bibfnamefont {J.-y.}\ \bibnamefont
  {Choi}}, \bibinfo {author} {\bibfnamefont {S.}~\bibnamefont {Hild}}, \bibinfo
  {author} {\bibfnamefont {J.}~\bibnamefont {Zeiher}}, \bibinfo {author}
  {\bibfnamefont {P.}~\bibnamefont {Schau{\ss}}}, \bibinfo {author}
  {\bibfnamefont {A.}~\bibnamefont {Rubio-Abadal}}, \bibinfo {author}
  {\bibfnamefont {T.}~\bibnamefont {Yefsah}}, \bibinfo {author} {\bibfnamefont
  {V.}~\bibnamefont {Khemani}}, \bibinfo {author} {\bibfnamefont {D.~A.}\
  \bibnamefont {Huse}}, \bibinfo {author} {\bibfnamefont {I.}~\bibnamefont
  {Bloch}}, \ and\ \bibinfo {author} {\bibfnamefont {C.}~\bibnamefont
  {Gross}},\ }\href {\doibase 10.1126/science.aaf8834} {\bibfield  {journal}
  {\bibinfo  {journal} {Science}\ }\textbf {\bibinfo {volume} {352}},\ \bibinfo
  {pages} {1547} (\bibinfo {year} {2016})}\BibitemShut {NoStop}%
\bibitem [{\citenamefont {Kondov}\ \emph {et~al.}(2015)\citenamefont {Kondov},
  \citenamefont {McGehee}, \citenamefont {Xu},\ and\ \citenamefont
  {DeMarco}}]{kondov2015disorder}%
  \BibitemOpen
  \bibfield  {author} {\bibinfo {author} {\bibfnamefont {S.}~\bibnamefont
  {Kondov}}, \bibinfo {author} {\bibfnamefont {W.}~\bibnamefont {McGehee}},
  \bibinfo {author} {\bibfnamefont {W.}~\bibnamefont {Xu}}, \ and\ \bibinfo
  {author} {\bibfnamefont {B.}~\bibnamefont {DeMarco}},\ }\href@noop {}
  {\bibfield  {journal} {\bibinfo  {journal} {Physical Review Letters}\
  }\textbf {\bibinfo {volume} {114}},\ \bibinfo {pages} {083002} (\bibinfo
  {year} {2015})}\BibitemShut {NoStop}%
\bibitem [{\citenamefont {Smith}\ \emph {et~al.}(2016)\citenamefont {Smith},
  \citenamefont {Lee}, \citenamefont {Richerme}, \citenamefont {Neyenhuis},
  \citenamefont {Hess}, \citenamefont {Hauke}, \citenamefont {Heyl},
  \citenamefont {Huse}, \citenamefont {Monroe}, \citenamefont {Fang} \emph
  {et~al.}}]{smith2016many}%
  \BibitemOpen
  \bibfield  {author} {\bibinfo {author} {\bibfnamefont {J.}~\bibnamefont
  {Smith}}, \bibinfo {author} {\bibfnamefont {A.}~\bibnamefont {Lee}}, \bibinfo
  {author} {\bibfnamefont {P.}~\bibnamefont {Richerme}}, \bibinfo {author}
  {\bibfnamefont {B.}~\bibnamefont {Neyenhuis}}, \bibinfo {author}
  {\bibfnamefont {P.}~\bibnamefont {Hess}}, \bibinfo {author} {\bibfnamefont
  {P.}~\bibnamefont {Hauke}}, \bibinfo {author} {\bibfnamefont
  {M.}~\bibnamefont {Heyl}}, \bibinfo {author} {\bibfnamefont {D.}~\bibnamefont
  {Huse}}, \bibinfo {author} {\bibfnamefont {C.}~\bibnamefont {Monroe}},
  \bibinfo {author} {\bibfnamefont {C.}~\bibnamefont {Fang}},  \emph {et~al.},\
  }\href@noop {} {\bibfield  {journal} {\bibinfo  {journal} {Nature Physics}\
  }\textbf {\bibinfo {volume} {12}},\ \bibinfo {pages} {907} (\bibinfo {year}
  {2016})}\BibitemShut {NoStop}%
\bibitem [{\citenamefont {Bordia}\ \emph {et~al.}(2016)\citenamefont {Bordia},
  \citenamefont {L{\"u}schen}, \citenamefont {Hodgman}, \citenamefont
  {Schreiber}, \citenamefont {Bloch},\ and\ \citenamefont
  {Schneider}}]{bordia15}%
  \BibitemOpen
  \bibfield  {author} {\bibinfo {author} {\bibfnamefont {P.}~\bibnamefont
  {Bordia}}, \bibinfo {author} {\bibfnamefont {H.~P.}\ \bibnamefont
  {L{\"u}schen}}, \bibinfo {author} {\bibfnamefont {S.~S.}\ \bibnamefont
  {Hodgman}}, \bibinfo {author} {\bibfnamefont {M.}~\bibnamefont {Schreiber}},
  \bibinfo {author} {\bibfnamefont {I.}~\bibnamefont {Bloch}}, \ and\ \bibinfo
  {author} {\bibfnamefont {U.}~\bibnamefont {Schneider}},\ }\href@noop {}
  {\bibfield  {journal} {\bibinfo  {journal} {Physical Review Letters}\
  }\textbf {\bibinfo {volume} {116}},\ \bibinfo {pages} {140401} (\bibinfo
  {year} {2016})}\BibitemShut {NoStop}%
\bibitem [{\citenamefont {L{\"u}schen}\ \emph {et~al.}(2017)\citenamefont
  {L{\"u}schen}, \citenamefont {Bordia}, \citenamefont {Hodgman}, \citenamefont
  {Schreiber}, \citenamefont {Sarkar}, \citenamefont {Daley}, \citenamefont
  {Fischer}, \citenamefont {Altman}, \citenamefont {Bloch},\ and\ \citenamefont
  {Schneider}}]{lueschen16}%
  \BibitemOpen
  \bibfield  {author} {\bibinfo {author} {\bibfnamefont {H.~P.}\ \bibnamefont
  {L{\"u}schen}}, \bibinfo {author} {\bibfnamefont {P.}~\bibnamefont {Bordia}},
  \bibinfo {author} {\bibfnamefont {S.~S.}\ \bibnamefont {Hodgman}}, \bibinfo
  {author} {\bibfnamefont {M.}~\bibnamefont {Schreiber}}, \bibinfo {author}
  {\bibfnamefont {S.}~\bibnamefont {Sarkar}}, \bibinfo {author} {\bibfnamefont
  {A.~J.}\ \bibnamefont {Daley}}, \bibinfo {author} {\bibfnamefont {M.~H.}\
  \bibnamefont {Fischer}}, \bibinfo {author} {\bibfnamefont {E.}~\bibnamefont
  {Altman}}, \bibinfo {author} {\bibfnamefont {I.}~\bibnamefont {Bloch}}, \
  and\ \bibinfo {author} {\bibfnamefont {U.}~\bibnamefont {Schneider}},\
  }\href@noop {} {\bibfield  {journal} {\bibinfo  {journal} {Phys. Rev. X}\
  }\textbf {\bibinfo {volume} {7}},\ \bibinfo {pages} {011034} (\bibinfo {year}
  {2017})}\BibitemShut {NoStop}%
\bibitem [{\citenamefont {Zhang}\ \emph
  {et~al.}(2017{\natexlab{a}})\citenamefont {Zhang}, \citenamefont {Hess},
  \citenamefont {Kyprianidis}, \citenamefont {Becker}, \citenamefont {Lee},
  \citenamefont {Smith}, \citenamefont {Pagano}, \citenamefont {Potirniche},
  \citenamefont {Potter}, \citenamefont {Vishwanath} \emph {et~al.}}]{Zhang17}%
  \BibitemOpen
  \bibfield  {author} {\bibinfo {author} {\bibfnamefont {J.}~\bibnamefont
  {Zhang}}, \bibinfo {author} {\bibfnamefont {P.}~\bibnamefont {Hess}},
  \bibinfo {author} {\bibfnamefont {A.}~\bibnamefont {Kyprianidis}}, \bibinfo
  {author} {\bibfnamefont {P.}~\bibnamefont {Becker}}, \bibinfo {author}
  {\bibfnamefont {A.}~\bibnamefont {Lee}}, \bibinfo {author} {\bibfnamefont
  {J.}~\bibnamefont {Smith}}, \bibinfo {author} {\bibfnamefont
  {G.}~\bibnamefont {Pagano}}, \bibinfo {author} {\bibfnamefont {I.-D.}\
  \bibnamefont {Potirniche}}, \bibinfo {author} {\bibfnamefont {A.~C.}\
  \bibnamefont {Potter}}, \bibinfo {author} {\bibfnamefont {A.}~\bibnamefont
  {Vishwanath}},  \emph {et~al.},\ }\href@noop {} {\bibfield  {journal}
  {\bibinfo  {journal} {Nature}\ }\textbf {\bibinfo {volume} {543}},\ \bibinfo
  {pages} {217} (\bibinfo {year} {2017}{\natexlab{a}})}\BibitemShut {NoStop}%
\bibitem [{\citenamefont {Bordia}\ \emph {et~al.}(2017)\citenamefont {Bordia},
  \citenamefont {L\"uschen}, \citenamefont {Scherg}, \citenamefont
  {Gopalakrishnan}, \citenamefont {Knap}, \citenamefont {Schneider},\ and\
  \citenamefont {Bloch}}]{Bordia2017a}%
  \BibitemOpen
  \bibfield  {author} {\bibinfo {author} {\bibfnamefont {P.}~\bibnamefont
  {Bordia}}, \bibinfo {author} {\bibfnamefont {H.}~\bibnamefont {L\"uschen}},
  \bibinfo {author} {\bibfnamefont {S.}~\bibnamefont {Scherg}}, \bibinfo
  {author} {\bibfnamefont {S.}~\bibnamefont {Gopalakrishnan}}, \bibinfo
  {author} {\bibfnamefont {M.}~\bibnamefont {Knap}}, \bibinfo {author}
  {\bibfnamefont {U.}~\bibnamefont {Schneider}}, \ and\ \bibinfo {author}
  {\bibfnamefont {I.}~\bibnamefont {Bloch}},\ }\href {\doibase
  10.1103/PhysRevX.7.041047} {\bibfield  {journal} {\bibinfo  {journal} {Phys.
  Rev. X}\ }\textbf {\bibinfo {volume} {7}},\ \bibinfo {pages} {041047}
  (\bibinfo {year} {2017})}\BibitemShut {NoStop}%
\bibitem [{\citenamefont {Rubio-Abadal}\ \emph {et~al.}(2019)\citenamefont
  {Rubio-Abadal}, \citenamefont {Choi}, \citenamefont {Zeiher}, \citenamefont
  {Hollerith}, \citenamefont {Rui}, \citenamefont {Bloch},\ and\ \citenamefont
  {Gross}}]{Rubio-Abadal2019a}%
  \BibitemOpen
  \bibfield  {author} {\bibinfo {author} {\bibfnamefont {A.}~\bibnamefont
  {Rubio-Abadal}}, \bibinfo {author} {\bibfnamefont {J.-y.}\ \bibnamefont
  {Choi}}, \bibinfo {author} {\bibfnamefont {J.}~\bibnamefont {Zeiher}},
  \bibinfo {author} {\bibfnamefont {S.}~\bibnamefont {Hollerith}}, \bibinfo
  {author} {\bibfnamefont {J.}~\bibnamefont {Rui}}, \bibinfo {author}
  {\bibfnamefont {I.}~\bibnamefont {Bloch}}, \ and\ \bibinfo {author}
  {\bibfnamefont {C.}~\bibnamefont {Gross}},\ }\href {\doibase
  10.1103/PhysRevX.9.041014} {\bibfield  {journal} {\bibinfo  {journal} {Phys.
  Rev. X}\ }\textbf {\bibinfo {volume} {9}},\ \bibinfo {pages} {041014}
  (\bibinfo {year} {2019})}\BibitemShut {NoStop}%
\bibitem [{\citenamefont {Choi}\ \emph
  {et~al.}(2017{\natexlab{a}})\citenamefont {Choi}, \citenamefont {Choi},
  \citenamefont {Kucsko}, \citenamefont {Maurer}, \citenamefont {Shields},
  \citenamefont {Sumiya}, \citenamefont {Onoda}, \citenamefont {Isoya},
  \citenamefont {Demler}, \citenamefont {Jelezko} \emph {et~al.}}]{Choi17a}%
  \BibitemOpen
  \bibfield  {author} {\bibinfo {author} {\bibfnamefont {J.}~\bibnamefont
  {Choi}}, \bibinfo {author} {\bibfnamefont {S.}~\bibnamefont {Choi}}, \bibinfo
  {author} {\bibfnamefont {G.}~\bibnamefont {Kucsko}}, \bibinfo {author}
  {\bibfnamefont {P.~C.}\ \bibnamefont {Maurer}}, \bibinfo {author}
  {\bibfnamefont {B.~J.}\ \bibnamefont {Shields}}, \bibinfo {author}
  {\bibfnamefont {H.}~\bibnamefont {Sumiya}}, \bibinfo {author} {\bibfnamefont
  {S.}~\bibnamefont {Onoda}}, \bibinfo {author} {\bibfnamefont
  {J.}~\bibnamefont {Isoya}}, \bibinfo {author} {\bibfnamefont
  {E.}~\bibnamefont {Demler}}, \bibinfo {author} {\bibfnamefont
  {F.}~\bibnamefont {Jelezko}},  \emph {et~al.},\ }\href@noop {} {\bibfield
  {journal} {\bibinfo  {journal} {Physical Review Letters}\ }\textbf {\bibinfo
  {volume} {118}},\ \bibinfo {pages} {093601} (\bibinfo {year}
  {2017}{\natexlab{a}})}\BibitemShut {NoStop}%
\bibitem [{\citenamefont {Kucsko}\ \emph {et~al.}(2016)\citenamefont {Kucsko},
  \citenamefont {Choi}, \citenamefont {Choi}, \citenamefont {Maurer},
  \citenamefont {Sumiya}, \citenamefont {Onoda}, \citenamefont {Isoya},
  \citenamefont {Jelezko}, \citenamefont {Demler}, \citenamefont {Yao} \emph
  {et~al.}}]{Choi16}%
  \BibitemOpen
  \bibfield  {author} {\bibinfo {author} {\bibfnamefont {G.}~\bibnamefont
  {Kucsko}}, \bibinfo {author} {\bibfnamefont {S.}~\bibnamefont {Choi}},
  \bibinfo {author} {\bibfnamefont {J.}~\bibnamefont {Choi}}, \bibinfo {author}
  {\bibfnamefont {P.~C.}\ \bibnamefont {Maurer}}, \bibinfo {author}
  {\bibfnamefont {H.}~\bibnamefont {Sumiya}}, \bibinfo {author} {\bibfnamefont
  {S.}~\bibnamefont {Onoda}}, \bibinfo {author} {\bibfnamefont
  {J.}~\bibnamefont {Isoya}}, \bibinfo {author} {\bibfnamefont
  {F.}~\bibnamefont {Jelezko}}, \bibinfo {author} {\bibfnamefont
  {E.}~\bibnamefont {Demler}}, \bibinfo {author} {\bibfnamefont {N.~Y.}\
  \bibnamefont {Yao}},  \emph {et~al.},\ }\href@noop {} {\bibfield  {journal}
  {\bibinfo  {journal} {arXiv:1609.08216}\ } (\bibinfo {year}
  {2016})}\BibitemShut {NoStop}%
\bibitem [{\citenamefont {Choi}\ \emph
  {et~al.}(2017{\natexlab{b}})\citenamefont {Choi}, \citenamefont {Choi},
  \citenamefont {Landig}, \citenamefont {Kucsko}, \citenamefont {Zhou},
  \citenamefont {Isoya}, \citenamefont {Jelezko}, \citenamefont {Onoda},
  \citenamefont {Sumiya}, \citenamefont {Khemani} \emph
  {et~al.}}]{choi2017observation}%
  \BibitemOpen
  \bibfield  {author} {\bibinfo {author} {\bibfnamefont {S.}~\bibnamefont
  {Choi}}, \bibinfo {author} {\bibfnamefont {J.}~\bibnamefont {Choi}}, \bibinfo
  {author} {\bibfnamefont {R.}~\bibnamefont {Landig}}, \bibinfo {author}
  {\bibfnamefont {G.}~\bibnamefont {Kucsko}}, \bibinfo {author} {\bibfnamefont
  {H.}~\bibnamefont {Zhou}}, \bibinfo {author} {\bibfnamefont {J.}~\bibnamefont
  {Isoya}}, \bibinfo {author} {\bibfnamefont {F.}~\bibnamefont {Jelezko}},
  \bibinfo {author} {\bibfnamefont {S.}~\bibnamefont {Onoda}}, \bibinfo
  {author} {\bibfnamefont {H.}~\bibnamefont {Sumiya}}, \bibinfo {author}
  {\bibfnamefont {V.}~\bibnamefont {Khemani}},  \emph {et~al.},\ }\href@noop {}
  {\bibfield  {journal} {\bibinfo  {journal} {Nature}\ }\textbf {\bibinfo
  {volume} {543}},\ \bibinfo {pages} {221} (\bibinfo {year}
  {2017}{\natexlab{b}})}\BibitemShut {NoStop}%
\bibitem [{\citenamefont {Ho}\ \emph {et~al.}(2017)\citenamefont {Ho},
  \citenamefont {Choi}, \citenamefont {Lukin},\ and\ \citenamefont
  {Abanin}}]{Ho17}%
  \BibitemOpen
  \bibfield  {author} {\bibinfo {author} {\bibfnamefont {W.~W.}\ \bibnamefont
  {Ho}}, \bibinfo {author} {\bibfnamefont {S.}~\bibnamefont {Choi}}, \bibinfo
  {author} {\bibfnamefont {M.~D.}\ \bibnamefont {Lukin}}, \ and\ \bibinfo
  {author} {\bibfnamefont {D.~A.}\ \bibnamefont {Abanin}},\ }\href@noop {}
  {\bibfield  {journal} {\bibinfo  {journal} {Physical Review Letters}\
  }\textbf {\bibinfo {volume} {119}},\ \bibinfo {pages} {010602} (\bibinfo
  {year} {2017})}\BibitemShut {NoStop}%
\bibitem [{\citenamefont {Roushan}\ \emph {et~al.}(2017)\citenamefont
  {Roushan}, \citenamefont {Neill}, \citenamefont {Tangpanitanon},
  \citenamefont {Bastidas}, \citenamefont {Megrant}, \citenamefont {Barends},
  \citenamefont {Chen}, \citenamefont {Chen}, \citenamefont {Chiaro},
  \citenamefont {Dunsworth} \emph {et~al.}}]{roushan17}%
  \BibitemOpen
  \bibfield  {author} {\bibinfo {author} {\bibfnamefont {P.}~\bibnamefont
  {Roushan}}, \bibinfo {author} {\bibfnamefont {C.}~\bibnamefont {Neill}},
  \bibinfo {author} {\bibfnamefont {J.}~\bibnamefont {Tangpanitanon}}, \bibinfo
  {author} {\bibfnamefont {V.}~\bibnamefont {Bastidas}}, \bibinfo {author}
  {\bibfnamefont {A.}~\bibnamefont {Megrant}}, \bibinfo {author} {\bibfnamefont
  {R.}~\bibnamefont {Barends}}, \bibinfo {author} {\bibfnamefont
  {Y.}~\bibnamefont {Chen}}, \bibinfo {author} {\bibfnamefont {Z.}~\bibnamefont
  {Chen}}, \bibinfo {author} {\bibfnamefont {B.}~\bibnamefont {Chiaro}},
  \bibinfo {author} {\bibfnamefont {A.}~\bibnamefont {Dunsworth}},  \emph
  {et~al.},\ }\href@noop {} {\bibfield  {journal} {\bibinfo  {journal}
  {Science}\ }\textbf {\bibinfo {volume} {358}},\ \bibinfo {pages} {1175}
  (\bibinfo {year} {2017})}\BibitemShut {NoStop}%
\bibitem [{\citenamefont {Chiaro}\ \emph {et~al.}(2019)\citenamefont {Chiaro},
  \citenamefont {Neill}, \citenamefont {Bohrdt}, \citenamefont {Filippone},
  \citenamefont {Arute}, \citenamefont {Arya}, \citenamefont {Babbush},
  \citenamefont {Bacon}, \citenamefont {Bardin}, \citenamefont {Barends},
  \citenamefont {Boixo}, \citenamefont {Buell}, \citenamefont {Burkett},
  \citenamefont {Chen}, \citenamefont {Chen}, \citenamefont {Collins},
  \citenamefont {Dunsworth}, \citenamefont {Farhi}, \citenamefont {Fowler},
  \citenamefont {Foxen}, \citenamefont {Gidney}, \citenamefont {Giustina},
  \citenamefont {Harrigan}, \citenamefont {Huang}, \citenamefont {Isakov},
  \citenamefont {Jeffrey}, \citenamefont {Jiang}, \citenamefont {Kafri},
  \citenamefont {Kechedzhi}, \citenamefont {Kelly}, \citenamefont {Klimov},
  \citenamefont {Korotkov}, \citenamefont {Kostritsa}, \citenamefont
  {Landhuis}, \citenamefont {Lucero}, \citenamefont {McClean}, \citenamefont
  {Mi}, \citenamefont {Megrant}, \citenamefont {Mohseni}, \citenamefont
  {Mutus}, \citenamefont {McEwen}, \citenamefont {Naaman}, \citenamefont
  {Neeley}, \citenamefont {Niu}, \citenamefont {Petukhov}, \citenamefont
  {Quintana}, \citenamefont {Rubin}, \citenamefont {Sank}, \citenamefont
  {Satzinger}, \citenamefont {Vainsencher}, \citenamefont {White},
  \citenamefont {Yao}, \citenamefont {Yeh}, \citenamefont {Zalcman},
  \citenamefont {Smelyanskiy}, \citenamefont {Neven}, \citenamefont
  {Gopalakrishnan}, \citenamefont {Abanin}, \citenamefont {Knap}, \citenamefont
  {Martinis},\ and\ \citenamefont {Roushan}}]{Chiaro2019a}%
  \BibitemOpen
  \bibfield  {author} {\bibinfo {author} {\bibfnamefont {B.}~\bibnamefont
  {Chiaro}}, \bibinfo {author} {\bibfnamefont {C.}~\bibnamefont {Neill}},
  \bibinfo {author} {\bibfnamefont {A.}~\bibnamefont {Bohrdt}}, \bibinfo
  {author} {\bibfnamefont {M.}~\bibnamefont {Filippone}}, \bibinfo {author}
  {\bibfnamefont {F.}~\bibnamefont {Arute}}, \bibinfo {author} {\bibfnamefont
  {K.}~\bibnamefont {Arya}}, \bibinfo {author} {\bibfnamefont {R.}~\bibnamefont
  {Babbush}}, \bibinfo {author} {\bibfnamefont {D.}~\bibnamefont {Bacon}},
  \bibinfo {author} {\bibfnamefont {J.}~\bibnamefont {Bardin}}, \bibinfo
  {author} {\bibfnamefont {R.}~\bibnamefont {Barends}}, \bibinfo {author}
  {\bibfnamefont {S.}~\bibnamefont {Boixo}}, \bibinfo {author} {\bibfnamefont
  {D.}~\bibnamefont {Buell}}, \bibinfo {author} {\bibfnamefont
  {B.}~\bibnamefont {Burkett}}, \bibinfo {author} {\bibfnamefont
  {Y.}~\bibnamefont {Chen}}, \bibinfo {author} {\bibfnamefont {Z.}~\bibnamefont
  {Chen}}, \bibinfo {author} {\bibfnamefont {R.}~\bibnamefont {Collins}},
  \bibinfo {author} {\bibfnamefont {A.}~\bibnamefont {Dunsworth}}, \bibinfo
  {author} {\bibfnamefont {E.}~\bibnamefont {Farhi}}, \bibinfo {author}
  {\bibfnamefont {A.}~\bibnamefont {Fowler}}, \bibinfo {author} {\bibfnamefont
  {B.}~\bibnamefont {Foxen}}, \bibinfo {author} {\bibfnamefont
  {C.}~\bibnamefont {Gidney}}, \bibinfo {author} {\bibfnamefont
  {M.}~\bibnamefont {Giustina}}, \bibinfo {author} {\bibfnamefont
  {M.}~\bibnamefont {Harrigan}}, \bibinfo {author} {\bibfnamefont
  {T.}~\bibnamefont {Huang}}, \bibinfo {author} {\bibfnamefont
  {S.}~\bibnamefont {Isakov}}, \bibinfo {author} {\bibfnamefont
  {E.}~\bibnamefont {Jeffrey}}, \bibinfo {author} {\bibfnamefont
  {Z.}~\bibnamefont {Jiang}}, \bibinfo {author} {\bibfnamefont
  {D.}~\bibnamefont {Kafri}}, \bibinfo {author} {\bibfnamefont
  {K.}~\bibnamefont {Kechedzhi}}, \bibinfo {author} {\bibfnamefont
  {J.}~\bibnamefont {Kelly}}, \bibinfo {author} {\bibfnamefont
  {P.}~\bibnamefont {Klimov}}, \bibinfo {author} {\bibfnamefont
  {A.}~\bibnamefont {Korotkov}}, \bibinfo {author} {\bibfnamefont
  {F.}~\bibnamefont {Kostritsa}}, \bibinfo {author} {\bibfnamefont
  {D.}~\bibnamefont {Landhuis}}, \bibinfo {author} {\bibfnamefont
  {E.}~\bibnamefont {Lucero}}, \bibinfo {author} {\bibfnamefont
  {J.}~\bibnamefont {McClean}}, \bibinfo {author} {\bibfnamefont
  {X.}~\bibnamefont {Mi}}, \bibinfo {author} {\bibfnamefont {A.}~\bibnamefont
  {Megrant}}, \bibinfo {author} {\bibfnamefont {M.}~\bibnamefont {Mohseni}},
  \bibinfo {author} {\bibfnamefont {J.}~\bibnamefont {Mutus}}, \bibinfo
  {author} {\bibfnamefont {M.}~\bibnamefont {McEwen}}, \bibinfo {author}
  {\bibfnamefont {O.}~\bibnamefont {Naaman}}, \bibinfo {author} {\bibfnamefont
  {M.}~\bibnamefont {Neeley}}, \bibinfo {author} {\bibfnamefont
  {M.}~\bibnamefont {Niu}}, \bibinfo {author} {\bibfnamefont {A.}~\bibnamefont
  {Petukhov}}, \bibinfo {author} {\bibfnamefont {C.}~\bibnamefont {Quintana}},
  \bibinfo {author} {\bibfnamefont {N.}~\bibnamefont {Rubin}}, \bibinfo
  {author} {\bibfnamefont {D.}~\bibnamefont {Sank}}, \bibinfo {author}
  {\bibfnamefont {K.}~\bibnamefont {Satzinger}}, \bibinfo {author}
  {\bibfnamefont {A.}~\bibnamefont {Vainsencher}}, \bibinfo {author}
  {\bibfnamefont {T.}~\bibnamefont {White}}, \bibinfo {author} {\bibfnamefont
  {Z.}~\bibnamefont {Yao}}, \bibinfo {author} {\bibfnamefont {P.}~\bibnamefont
  {Yeh}}, \bibinfo {author} {\bibfnamefont {A.}~\bibnamefont {Zalcman}},
  \bibinfo {author} {\bibfnamefont {V.}~\bibnamefont {Smelyanskiy}}, \bibinfo
  {author} {\bibfnamefont {H.}~\bibnamefont {Neven}}, \bibinfo {author}
  {\bibfnamefont {S.}~\bibnamefont {Gopalakrishnan}}, \bibinfo {author}
  {\bibfnamefont {D.}~\bibnamefont {Abanin}}, \bibinfo {author} {\bibfnamefont
  {M.}~\bibnamefont {Knap}}, \bibinfo {author} {\bibfnamefont {J.}~\bibnamefont
  {Martinis}}, \ and\ \bibinfo {author} {\bibfnamefont {P.}~\bibnamefont
  {Roushan}},\ }\href@noop {} {\  (\bibinfo {year} {2019})},\ \Eprint
  {http://arxiv.org/abs/1910.06024} {arXiv:1910.06024 [cond-mat.dis-nn]}
  \BibitemShut {NoStop}%
\bibitem [{\citenamefont {Ovadyahu}(2012)}]{ovadyahu1}%
  \BibitemOpen
  \bibfield  {author} {\bibinfo {author} {\bibfnamefont {Z.}~\bibnamefont
  {Ovadyahu}},\ }\href@noop {} {\bibfield  {journal} {\bibinfo  {journal}
  {Physical Review Letters}\ }\textbf {\bibinfo {volume} {108}},\ \bibinfo
  {pages} {156602} (\bibinfo {year} {2012})}\BibitemShut {NoStop}%
\bibitem [{\citenamefont {Ovadyahu}(2015)}]{ovadyahu2}%
  \BibitemOpen
  \bibfield  {author} {\bibinfo {author} {\bibfnamefont {Z.}~\bibnamefont
  {Ovadyahu}},\ }\href@noop {} {\bibfield  {journal} {\bibinfo  {journal}
  {Phys. Rev. B}\ }\textbf {\bibinfo {volume} {91}},\ \bibinfo {pages} {035113}
  (\bibinfo {year} {2015})}\BibitemShut {NoStop}%
\bibitem [{\citenamefont {Ovadia}\ \emph {et~al.}(2015)\citenamefont {Ovadia},
  \citenamefont {Kalok}, \citenamefont {Tamir}, \citenamefont {Mitra},
  \citenamefont {Sac{\'e}p{\'e}},\ and\ \citenamefont
  {Shahar}}]{ovadia2015evidence}%
  \BibitemOpen
  \bibfield  {author} {\bibinfo {author} {\bibfnamefont {M.}~\bibnamefont
  {Ovadia}}, \bibinfo {author} {\bibfnamefont {D.}~\bibnamefont {Kalok}},
  \bibinfo {author} {\bibfnamefont {I.}~\bibnamefont {Tamir}}, \bibinfo
  {author} {\bibfnamefont {S.}~\bibnamefont {Mitra}}, \bibinfo {author}
  {\bibfnamefont {B.}~\bibnamefont {Sac{\'e}p{\'e}}}, \ and\ \bibinfo {author}
  {\bibfnamefont {D.}~\bibnamefont {Shahar}},\ }\href@noop {} {\bibfield
  {journal} {\bibinfo  {journal} {Scientific Reports}\ }\textbf {\bibinfo
  {volume} {5}} (\bibinfo {year} {2015})}\BibitemShut {NoStop}%
\bibitem [{\citenamefont {Anderson}(1958)}]{anderson58}%
  \BibitemOpen
  \bibfield  {author} {\bibinfo {author} {\bibfnamefont {P.~W.}\ \bibnamefont
  {Anderson}},\ }\href@noop {} {\bibfield  {journal} {\bibinfo  {journal}
  {Phys. Rev.}\ }\textbf {\bibinfo {volume} {109}},\ \bibinfo {pages} {1492}
  (\bibinfo {year} {1958})}\BibitemShut {NoStop}%
\bibitem [{\citenamefont {Evers}\ and\ \citenamefont {Mirlin}(2008)}]{evers08}%
  \BibitemOpen
  \bibfield  {author} {\bibinfo {author} {\bibfnamefont {F.}~\bibnamefont
  {Evers}}\ and\ \bibinfo {author} {\bibfnamefont {A.~D.}\ \bibnamefont
  {Mirlin}},\ }\href@noop {} {\bibfield  {journal} {\bibinfo  {journal}
  {Reviews of Modern Physics}\ }\textbf {\bibinfo {volume} {80}},\ \bibinfo
  {pages} {1355} (\bibinfo {year} {2008})}\BibitemShut {NoStop}%
\bibitem [{\citenamefont {Chalker}(1990)}]{chalker90scaling}%
  \BibitemOpen
  \bibfield  {author} {\bibinfo {author} {\bibfnamefont {J.}~\bibnamefont
  {Chalker}},\ }\href {\doibase https://doi.org/10.1016/0378-4371(90)90056-X}
  {\bibfield  {journal} {\bibinfo  {journal} {Physica A: Statistical Mechanics
  and its Applications}\ }\textbf {\bibinfo {volume} {167}},\ \bibinfo {pages}
  {253 } (\bibinfo {year} {1990})}\BibitemShut {NoStop}%
\bibitem [{\citenamefont {Mirlin}(2000)}]{mirlin00}%
  \BibitemOpen
  \bibfield  {author} {\bibinfo {author} {\bibfnamefont {A.~D.}\ \bibnamefont
  {Mirlin}},\ }\href@noop {} {\bibfield  {journal} {\bibinfo  {journal}
  {Physics Reports}\ }\textbf {\bibinfo {volume} {326}},\ \bibinfo {pages}
  {259} (\bibinfo {year} {2000})}\BibitemShut {NoStop}%
\bibitem [{\citenamefont {Cuevas}\ and\ \citenamefont
  {Kravtsov}(2007)}]{cuevas07multifrac}%
  \BibitemOpen
  \bibfield  {author} {\bibinfo {author} {\bibfnamefont {E.}~\bibnamefont
  {Cuevas}}\ and\ \bibinfo {author} {\bibfnamefont {V.~E.}\ \bibnamefont
  {Kravtsov}},\ }\href {\doibase 10.1103/PhysRevB.76.235119} {\bibfield
  {journal} {\bibinfo  {journal} {Phys. Rev. B}\ }\textbf {\bibinfo {volume}
  {76}},\ \bibinfo {pages} {235119} (\bibinfo {year} {2007})}\BibitemShut
  {NoStop}%
\bibitem [{\citenamefont {Serbyn}\ \emph {et~al.}(2017)\citenamefont {Serbyn},
  \citenamefont {Papi{\'c}},\ and\ \citenamefont
  {Abanin}}]{serbyn2017thouless}%
  \BibitemOpen
  \bibfield  {author} {\bibinfo {author} {\bibfnamefont {M.}~\bibnamefont
  {Serbyn}}, \bibinfo {author} {\bibfnamefont {Z.}~\bibnamefont {Papi{\'c}}}, \
  and\ \bibinfo {author} {\bibfnamefont {D.~A.}\ \bibnamefont {Abanin}},\
  }\href@noop {} {\bibfield  {journal} {\bibinfo  {journal} {Physical Review
  B}\ }\textbf {\bibinfo {volume} {96}},\ \bibinfo {pages} {104201} (\bibinfo
  {year} {2017})}\BibitemShut {NoStop}%
\bibitem [{\citenamefont {Altshuler}\ \emph {et~al.}(1997)\citenamefont
  {Altshuler}, \citenamefont {Gefen}, \citenamefont {Kamenev},\ and\
  \citenamefont {Levitov}}]{altshuler1997quasiparticle}%
  \BibitemOpen
  \bibfield  {author} {\bibinfo {author} {\bibfnamefont {B.~L.}\ \bibnamefont
  {Altshuler}}, \bibinfo {author} {\bibfnamefont {Y.}~\bibnamefont {Gefen}},
  \bibinfo {author} {\bibfnamefont {A.}~\bibnamefont {Kamenev}}, \ and\
  \bibinfo {author} {\bibfnamefont {L.~S.}\ \bibnamefont {Levitov}},\
  }\href@noop {} {\bibfield  {journal} {\bibinfo  {journal} {Physical Review
  Letters}\ }\textbf {\bibinfo {volume} {78}},\ \bibinfo {pages} {2803}
  (\bibinfo {year} {1997})}\BibitemShut {NoStop}%
\bibitem [{\citenamefont {Biroli}\ \emph {et~al.}(2012)\citenamefont {Biroli},
  \citenamefont {Ribeiro-Teixeira},\ and\ \citenamefont
  {Tarzia}}]{biroli2012difference}%
  \BibitemOpen
  \bibfield  {author} {\bibinfo {author} {\bibfnamefont {G.}~\bibnamefont
  {Biroli}}, \bibinfo {author} {\bibfnamefont {A.}~\bibnamefont
  {Ribeiro-Teixeira}}, \ and\ \bibinfo {author} {\bibfnamefont
  {M.}~\bibnamefont {Tarzia}},\ }\href@noop {} {\bibfield  {journal} {\bibinfo
  {journal} {arXiv:1211.7334}\ } (\bibinfo {year} {2012})}\BibitemShut
  {NoStop}%
\bibitem [{\citenamefont {De~Luca}\ \emph {et~al.}(2014)\citenamefont
  {De~Luca}, \citenamefont {Altshuler}, \citenamefont {Kravtsov},\ and\
  \citenamefont {Scardicchio}}]{de2014anderson}%
  \BibitemOpen
  \bibfield  {author} {\bibinfo {author} {\bibfnamefont {A.}~\bibnamefont
  {De~Luca}}, \bibinfo {author} {\bibfnamefont {B.}~\bibnamefont {Altshuler}},
  \bibinfo {author} {\bibfnamefont {V.}~\bibnamefont {Kravtsov}}, \ and\
  \bibinfo {author} {\bibfnamefont {A.}~\bibnamefont {Scardicchio}},\
  }\href@noop {} {\bibfield  {journal} {\bibinfo  {journal} {Physical Review
  Letters}\ }\textbf {\bibinfo {volume} {113}},\ \bibinfo {pages} {046806}
  (\bibinfo {year} {2014})}\BibitemShut {NoStop}%
\bibitem [{\citenamefont {Tikhonov}\ \emph {et~al.}(2016)\citenamefont
  {Tikhonov}, \citenamefont {Mirlin},\ and\ \citenamefont
  {Skvortsov}}]{tikhonov2016anderson}%
  \BibitemOpen
  \bibfield  {author} {\bibinfo {author} {\bibfnamefont {K.}~\bibnamefont
  {Tikhonov}}, \bibinfo {author} {\bibfnamefont {A.}~\bibnamefont {Mirlin}}, \
  and\ \bibinfo {author} {\bibfnamefont {M.}~\bibnamefont {Skvortsov}},\
  }\href@noop {} {\bibfield  {journal} {\bibinfo  {journal} {Phys. Rev. B}\
  }\textbf {\bibinfo {volume} {94}},\ \bibinfo {pages} {220203} (\bibinfo
  {year} {2016})}\BibitemShut {NoStop}%
\bibitem [{\citenamefont {Garc{\'\i}a-Mata}\ \emph {et~al.}(2017)\citenamefont
  {Garc{\'\i}a-Mata}, \citenamefont {Giraud}, \citenamefont {Georgeot},
  \citenamefont {Martin}, \citenamefont {Dubertrand},\ and\ \citenamefont
  {Lemari{\'e}}}]{garcia-mata17}%
  \BibitemOpen
  \bibfield  {author} {\bibinfo {author} {\bibfnamefont {I.}~\bibnamefont
  {Garc{\'\i}a-Mata}}, \bibinfo {author} {\bibfnamefont {O.}~\bibnamefont
  {Giraud}}, \bibinfo {author} {\bibfnamefont {B.}~\bibnamefont {Georgeot}},
  \bibinfo {author} {\bibfnamefont {J.}~\bibnamefont {Martin}}, \bibinfo
  {author} {\bibfnamefont {R.}~\bibnamefont {Dubertrand}}, \ and\ \bibinfo
  {author} {\bibfnamefont {G.}~\bibnamefont {Lemari{\'e}}},\ }\href@noop {}
  {\bibfield  {journal} {\bibinfo  {journal} {Physical Review Letters}\
  }\textbf {\bibinfo {volume} {118}},\ \bibinfo {pages} {166801} (\bibinfo
  {year} {2017})}\BibitemShut {NoStop}%
\bibitem [{\citenamefont {Metz}\ and\ \citenamefont
  {Castillo}(2017{\natexlab{a}})}]{metz2017level}%
  \BibitemOpen
  \bibfield  {author} {\bibinfo {author} {\bibfnamefont {F.~L.}\ \bibnamefont
  {Metz}}\ and\ \bibinfo {author} {\bibfnamefont {I.~P.}\ \bibnamefont
  {Castillo}},\ }\href@noop {} {\bibfield  {journal} {\bibinfo  {journal}
  {Phys. Rev. B}\ }\textbf {\bibinfo {volume} {96}},\ \bibinfo {pages} {064202}
  (\bibinfo {year} {2017}{\natexlab{a}})}\BibitemShut {NoStop}%
\bibitem [{\citenamefont {Biroli}\ and\ \citenamefont
  {Tarzia}(2017)}]{Biroli2017}%
  \BibitemOpen
  \bibfield  {author} {\bibinfo {author} {\bibfnamefont {G.}~\bibnamefont
  {Biroli}}\ and\ \bibinfo {author} {\bibfnamefont {M.}~\bibnamefont
  {Tarzia}},\ }\href@noop {} {\bibfield  {journal} {\bibinfo  {journal} {Phys.
  Rev. B}\ }\textbf {\bibinfo {volume} {96}},\ \bibinfo {pages} {201114}
  (\bibinfo {year} {2017})}\BibitemShut {NoStop}%
\bibitem [{\citenamefont {Kravtsov}\ \emph {et~al.}(2018)\citenamefont
  {Kravtsov}, \citenamefont {Altshuler},\ and\ \citenamefont
  {Ioffe}}]{kravtsov2018non}%
  \BibitemOpen
  \bibfield  {author} {\bibinfo {author} {\bibfnamefont {V.}~\bibnamefont
  {Kravtsov}}, \bibinfo {author} {\bibfnamefont {B.}~\bibnamefont {Altshuler}},
  \ and\ \bibinfo {author} {\bibfnamefont {L.}~\bibnamefont {Ioffe}},\
  }\href@noop {} {\bibfield  {journal} {\bibinfo  {journal} {Annals of
  Physics}\ }\textbf {\bibinfo {volume} {389}},\ \bibinfo {pages} {148}
  (\bibinfo {year} {2018})}\BibitemShut {NoStop}%
\bibitem [{\citenamefont {Biroli}\ and\ \citenamefont
  {Tarzia}(2018)}]{biroli2018}%
  \BibitemOpen
  \bibfield  {author} {\bibinfo {author} {\bibfnamefont {G.}~\bibnamefont
  {Biroli}}\ and\ \bibinfo {author} {\bibfnamefont {M.}~\bibnamefont
  {Tarzia}},\ }\href@noop {} {\bibfield  {journal} {\bibinfo  {journal}
  {arXiv:1810.07545}\ } (\bibinfo {year} {2018})}\BibitemShut {NoStop}%
\bibitem [{\citenamefont {Tikhonov}\ and\ \citenamefont
  {Mirlin}(2019{\natexlab{a}})}]{tikhonov19statistics}%
  \BibitemOpen
  \bibfield  {author} {\bibinfo {author} {\bibfnamefont {K.~S.}\ \bibnamefont
  {Tikhonov}}\ and\ \bibinfo {author} {\bibfnamefont {A.~D.}\ \bibnamefont
  {Mirlin}},\ }\href {\doibase 10.1103/PhysRevB.99.024202} {\bibfield
  {journal} {\bibinfo  {journal} {Phys. Rev. B}\ }\textbf {\bibinfo {volume}
  {99}},\ \bibinfo {pages} {024202} (\bibinfo {year}
  {2019}{\natexlab{a}})}\BibitemShut {NoStop}%
\bibitem [{\citenamefont {Tikhonov}\ and\ \citenamefont
  {Mirlin}(2019{\natexlab{b}})}]{tikhonov19critical}%
  \BibitemOpen
  \bibfield  {author} {\bibinfo {author} {\bibfnamefont {K.~S.}\ \bibnamefont
  {Tikhonov}}\ and\ \bibinfo {author} {\bibfnamefont {A.~D.}\ \bibnamefont
  {Mirlin}},\ }\href {\doibase 10.1103/PhysRevB.99.214202} {\bibfield
  {journal} {\bibinfo  {journal} {Phys. Rev. B}\ }\textbf {\bibinfo {volume}
  {99}},\ \bibinfo {pages} {214202} (\bibinfo {year}
  {2019}{\natexlab{b}})}\BibitemShut {NoStop}%
\bibitem [{\citenamefont {Garc\'{\i}a-Mata}\ \emph {et~al.}(2020)\citenamefont
  {Garc\'{\i}a-Mata}, \citenamefont {Martin}, \citenamefont {Dubertrand},
  \citenamefont {Giraud}, \citenamefont {Georgeot},\ and\ \citenamefont
  {Lemari\'e}}]{PhysRevResearch.2.012020}%
  \BibitemOpen
  \bibfield  {author} {\bibinfo {author} {\bibfnamefont {I.}~\bibnamefont
  {Garc\'{\i}a-Mata}}, \bibinfo {author} {\bibfnamefont {J.}~\bibnamefont
  {Martin}}, \bibinfo {author} {\bibfnamefont {R.}~\bibnamefont {Dubertrand}},
  \bibinfo {author} {\bibfnamefont {O.}~\bibnamefont {Giraud}}, \bibinfo
  {author} {\bibfnamefont {B.}~\bibnamefont {Georgeot}}, \ and\ \bibinfo
  {author} {\bibfnamefont {G.}~\bibnamefont {Lemari\'e}},\ }\href {\doibase
  10.1103/PhysRevResearch.2.012020} {\bibfield  {journal} {\bibinfo  {journal}
  {Phys. Rev. Research}\ }\textbf {\bibinfo {volume} {2}},\ \bibinfo {pages}
  {012020} (\bibinfo {year} {2020})}\BibitemShut {NoStop}%
\bibitem [{\citenamefont {Mac\'e}\ \emph {et~al.}(2019)\citenamefont {Mac\'e},
  \citenamefont {Alet},\ and\ \citenamefont
  {Laflorencie}}]{mace19multifractal}%
  \BibitemOpen
  \bibfield  {author} {\bibinfo {author} {\bibfnamefont {N.}~\bibnamefont
  {Mac\'e}}, \bibinfo {author} {\bibfnamefont {F.}~\bibnamefont {Alet}}, \ and\
  \bibinfo {author} {\bibfnamefont {N.}~\bibnamefont {Laflorencie}},\ }\href
  {\doibase 10.1103/PhysRevLett.123.180601} {\bibfield  {journal} {\bibinfo
  {journal} {Phys. Rev. Lett.}\ }\textbf {\bibinfo {volume} {123}},\ \bibinfo
  {pages} {180601} (\bibinfo {year} {2019})}\BibitemShut {NoStop}%
\bibitem [{\citenamefont {Mirlin}\ and\ \citenamefont
  {Fyodorov}(1991{\natexlab{a}})}]{mirlin1991universality}%
  \BibitemOpen
  \bibfield  {author} {\bibinfo {author} {\bibfnamefont {A.}~\bibnamefont
  {Mirlin}}\ and\ \bibinfo {author} {\bibfnamefont {Y.~V.}\ \bibnamefont
  {Fyodorov}},\ }\href@noop {} {\bibfield  {journal} {\bibinfo  {journal}
  {Journal of Physics A: Mathematical and General}\ }\textbf {\bibinfo {volume}
  {24}},\ \bibinfo {pages} {2273} (\bibinfo {year}
  {1991}{\natexlab{a}})}\BibitemShut {NoStop}%
\bibitem [{\citenamefont {Fyodorov}\ and\ \citenamefont
  {Mirlin}(1991)}]{fyodorov1991localization}%
  \BibitemOpen
  \bibfield  {author} {\bibinfo {author} {\bibfnamefont {Y.~V.}\ \bibnamefont
  {Fyodorov}}\ and\ \bibinfo {author} {\bibfnamefont {A.~D.}\ \bibnamefont
  {Mirlin}},\ }\href@noop {} {\bibfield  {journal} {\bibinfo  {journal}
  {Physical Review Letters}\ }\textbf {\bibinfo {volume} {67}},\ \bibinfo
  {pages} {2049} (\bibinfo {year} {1991})}\BibitemShut {NoStop}%
\bibitem [{\citenamefont {Fyodorov}\ \emph {et~al.}(1992)\citenamefont
  {Fyodorov}, \citenamefont {Mirlin},\ and\ \citenamefont
  {Sommers}}]{fyodorov1992novel}%
  \BibitemOpen
  \bibfield  {author} {\bibinfo {author} {\bibfnamefont {Y.~V.}\ \bibnamefont
  {Fyodorov}}, \bibinfo {author} {\bibfnamefont {A.~D.}\ \bibnamefont
  {Mirlin}}, \ and\ \bibinfo {author} {\bibfnamefont {H.-J.}\ \bibnamefont
  {Sommers}},\ }\href@noop {} {\bibfield  {journal} {\bibinfo  {journal}
  {Journal de Physique I}\ }\textbf {\bibinfo {volume} {2}},\ \bibinfo {pages}
  {1571} (\bibinfo {year} {1992})}\BibitemShut {NoStop}%
\bibitem [{\citenamefont {Metz}\ and\ \citenamefont
  {Castillo}(2017{\natexlab{b}})}]{metz17}%
  \BibitemOpen
  \bibfield  {author} {\bibinfo {author} {\bibfnamefont {F.~L.}\ \bibnamefont
  {Metz}}\ and\ \bibinfo {author} {\bibfnamefont {I.~P.}\ \bibnamefont
  {Castillo}},\ }\href@noop {} {\bibfield  {journal} {\bibinfo  {journal}
  {Phys. Rev. B}\ }\textbf {\bibinfo {volume} {96}},\ \bibinfo {pages} {064202}
  (\bibinfo {year} {2017}{\natexlab{b}})}\BibitemShut {NoStop}%
\bibitem [{\citenamefont {Zhang}\ \emph
  {et~al.}(2017{\natexlab{b}})\citenamefont {Zhang}, \citenamefont {Pagano},
  \citenamefont {Hess}, \citenamefont {Kyprianidis}, \citenamefont {Becker},
  \citenamefont {Kaplan}, \citenamefont {Gorshkov}, \citenamefont {Gong},\ and\
  \citenamefont {Monroe}}]{zhang2017observation}%
  \BibitemOpen
  \bibfield  {author} {\bibinfo {author} {\bibfnamefont {J.}~\bibnamefont
  {Zhang}}, \bibinfo {author} {\bibfnamefont {G.}~\bibnamefont {Pagano}},
  \bibinfo {author} {\bibfnamefont {P.~W.}\ \bibnamefont {Hess}}, \bibinfo
  {author} {\bibfnamefont {A.}~\bibnamefont {Kyprianidis}}, \bibinfo {author}
  {\bibfnamefont {P.}~\bibnamefont {Becker}}, \bibinfo {author} {\bibfnamefont
  {H.}~\bibnamefont {Kaplan}}, \bibinfo {author} {\bibfnamefont {A.~V.}\
  \bibnamefont {Gorshkov}}, \bibinfo {author} {\bibfnamefont {Z.-X.}\
  \bibnamefont {Gong}}, \ and\ \bibinfo {author} {\bibfnamefont
  {C.}~\bibnamefont {Monroe}},\ }\href@noop {} {\bibfield  {journal} {\bibinfo
  {journal} {Nature}\ }\textbf {\bibinfo {volume} {551}},\ \bibinfo {pages}
  {601} (\bibinfo {year} {2017}{\natexlab{b}})}\BibitemShut {NoStop}%
\bibitem [{\citenamefont {Bernien}\ \emph {et~al.}(2017)\citenamefont
  {Bernien}, \citenamefont {Schwartz}, \citenamefont {Keesling}, \citenamefont
  {Levine}, \citenamefont {Omran}, \citenamefont {Pichler}, \citenamefont
  {Choi}, \citenamefont {Zibrov}, \citenamefont {Endres}, \citenamefont
  {Greiner} \emph {et~al.}}]{bernien2017probing}%
  \BibitemOpen
  \bibfield  {author} {\bibinfo {author} {\bibfnamefont {H.}~\bibnamefont
  {Bernien}}, \bibinfo {author} {\bibfnamefont {S.}~\bibnamefont {Schwartz}},
  \bibinfo {author} {\bibfnamefont {A.}~\bibnamefont {Keesling}}, \bibinfo
  {author} {\bibfnamefont {H.}~\bibnamefont {Levine}}, \bibinfo {author}
  {\bibfnamefont {A.}~\bibnamefont {Omran}}, \bibinfo {author} {\bibfnamefont
  {H.}~\bibnamefont {Pichler}}, \bibinfo {author} {\bibfnamefont
  {S.}~\bibnamefont {Choi}}, \bibinfo {author} {\bibfnamefont {A.~S.}\
  \bibnamefont {Zibrov}}, \bibinfo {author} {\bibfnamefont {M.}~\bibnamefont
  {Endres}}, \bibinfo {author} {\bibfnamefont {M.}~\bibnamefont {Greiner}},
  \emph {et~al.},\ }\href@noop {} {\bibfield  {journal} {\bibinfo  {journal}
  {Nature}\ }\textbf {\bibinfo {volume} {551}},\ \bibinfo {pages} {579}
  (\bibinfo {year} {2017})}\BibitemShut {NoStop}%
\bibitem [{\citenamefont {Arute}\ \emph {et~al.}(2019)\citenamefont {Arute},
  \citenamefont {Arya}, \citenamefont {Babbush}, \citenamefont {Bacon},
  \citenamefont {Bardin}, \citenamefont {Barends}, \citenamefont {Biswas},
  \citenamefont {Boixo}, \citenamefont {Brandao}, \citenamefont {Buell},
  \citenamefont {Burkett}, \citenamefont {Chen}, \citenamefont {Chen},
  \citenamefont {Chiaro}, \citenamefont {Collins}, \citenamefont {Courtney},
  \citenamefont {Dunsworth}, \citenamefont {Farhi}, \citenamefont {Foxen},
  \citenamefont {Fowler}, \citenamefont {Gidney}, \citenamefont {Giustina},
  \citenamefont {Graff}, \citenamefont {Guerin}, \citenamefont {Habegger},
  \citenamefont {Harrigan}, \citenamefont {Hartmann}, \citenamefont {Ho},
  \citenamefont {Hoffmann}, \citenamefont {Huang}, \citenamefont {Humble},
  \citenamefont {Isakov}, \citenamefont {Jeffrey}, \citenamefont {Jiang},
  \citenamefont {Kafri}, \citenamefont {Kechedzhi}, \citenamefont {Kelly},
  \citenamefont {Klimov}, \citenamefont {Knysh}, \citenamefont {Korotkov},
  \citenamefont {Kostritsa}, \citenamefont {Landhuis}, \citenamefont
  {Lindmark}, \citenamefont {Lucero}, \citenamefont {Lyakh}, \citenamefont
  {Mandr{\`a}}, \citenamefont {McClean}, \citenamefont {McEwen}, \citenamefont
  {Megrant}, \citenamefont {Mi}, \citenamefont {Michielsen}, \citenamefont
  {Mohseni}, \citenamefont {Mutus}, \citenamefont {Naaman}, \citenamefont
  {Neeley}, \citenamefont {Neill}, \citenamefont {Niu}, \citenamefont {Ostby},
  \citenamefont {Petukhov}, \citenamefont {Platt}, \citenamefont {Quintana},
  \citenamefont {Rieffel}, \citenamefont {Roushan}, \citenamefont {Rubin},
  \citenamefont {Sank}, \citenamefont {Satzinger}, \citenamefont {Smelyanskiy},
  \citenamefont {Sung}, \citenamefont {Trevithick}, \citenamefont
  {Vainsencher}, \citenamefont {Villalonga}, \citenamefont {White},
  \citenamefont {Yao}, \citenamefont {Yeh}, \citenamefont {Zalcman},
  \citenamefont {Neven},\ and\ \citenamefont
  {Martinis}}]{google_quantum_supremacy}%
  \BibitemOpen
  \bibfield  {author} {\bibinfo {author} {\bibfnamefont {F.}~\bibnamefont
  {Arute}}, \bibinfo {author} {\bibfnamefont {K.}~\bibnamefont {Arya}},
  \bibinfo {author} {\bibfnamefont {R.}~\bibnamefont {Babbush}}, \bibinfo
  {author} {\bibfnamefont {D.}~\bibnamefont {Bacon}}, \bibinfo {author}
  {\bibfnamefont {J.~C.}\ \bibnamefont {Bardin}}, \bibinfo {author}
  {\bibfnamefont {R.}~\bibnamefont {Barends}}, \bibinfo {author} {\bibfnamefont
  {R.}~\bibnamefont {Biswas}}, \bibinfo {author} {\bibfnamefont
  {S.}~\bibnamefont {Boixo}}, \bibinfo {author} {\bibfnamefont {F.~G. S.~L.}\
  \bibnamefont {Brandao}}, \bibinfo {author} {\bibfnamefont {D.~A.}\
  \bibnamefont {Buell}}, \bibinfo {author} {\bibfnamefont {B.}~\bibnamefont
  {Burkett}}, \bibinfo {author} {\bibfnamefont {Y.}~\bibnamefont {Chen}},
  \bibinfo {author} {\bibfnamefont {Z.}~\bibnamefont {Chen}}, \bibinfo {author}
  {\bibfnamefont {B.}~\bibnamefont {Chiaro}}, \bibinfo {author} {\bibfnamefont
  {R.}~\bibnamefont {Collins}}, \bibinfo {author} {\bibfnamefont
  {W.}~\bibnamefont {Courtney}}, \bibinfo {author} {\bibfnamefont
  {A.}~\bibnamefont {Dunsworth}}, \bibinfo {author} {\bibfnamefont
  {E.}~\bibnamefont {Farhi}}, \bibinfo {author} {\bibfnamefont
  {B.}~\bibnamefont {Foxen}}, \bibinfo {author} {\bibfnamefont
  {A.}~\bibnamefont {Fowler}}, \bibinfo {author} {\bibfnamefont
  {C.}~\bibnamefont {Gidney}}, \bibinfo {author} {\bibfnamefont
  {M.}~\bibnamefont {Giustina}}, \bibinfo {author} {\bibfnamefont
  {R.}~\bibnamefont {Graff}}, \bibinfo {author} {\bibfnamefont
  {K.}~\bibnamefont {Guerin}}, \bibinfo {author} {\bibfnamefont
  {S.}~\bibnamefont {Habegger}}, \bibinfo {author} {\bibfnamefont {M.~P.}\
  \bibnamefont {Harrigan}}, \bibinfo {author} {\bibfnamefont {M.~J.}\
  \bibnamefont {Hartmann}}, \bibinfo {author} {\bibfnamefont {A.}~\bibnamefont
  {Ho}}, \bibinfo {author} {\bibfnamefont {M.}~\bibnamefont {Hoffmann}},
  \bibinfo {author} {\bibfnamefont {T.}~\bibnamefont {Huang}}, \bibinfo
  {author} {\bibfnamefont {T.~S.}\ \bibnamefont {Humble}}, \bibinfo {author}
  {\bibfnamefont {S.~V.}\ \bibnamefont {Isakov}}, \bibinfo {author}
  {\bibfnamefont {E.}~\bibnamefont {Jeffrey}}, \bibinfo {author} {\bibfnamefont
  {Z.}~\bibnamefont {Jiang}}, \bibinfo {author} {\bibfnamefont
  {D.}~\bibnamefont {Kafri}}, \bibinfo {author} {\bibfnamefont
  {K.}~\bibnamefont {Kechedzhi}}, \bibinfo {author} {\bibfnamefont
  {J.}~\bibnamefont {Kelly}}, \bibinfo {author} {\bibfnamefont {P.~V.}\
  \bibnamefont {Klimov}}, \bibinfo {author} {\bibfnamefont {S.}~\bibnamefont
  {Knysh}}, \bibinfo {author} {\bibfnamefont {A.}~\bibnamefont {Korotkov}},
  \bibinfo {author} {\bibfnamefont {F.}~\bibnamefont {Kostritsa}}, \bibinfo
  {author} {\bibfnamefont {D.}~\bibnamefont {Landhuis}}, \bibinfo {author}
  {\bibfnamefont {M.}~\bibnamefont {Lindmark}}, \bibinfo {author}
  {\bibfnamefont {E.}~\bibnamefont {Lucero}}, \bibinfo {author} {\bibfnamefont
  {D.}~\bibnamefont {Lyakh}}, \bibinfo {author} {\bibfnamefont
  {S.}~\bibnamefont {Mandr{\`a}}}, \bibinfo {author} {\bibfnamefont {J.~R.}\
  \bibnamefont {McClean}}, \bibinfo {author} {\bibfnamefont {M.}~\bibnamefont
  {McEwen}}, \bibinfo {author} {\bibfnamefont {A.}~\bibnamefont {Megrant}},
  \bibinfo {author} {\bibfnamefont {X.}~\bibnamefont {Mi}}, \bibinfo {author}
  {\bibfnamefont {K.}~\bibnamefont {Michielsen}}, \bibinfo {author}
  {\bibfnamefont {M.}~\bibnamefont {Mohseni}}, \bibinfo {author} {\bibfnamefont
  {J.}~\bibnamefont {Mutus}}, \bibinfo {author} {\bibfnamefont
  {O.}~\bibnamefont {Naaman}}, \bibinfo {author} {\bibfnamefont
  {M.}~\bibnamefont {Neeley}}, \bibinfo {author} {\bibfnamefont
  {C.}~\bibnamefont {Neill}}, \bibinfo {author} {\bibfnamefont {M.~Y.}\
  \bibnamefont {Niu}}, \bibinfo {author} {\bibfnamefont {E.}~\bibnamefont
  {Ostby}}, \bibinfo {author} {\bibfnamefont {A.}~\bibnamefont {Petukhov}},
  \bibinfo {author} {\bibfnamefont {J.~C.}\ \bibnamefont {Platt}}, \bibinfo
  {author} {\bibfnamefont {C.}~\bibnamefont {Quintana}}, \bibinfo {author}
  {\bibfnamefont {E.~G.}\ \bibnamefont {Rieffel}}, \bibinfo {author}
  {\bibfnamefont {P.}~\bibnamefont {Roushan}}, \bibinfo {author} {\bibfnamefont
  {N.~C.}\ \bibnamefont {Rubin}}, \bibinfo {author} {\bibfnamefont
  {D.}~\bibnamefont {Sank}}, \bibinfo {author} {\bibfnamefont {K.~J.}\
  \bibnamefont {Satzinger}}, \bibinfo {author} {\bibfnamefont {V.}~\bibnamefont
  {Smelyanskiy}}, \bibinfo {author} {\bibfnamefont {K.~J.}\ \bibnamefont
  {Sung}}, \bibinfo {author} {\bibfnamefont {M.~D.}\ \bibnamefont
  {Trevithick}}, \bibinfo {author} {\bibfnamefont {A.}~\bibnamefont
  {Vainsencher}}, \bibinfo {author} {\bibfnamefont {B.}~\bibnamefont
  {Villalonga}}, \bibinfo {author} {\bibfnamefont {T.}~\bibnamefont {White}},
  \bibinfo {author} {\bibfnamefont {Z.~J.}\ \bibnamefont {Yao}}, \bibinfo
  {author} {\bibfnamefont {P.}~\bibnamefont {Yeh}}, \bibinfo {author}
  {\bibfnamefont {A.}~\bibnamefont {Zalcman}}, \bibinfo {author} {\bibfnamefont
  {H.}~\bibnamefont {Neven}}, \ and\ \bibinfo {author} {\bibfnamefont {J.~M.}\
  \bibnamefont {Martinis}},\ }\href {\doibase 10.1038/s41586-019-1666-5}
  {\bibfield  {journal} {\bibinfo  {journal} {Nature}\ }\textbf {\bibinfo
  {volume} {574}},\ \bibinfo {pages} {505} (\bibinfo {year}
  {2019})}\BibitemShut {NoStop}%
\bibitem [{SM()}]{SM}%
  \BibitemOpen
  \href@noop {} {}\bibinfo {note} {{See Supplemental Material at [URL will be
  inserted by publisher] for further details of the population-dynamics
  analysis and of the resonance counting in the localized phase.}}\BibitemShut
  {Stop}%
\bibitem [{\citenamefont {Biroli}\ \emph {et~al.}(2010)\citenamefont {Biroli},
  \citenamefont {Semerjian},\ and\ \citenamefont
  {Tarzia}}]{biroli2010anderson}%
  \BibitemOpen
  \bibfield  {author} {\bibinfo {author} {\bibfnamefont {G.}~\bibnamefont
  {Biroli}}, \bibinfo {author} {\bibfnamefont {G.}~\bibnamefont {Semerjian}}, \
  and\ \bibinfo {author} {\bibfnamefont {M.}~\bibnamefont {Tarzia}},\
  }\href@noop {} {\bibfield  {journal} {\bibinfo  {journal} {Progress of
  Theoretical Physics Supplement}\ }\textbf {\bibinfo {volume} {184}},\
  \bibinfo {pages} {187} (\bibinfo {year} {2010})}\BibitemShut {NoStop}%
\bibitem [{\citenamefont {Zirnbauer}(1986)}]{zirnbauer1986localization}%
  \BibitemOpen
  \bibfield  {author} {\bibinfo {author} {\bibfnamefont {M.~R.}\ \bibnamefont
  {Zirnbauer}},\ }\href@noop {} {\bibfield  {journal} {\bibinfo  {journal}
  {Phys. Rev. B}\ }\textbf {\bibinfo {volume} {34}},\ \bibinfo {pages} {6394}
  (\bibinfo {year} {1986})}\BibitemShut {NoStop}%
\bibitem [{\citenamefont {Mirlin}\ and\ \citenamefont
  {Fyodorov}(1991{\natexlab{b}})}]{mirlin1991localization}%
  \BibitemOpen
  \bibfield  {author} {\bibinfo {author} {\bibfnamefont {A.~D.}\ \bibnamefont
  {Mirlin}}\ and\ \bibinfo {author} {\bibfnamefont {Y.~V.}\ \bibnamefont
  {Fyodorov}},\ }\href@noop {} {\bibfield  {journal} {\bibinfo  {journal}
  {Nuclear Physics B}\ }\textbf {\bibinfo {volume} {366}},\ \bibinfo {pages}
  {507} (\bibinfo {year} {1991}{\natexlab{b}})}\BibitemShut {NoStop}%
\bibitem [{\citenamefont {Oganesyan}\ and\ \citenamefont
  {Huse}(2007{\natexlab{b}})}]{Oganesyan2007}%
  \BibitemOpen
  \bibfield  {author} {\bibinfo {author} {\bibfnamefont {V.}~\bibnamefont
  {Oganesyan}}\ and\ \bibinfo {author} {\bibfnamefont {D.~A.}\ \bibnamefont
  {Huse}},\ }\href {\doibase 10.1103/PhysRevB.75.155111} {\bibfield  {journal}
  {\bibinfo  {journal} {Phys. Rev. B}\ }\textbf {\bibinfo {volume} {75}},\
  \bibinfo {pages} {155111} (\bibinfo {year} {2007}{\natexlab{b}})}\BibitemShut
  {NoStop}%
\bibitem [{\citenamefont {Pal}\ and\ \citenamefont
  {Huse}(2010)}]{PhysRevB.82.174411}%
  \BibitemOpen
  \bibfield  {author} {\bibinfo {author} {\bibfnamefont {A.}~\bibnamefont
  {Pal}}\ and\ \bibinfo {author} {\bibfnamefont {D.~A.}\ \bibnamefont {Huse}},\
  }\href {\doibase 10.1103/PhysRevB.82.174411} {\bibfield  {journal} {\bibinfo
  {journal} {Phys. Rev. B}\ }\textbf {\bibinfo {volume} {82}},\ \bibinfo
  {pages} {174411} (\bibinfo {year} {2010})}\BibitemShut {NoStop}%
\bibitem [{\citenamefont {Abanin}\ \emph
  {et~al.}(2019{\natexlab{b}})\citenamefont {Abanin}, \citenamefont
  {Bardarson}, \citenamefont {De~Tomasi}, \citenamefont {Gopalakrishnan},
  \citenamefont {Khemani}, \citenamefont {Parameswaran}, \citenamefont
  {Pollmann}, \citenamefont {Potter}, \citenamefont {Serbyn},\ and\
  \citenamefont {Vasseur}}]{abanin2019distinguishing}%
  \BibitemOpen
  \bibfield  {author} {\bibinfo {author} {\bibfnamefont {D.}~\bibnamefont
  {Abanin}}, \bibinfo {author} {\bibfnamefont {J.}~\bibnamefont {Bardarson}},
  \bibinfo {author} {\bibfnamefont {G.}~\bibnamefont {De~Tomasi}}, \bibinfo
  {author} {\bibfnamefont {S.}~\bibnamefont {Gopalakrishnan}}, \bibinfo
  {author} {\bibfnamefont {V.}~\bibnamefont {Khemani}}, \bibinfo {author}
  {\bibfnamefont {S.}~\bibnamefont {Parameswaran}}, \bibinfo {author}
  {\bibfnamefont {F.}~\bibnamefont {Pollmann}}, \bibinfo {author}
  {\bibfnamefont {A.}~\bibnamefont {Potter}}, \bibinfo {author} {\bibfnamefont
  {M.}~\bibnamefont {Serbyn}}, \ and\ \bibinfo {author} {\bibfnamefont
  {R.}~\bibnamefont {Vasseur}},\ }\href@noop {} {\bibfield  {journal} {\bibinfo
   {journal} {arXiv preprint arXiv:1911.04501}\ } (\bibinfo {year}
  {2019}{\natexlab{b}})}\BibitemShut {NoStop}%
\bibitem [{\citenamefont {Luitz}\ \emph {et~al.}(2016)\citenamefont {Luitz},
  \citenamefont {Laflorencie},\ and\ \citenamefont {Alet}}]{Luitz2016a}%
  \BibitemOpen
  \bibfield  {author} {\bibinfo {author} {\bibfnamefont {D.~J.}\ \bibnamefont
  {Luitz}}, \bibinfo {author} {\bibfnamefont {N.}~\bibnamefont {Laflorencie}},
  \ and\ \bibinfo {author} {\bibfnamefont {F.}~\bibnamefont {Alet}},\ }\href
  {\doibase 10.1103/PhysRevB.93.060201} {\bibfield  {journal} {\bibinfo
  {journal} {Phys. Rev. B}\ }\textbf {\bibinfo {volume} {93}},\ \bibinfo
  {pages} {060201(R)} (\bibinfo {year} {2016})}\BibitemShut {NoStop}%
\bibitem [{\citenamefont {Gopalakrishnan}\ \emph {et~al.}(2015)\citenamefont
  {Gopalakrishnan}, \citenamefont {M{\"u}ller}, \citenamefont {Khemani},
  \citenamefont {Knap}, \citenamefont {Demler},\ and\ \citenamefont
  {Huse}}]{gopalakrishnan2015low}%
  \BibitemOpen
  \bibfield  {author} {\bibinfo {author} {\bibfnamefont {S.}~\bibnamefont
  {Gopalakrishnan}}, \bibinfo {author} {\bibfnamefont {M.}~\bibnamefont
  {M{\"u}ller}}, \bibinfo {author} {\bibfnamefont {V.}~\bibnamefont {Khemani}},
  \bibinfo {author} {\bibfnamefont {M.}~\bibnamefont {Knap}}, \bibinfo {author}
  {\bibfnamefont {E.}~\bibnamefont {Demler}}, \ and\ \bibinfo {author}
  {\bibfnamefont {D.~A.}\ \bibnamefont {Huse}},\ }\href@noop {} {\bibfield
  {journal} {\bibinfo  {journal} {Phys. Rev. B}\ }\textbf {\bibinfo {volume}
  {92}},\ \bibinfo {pages} {104202} (\bibinfo {year} {2015})}\BibitemShut
  {NoStop}%
\bibitem [{\citenamefont {Bera}\ \emph {et~al.}(2017)\citenamefont {Bera},
  \citenamefont {De~Tomasi}, \citenamefont {Weiner},\ and\ \citenamefont
  {Evers}}]{bera2017density}%
  \BibitemOpen
  \bibfield  {author} {\bibinfo {author} {\bibfnamefont {S.}~\bibnamefont
  {Bera}}, \bibinfo {author} {\bibfnamefont {G.}~\bibnamefont {De~Tomasi}},
  \bibinfo {author} {\bibfnamefont {F.}~\bibnamefont {Weiner}}, \ and\ \bibinfo
  {author} {\bibfnamefont {F.}~\bibnamefont {Evers}},\ }\href@noop {}
  {\bibfield  {journal} {\bibinfo  {journal} {Physical Review Letters}\
  }\textbf {\bibinfo {volume} {118}},\ \bibinfo {pages} {196801} (\bibinfo
  {year} {2017})}\BibitemShut {NoStop}%
\bibitem [{\citenamefont {Sierant}\ \emph {et~al.}(2020)\citenamefont
  {Sierant}, \citenamefont {Lewenstein},\ and\ \citenamefont
  {Zakrzewski}}]{sierant2020polynomially}%
  \BibitemOpen
  \bibfield  {author} {\bibinfo {author} {\bibfnamefont {P.}~\bibnamefont
  {Sierant}}, \bibinfo {author} {\bibfnamefont {M.}~\bibnamefont {Lewenstein}},
  \ and\ \bibinfo {author} {\bibfnamefont {J.}~\bibnamefont {Zakrzewski}},\
  }\href@noop {} {\enquote {\bibinfo {title} {Polynomially filtered exact
  diagonalization approach to many-body localization},}\ } (\bibinfo {year}
  {2020}),\ \Eprint {http://arxiv.org/abs/2005.09534} {arXiv:2005.09534
  [cond-mat.dis-nn]} \BibitemShut {NoStop}%
\bibitem [{\citenamefont {Sels}\ and\ \citenamefont
  {Polkovnikov}(2020)}]{sels2020dynamical}%
  \BibitemOpen
  \bibfield  {author} {\bibinfo {author} {\bibfnamefont {D.}~\bibnamefont
  {Sels}}\ and\ \bibinfo {author} {\bibfnamefont {A.}~\bibnamefont
  {Polkovnikov}},\ }\href@noop {} {\bibfield  {journal} {\bibinfo  {journal}
  {arXiv preprint arXiv:2009.04501}\ } (\bibinfo {year} {2020})}\BibitemShut
  {NoStop}%
\bibitem [{\citenamefont {Metz}\ and\ \citenamefont
  {P\'erez~Castillo}(2016)}]{PhysRevLett.117.104101}%
  \BibitemOpen
  \bibfield  {author} {\bibinfo {author} {\bibfnamefont {F.~L.}\ \bibnamefont
  {Metz}}\ and\ \bibinfo {author} {\bibfnamefont {I.}~\bibnamefont
  {P\'erez~Castillo}},\ }\href {\doibase 10.1103/PhysRevLett.117.104101}
  {\bibfield  {journal} {\bibinfo  {journal} {Phys. Rev. Lett.}\ }\textbf
  {\bibinfo {volume} {117}},\ \bibinfo {pages} {104101} (\bibinfo {year}
  {2016})}\BibitemShut {NoStop}%
\end{thebibliography}%
\newpage
\onecolumngrid

\begin{center}
{ \bf \Large Supplemental Material\\[0.2cm] to the article ``Eigenstate correlations around many-body localization transition''\\[0.2cm]  by K.S. Tikhonov and A.D. Mirlin}
\end{center}

\setcounter{figure}{0}
\makeatletter
\renewcommand{\thefigure}{S\@arabic\c@figure}
\makeatother

\vskip1cm

In this Supplemental Material, we provide some additional information to the analysis performed in the main part of the paper.

\section*{Population dynamics for correlation of eigenstates in the localized phase}

\begin{figure}[tbp]
\includegraphics[width=0.5\textwidth]{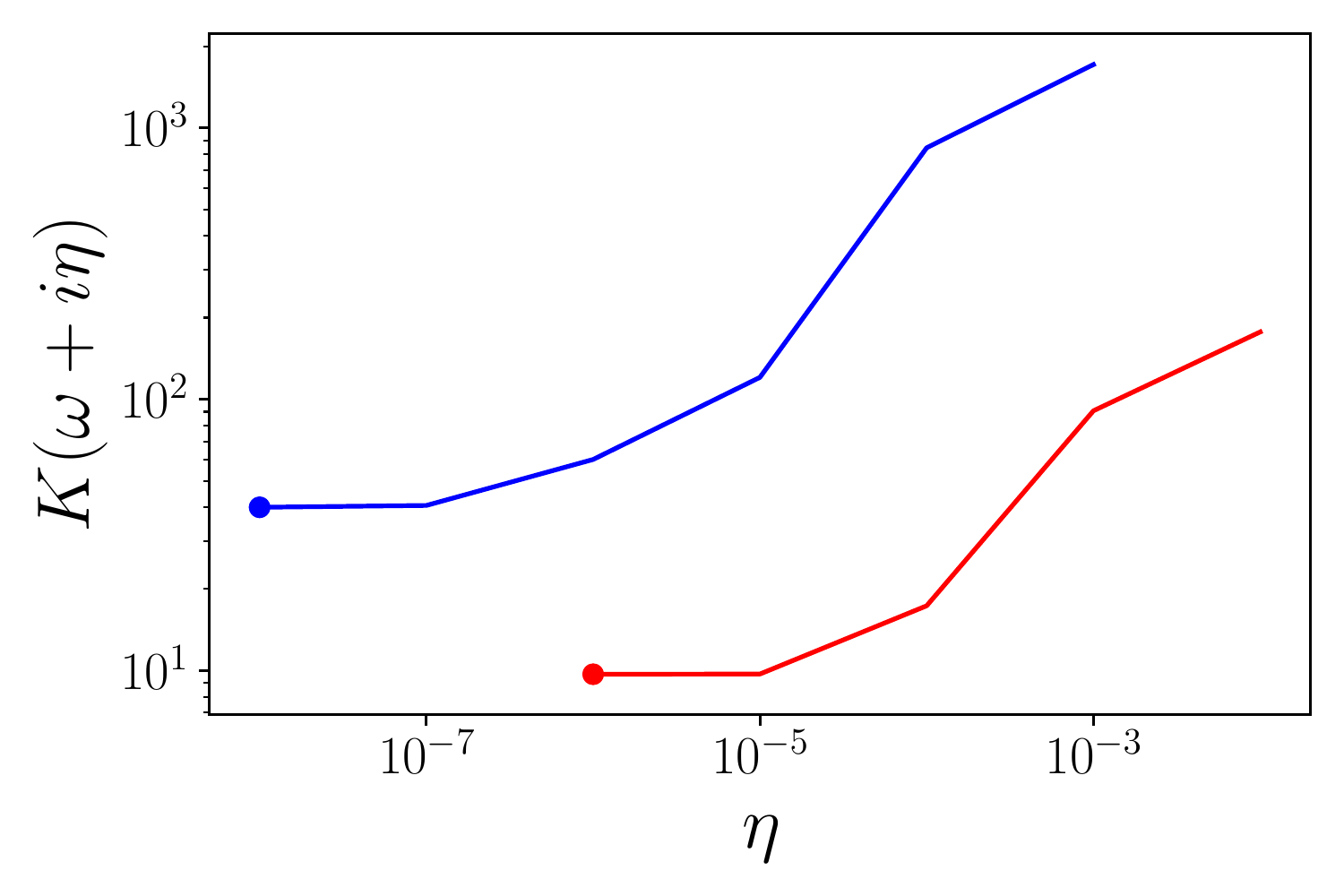}
\caption{Population dynamics results for $K(\omega+i\eta)$ at $W=24$ and frequencies $\omega=0.01$ (red) and $0.001$ (blue). The limiting $\eta \to 0$ values yields $N^2\beta(\omega)$, see Eq.~(\ref{betawithK}) of the main text. The corresponding results are shown by dots in Fig.~\ref{beta_rrg} of the main text. 
}
\label{rrg_pool_omegadep}
\end{figure}

As an alternative (to exact diagonalization) way to calculate $\beta(\omega)$ in the RRG model,  one can use the field-theoretical approach \cite{tikhonov19statistics}.
The correlation function $\beta(\omega)$ is given by Eqs. (\ref{Komega}), (\ref{rhoepsilon}), (\ref{betawithK}). To determine $K(\omega)$, one has to 
solve numerically the saddle-point equation to the effective action characterizing the problem. This equation is equivalent to the self-consistency equation for the joint distribution function of Green functions at two energies, $u=G_R(i,i,E+\omega/2)$ and $v=G_A(i,i,E-\omega/2)$, on the infinite Bethe lattice.  Here
$$
G_{R,A} (j,j,E) = \langle j | (E- {\cal H} \pm i\eta)^{-1} | j \rangle.
$$ 
The equations for the joint distribution function reads
\begin{eqnarray}
f^{(m)}(u,v) &=& \int d\epsilon \: \gamma(\epsilon)\int \left( \prod_{r=1}^{m} du_r \,  dv_r \, f^{(m)}(u_r,v_r)  \right) \nonumber \\
& \times & \delta\left[ u - \frac{1}{E+\frac{\omega}{2}+i\eta -\epsilon - \sum_{r=1}^{m} {u_r}} \right]    
\delta\left[ v - \frac{1}{E-\frac{\omega}{2}-i\eta -\epsilon  - \sum_{r=1}^{m}{v_r}} \right]  ;
\label{2tra}  \\
f^{(m+1)}(u,v) &=& \int d\epsilon \: \gamma(\epsilon) \int \left( \prod_{r=1}^{m+1} du_r \,  dv_r \, f^{(m)}(u_r,v_r) \right)  \nonumber  \\
& \times &  \delta\left[ u - \frac{1}{E+\frac{\omega}{2}+i\eta -\epsilon - \sum_{r=1}^{m+1} {u_r}} \right]    
\delta\left[ v - \frac{1}{E- \frac{\omega}{2}-i\eta -\epsilon  - \sum_{r=1}^{m+1}{v_r}} \right].
\label{1tra}
\end{eqnarray}
Equation (\ref{2tra}) is a self-consistency equation for the function $f^{(m)}(u,v)$, while Eq.~(\ref{1tra}) expresses the required distribution function $f^{(m+1)}(u,v)$ through $f^{(m)}(u,v)$.  The correlation function $K(\omega)$ is given by
\be
K(\omega) = - \frac{1}{\pi^2} \langle \Im u \: \Im v \rangle \,,
\ee
where the averaging is performed with the function $f^{(m+1)}(u,v)$.

Equations (\ref{2tra}), (\ref{1tra}) have been used to study the level number variance on RRG in  Refs.~\cite{PhysRevLett.117.104101,metz17} and to explore eigenstate correlations in the delocalized phase of the RRG model in Ref.~\cite{tikhonov19statistics}.

We have solved the self-consistency equations (\ref{2tra}), (\ref{1tra}) by using the pool size $M=2^{26}$ and the broadening $\eta$ from $\eta=10^{-2}$ down to $\eta=10^{-8}$. The results for the local-DOS correlation function $K(\omega)$ are shown in Fig. \ref{rrg_pool_omegadep}. The values for the correlation function $\beta(\omega)$ derived from the data in this figure are shown in Fig. \ref{beta_rrg} of the main text together with exact-diagonalization results.

\section*{Overlap of eigenfunctions in the localized phase of RRG via resonance counting}

In this section, we describe an analytical approach to evaluation of the frequency-dependent eigenfunction correlation function $\beta(\omega)$, see Eq. (\ref{sigmadef}), in the localized phase of the Anderson model (\ref{H}) on a RRG.  These correlations are expected to be controlled by Mott-type resonances between distant localized states.
In the main text of the paper, a simplified derivation of Eq.~(\ref{betafit}) is presented that discards strong fluctuations of eigenstates on tree-like graphs. Here we present a more accurate analysis that takes into account these fluctuations.

To calculate the probability of resonances [and, in this way, to evaluate the correlation function $\beta(\omega)$], we make use of a resonance counting approach in the spirit of Ref. \cite{altshuler1997quasiparticle}, which is expected to be valid in the strong--disorder regime $W\gg 1$. The starting point are eigenstates at $W \to \infty$ that are localized at individual sites of the graph.
Let us pick up a certain site $\ket{0}$ with local energy equal to $\epsilon_0$ (generation 0) and consider a tree formed by the graph around this node. Sites that are separated by distance $n$ from the site $\ket{0}$ form the generation $n$.  The number of sites in the generation $n$ grows as $m^n$. (We discard the loops, which is certainly justified as along as $n$ is smaller than the linear size $L \simeq \log_m N$ of the graph.)
 For a given site $\ket{j}$ of the generation $n+1$, let us compute the probability that this site is in resonance with the site $\ket{0}$ with resonance splitting $\sim\omega$. 
 The resonance conditions for the local energies and the \emph{effective} matrix element $M_{0j}$ reads $|\epsilon_0-\epsilon_j|\lesssim M_{0j}$. 
We focus on resonances with  $|\epsilon_0-\epsilon_j|\sim M_{0j}$, since they are obviously much more likely than those with $|\epsilon_0-\epsilon_j| \ll M_{0j}$ and thus will determine the disorder-averaged correlation function.
Therefore, the conditions for the resonance with splitting of order $\omega$ are as follows:
\be
|\epsilon_j-\epsilon_0|\in[\omega,2\omega]
\label{res-cond-1}
\ee
and
\be
|M_{0j}|\in[\omega,2\omega] \,.
\label{res-cond-2}
\ee
The matrix element $M_{0j}$ appears in the $n+1$--th order of perturbation theory over the hopping matrix element $V$ (in the main text of the paper we set $V=1$ but here we restore it for clarity): 
\be
M_{0j} = V\prod_{i=1}^{n}\frac{V}{\epsilon_0-\epsilon_i}\equiv V A_n,
\ee
where $\epsilon_1,...\epsilon_n$ are local energies encountered on the (unique) path from $\ket{0}$ to $\ket{j}$. The resonance probability equals 
$$P_n(\omega) = P^{(1)}_n(\omega) P^{(2)}_n(\omega) \,,$$ 
where probabilities $P_n^{(1)}(\omega)$ and $P_n^{(2)}(\omega)$ correspond to conditions (\ref{res-cond-1}) and (\ref{res-cond-2}), respectively. For the first of them, we clearly have 
\be
P^{(1)}_n(\omega) \sim \frac{\omega}{W}.
\ee
To find $P_n^{(2)}(\omega)$, we use Eq. (9) of Ref. \cite{altshuler1997quasiparticle} for the distribution of $|A_n|$: 
\be
\mathcal{P}(|A_n|)\sim\frac{\left[\ln |A_n|(W/V)^n \right]^{n-1}}{(W/V)^{n}(n-1)!|A_n|^2},
\ee
 valid for $|A_n|>(V/W)^n$. This inequality limits applicability of our analysis to $\omega>V(V/\omega)^n\equiv \omega_{\textrm{min}}$. This limitation is however not essential, since for $W>W_c$, one has $\omega_{\textrm{min}} \ll \Delta_n$, where $\Delta_n$ is the level spacing of a RRG with a linear size $n$. As a result, we find
\be
\quad P^{(2)}_n(\omega) = \int_{\omega/V}^{2\omega/V} d|A_n|\mathcal{P} (|A_n|)\sim \frac{\omega}{V}\mathcal{P} \left(|A_n|\sim\frac{\omega}{V} \right)\sim \frac{V}{\omega} \left(\frac{V}{W} \right)^n\frac{\ln^{n-1} \left [ (\omega/V)(W/V)^n \right]}{(n-1)!}.
\ee
Hence, the resonance probability equals 
\be
P_n(\omega)\sim   \left(\frac{V}{W} \right)^{n+1}\frac{\ln^{n-1} \left [ (\omega/V)(W/V)^n \right]}{(n-1)!}.
\ee
The average number of resonances that the initial state $\ket{0}$ encounters in the frequency interval $[\omega,2\omega]$ at distance $n+1$ is thus
\be
N_n(\omega) \sim m^n P_n(\omega) \sim \left[\frac{meV}{W}\left(\ln\frac{W}{V}+\frac{1}{n}\ln\frac{\omega}{V}\right)\right]^n \,,
\ee
where we have discarded a prefactor which does not scale with $n$. The total number of such resonances at all distances reads
\be
\label{sumres}
N_{\mathrm{tot}}^{\mathrm{res}}(\omega)=\sum_n N_n(\omega) = \sum_n \exp\left[-\frac{n}{\zeta}+n\ln(1-y/n)\right]
\ee
where $\zeta$ is the localization length given by
\be
\frac{1}{\zeta}=-\ln\left(\frac{emV}{W}\ln\frac{W}{V}\right)
\ee
and
\be
y=\frac{\ln(V/\omega)}{\ln(W/V)}.
\ee
In the relevant range of frequencies $\omega_{\mathrm{min}}<\omega<V$ we have $0<y<n$. The sum in Eq. (\ref{sumres}) is dominated by the vicinity of the stationary point
$n_*=y/x_*(\zeta)$ where $x_*(\zeta)$ is determined by equation
\be
f(x_*)=\zeta^{-1},\quad\textrm{with}\quad f(x)=\frac{x}{1-x}+\ln(1-x).
\label{eq-x-star}
\ee
Discarding the prefactor, we find the leading behavior
\be
\label{N_tot_res}
N_{\mathrm{tot}}^{\mathrm{res}}(\omega)\sim (\omega/V)^z
\ee
with
\be
z=\frac{1}{[1-x_*(\zeta)] \ln(c W/V)} \,.
\label{rrg-z}
\ee
In Eq.~(\ref{rrg-z}), we have restored a numerical constant $c$ (of order unity) under the argument of the logarithm. This constant was discarded in the approximation used above, so that its evaluation requires a more accurate analysis. 

The total number of states in the interval $[\omega,2\omega]$ on the RRG of volume $N \sim L^m$ equals
\be
\label{N_tot}
N_{\mathrm{tot}}(\omega)\sim\frac{\omega}{\Delta}\sim\frac{\omega N}{W}.
\ee
Combining Eqs. (\ref{N_tot_res}) and (\ref{N_tot}), we find that the probability that the fraction of the basis states $\ket{j}$ on RRG  in the range $[\omega,2\omega]$ that are in resonance with the state $\ket{0}$ equals
\be
p_{\mathrm{res}}(\omega)=\frac{N_{\mathrm{tot}}^{\mathrm{res}}(\omega)}{N_{\mathrm{tot}}(\omega)}\sim\left(\frac{\omega}{V}\right)^{-1+z}\frac{1}{N}.
\ee
If such a resonance takes place, the states $\ket{0}$ and $\ket{j}$ get strongly hybridized, and for two resulting states $\ket{\psi_1}$ and $\ket{\psi_2}$ we get the overlap
\be
\sum_{j'} |\psi_1(j')|^2 |\psi_2(j')|^2 \sim 1 \,,
\ee
where the sum goes over the RRG sites $j'$. 
We expect that the correlation function $\beta(\omega)$ is determined by such resonances, which implies
\be
N\beta(\omega)\sim p_{\mathrm{res}}(\omega)\sim\left(\frac{\omega}{V}\right)^{-1+z}\frac{1}{N} \,,
\ee
and finally
\be
N^2\beta(\omega) \sim(\omega/V)^{-\mu}
\ee
with 
\be
\mu = 1-z \,.
\label{rrg-mu-z}
\ee
Thus, we have derived Eq.~(\ref{betafit}) of the main text, which is the main goal of this section of the Supplemental Material. 

This computation can be trusted only as long as $n_*<L$ where $L$ stays for the linear size of the graph. One may check that this inequality implies that our approximation fails for $W \to W_c$ (when the localization length $\zeta$ is large). In the opposite limit, $W\gg W_c$, when $\zeta\ll 1$, the approximation is controllable. To determine the asymptotic behavior of the exponent $z$, Eq.~(\ref{rrg-z}), we inspect Eq.~(\ref{eq-x-star}) for small $\zeta$ and find
\be
\frac{1}{1-x_*} = \zeta^{-1} + \ln(\zeta^{-1}) + 1 + O \left(\frac{1}{\ln(\zeta^{-1})}\right) \,.
\ee
Substituting this in Eq.~(\ref{rrg-z}), we obtain
\be
z \approx \frac{\ln (W/V)-\ln m}{\ln ( c W/V)} \approx 1-\frac{\ln (c m)}{\ln W/V} \,.
\ee
Therefore, according to Eq.~(\ref{rrg-mu-z}), we have in the limit of strong disorder
\be
\mu\sim \frac{1}{\ln (W/V)} \,,
\label{mu-strong-disorder-suppl}
\ee
with proportionality constant of order unity (its exact value is beyond our accuracy). Note that this is the same logarithmic behavior that was found in the simplified approximation used in the main text, see Eq.~(\ref{mu-strong-disorder}). 

As we have pointed out above, this analysis is insufficient to determine the behavior of the exponent $\mu(W)$ in the limit $W \to W_c+0$.  We know, however, from the matching with the behavior at not too low frequencies on the delocalized side of the transition, see second line of  Eq.~\eqref{betascres}, that $\mu(W_c+0) = 1$.   This behavior has also emerged from the simplified approximation used in the main text, see Eq.~\eqref{mu-critical}.

\end{document}